\documentclass[iop]{emulateapj}

\newcommand\ione[2]{#1$\;${\sc{#2}}}%
\usepackage{epsfig}

\shorttitle{High Energy flares of FSRQs}
\begin{document}
\title{Exploring the blazar zone in High Energy flares of FSRQs}

\author{L. Pacciani\altaffilmark{1,3}, F. Tavecchio\altaffilmark{2}, I. Donnarumma\altaffilmark{1,3}, A. Stamerra\altaffilmark{4},
 L. Carrasco\altaffilmark{5}, E. Recillas\altaffilmark{5}, A. Porras\altaffilmark{5},
M.~Uemura\altaffilmark{6}}
\affil{$^{1}$INAF-Istituto di Astrofisica e Planetologia Spaziale, Via Fosso del Cavaliere 100, I-00133 Rome, Italy}
\affil{$^{2}$INAF-Osservatorio Astronomico di Brera, via E. Bianchi 46, I-23807, Merate, Italy}
\affil{$^{3}$INFN sez. Tor Vergata, Via della Ricerca Scientifica 1, I-00133 Rome, Italy}
\affil{$^{4}$INAF-Osservatorio Astrofisico di Torino, via P. Giuria 1, I-10125 Torino, Italy}
\affil{$^{5}$Instituto Nacional de Astrofisica, Optica y Electronica, Mexico, Luis E, Erro 1, Sta. Maria Tonantzintla, Puebla, CP 72840, Mexico}
\affil{$^{6}$Hiroshima Astrophysical Science Center, Hiroshima University  1-3-1 Kagamiyama, Higashi-Hiroshima, 739-8526, Japan}

\email{EMAIL: luigi.pacciani@iaps.inaf.it}

\keywords{ galaxies: active - galaxies: quasars: general - galaxies: Jets - radiation mechanism: non thermal}

\begin{abstract}

The gamma-ray emission offers a powerful diagnostic tool to probe jets and their surroundings in flat spectrum radio quasars (FSRQ).
In particular, sources emitting at high energies ($>10$ GeV) give us the strongest constraints. This motivates us to start a systematic study of flares with bright emission above 10 GeV, examining
archival data of {\it Fermi}-LAT gamma-ray telescope.
At the same time, we began to trigger Target of Opportunity observations to the {\it Swift} observatory at the
occurrence of high-energy flares, obtaining a wide coverage of the
spectral energy distributions for several FSRQs during flares. 
Among the others we investigate the SED of a peculiar flare of 3C 454.3, showing a remarkable hard
gamma-ray spectrum, quite different from the brightest flares of this source, and a bright flare of CTA 102.
We modeled the SED in the framework of the one--zone leptonic model, using also archival optical spectroscopic data
to derive the luminosity of the broad lines and thus estimate the disk luminosity, from which 
the structural parameters of the FSRQ nucleus can be inferred.\\
The model allowed us to evaluate the magnetic field intensity in the blazar zone,
and to locate the emitting region of gamma rays in the particular case in which gamma-ray spectra show
neither absorption from the BLR, nor the Klein-Nishina curvature expected in leptonic models assuming the BLR as source of seed photons for the External Compton. For FSRQs bright above 10 GeV, we where able
to identify short periods lasting less than 1 day characterized by high rate of high energy gamma rays, and hard gamma-ray spectra.\\
We discussed the observed spectra and variability timescales in terms of injection and  cooling of energetic particles, arguing that these flares could be triggered by magnetic reconnections events
or turbulence in the flow.
\end{abstract}

\section{Introduction}
In the framework of the unification scheme \citep{urry} of Active Galactic Nuclei (AGN), blazars are the radio--loud AGNs with jets oriented close to the line of sight of the observer.
Their emission encompasses the whole electromagnetic spectrum, from radio band to gamma-ray energies. A small fraction ($\sim$50 objects) have been detected at TeV energies with Cherenkov detectors (see, e.g., \citealt{holder2012}).\\
The Spectral Energy Distribution (SED) of blazars  shows two humps, whose origin is believed to be the boosted non-thermal emission from the relativistic jet, overwhelming the thermal components.
In general, the synchrotron emission from energetic electrons in a tangled magnetic field accounts for the low energy bump of the SED. The emission mechanisms responsible for
the high-energy bump, peaking in gamma-rays are still matter of debate. Both hadronic and leptonic models can explain the observations (e.g., \citealt{boettcher2013}). The high-energy emission is
explained in leptonic models as due to the Inverse--Compton (IC) scattering of relativistic electrons of
the jet with a seed photon field for both Bl Lac objects and for Flat Spectrum Radio Quasars (FSRQs).
Photon field for the inverse-Compton scattering can originate from the synchrotron emission itself (Synchrotron Self Compton, SSC,  \citealt{maraschi,marscher}).
The SED of BL Lac objects is usually explained with SC and SSC emissions.\\
The high energy bump of FSRQs is usually modeled in the
External Compton (EC) scenario, with photon fields for the IC
originating from a source external to the jet.  There are several sources of external photon fields that can play a role:
the direct thermal radiation from the disk, the reprocessed disk emission from the broad line region (BLR) or from the molecular torus \citep{Blazejowski2000,Arbeiter2002,sikora2002}, the thermal radiation from an hot corona \citep{sikora,dermer1993,ghise_tav}.

It is generally assumed (e.g., \citealt{ghise_tav}, \citealt{sikora2}) that the intensity of each external photon field depends on the distance of the emitting region from the Super Massive Black Hole (SMBH)
and on the accretion disk luminosity.
Particularly important,
these external radiation fields can also absorb the
gamma-ray photons, through the pair creation reaction $\gamma \gamma \rightarrow e^{\pm}$. In particular, the intense emission ( e.g., \citealt{poutanen}) from the BLR 
can partially absorb gamma rays at least for FSRQs with the most luminous accretion disk
(L$_{disk}$ $\sim$ 10$^{45}$ - 10$^{46}$ erg s$^{-1}$).
Above 20 GeV/(1+z) a spherical shell BLR is virtually opaque to gamma rays emitted in the center,
mainly due to luminous H Ly$\alpha$ and continuum emission of the BLR.
\cite{liu2006}  computed the BLR optical depth ($\tau_{\gamma\gamma}$) as a function of the location of the gamma-ray dissipation region, assuming a BLR luminosity of $L_{BLR}$=2.3$\times 10^{45}$ erg/s,
and a spherical shell geometry for the BLR with internal radius $R_{BLR}$, and external radius $R_{BLR}^{ext}$.
They evaluated $\tau_{\gamma\gamma} \sim 2.6$ ($\sim 13$)  at 35 GeV, and  $\sim$3.3 ($\sim 16$) at 50 GeV for gamma-ray photons emitted
at the mid point $R_{BLR}^M$ between internal and external radius (at the internal radius) of a BLR with luminosity $L_{BLR}=2.3\times10^{45}$ erg/s.
\cite{liu2008} also evaluated the optical opacity for the case of 3C 279 with a fainter BLR ($L_{BLR}=2.6\times10^{44}$ erg/s).
They obtained $\tau_{\gamma\gamma}$ $\sim$1 (7) at 35 GeV, and 0.6 (5) at 50 GeV for a gamma-ray emitting region at $R_{BLR}^M$ (at $R_{BLR}$) of the BLR shell.\\
In the standard view, high energy emission from
FSRQ has been located inside the BLR, whose intense
emission provides the ideal environment for
a powerful IC emission.
\cite{isler2013} and \cite{leontavares2013} found marginal evidence
that the gamma-emitting region is located within the BLR during at least
two flares of 3C 454.3.
However, there is growing
evidence that, at least in some occasions or in some
sources, the emission can occur much farther (up
to few pc) from the central SMBH.
The SED modeling of FSRQs flares bounds the dissipation region at the
edge, or outside the BLR in a few cases:
for PKS 1222+216 (dissipation region $>$ 0.1 pc, \citealt{aleksic,tavecchio1}),
3C 279 (dissipation region at 0.07--0.26 pc, \citealt{abdo2010,hayashida}),
PMN J2345-1555 (0.1 pc, \citealt{ghiseblue}),
PKS 1510-089 (0.07 -- 3.2 pc, \citealt{nalewajko}),
GB6 J1239+0443 (0.2 -- 7 pc, \citealt{pacciani}),
PKS B1424-418 (7 pc, \citealt{tavecchio2013}).\\
The study of the time-dependent polarimetric radio images at 43 and 86 GHz, the optical polarimetry, and the light curves from radio to gamma-rays
for the FSRQ 3C 454.3 \citep{jorstad,jorstad2013}, and for the BL Lac objects OJ287 and AO 0235+164 \citep{agudo1,agudo2}
suggest that the low and high energy emission is located close to the 43 GHz core, at a distance of the order of tens of parsec from the SMBH.
This localization of the blazar zone for 3C 454.3 is in contrast with the results of \cite{isler2013}. The gamma-ray emitting region for this FSRQ
is up for debate.\\
While the evidence for emission events at large distances seems to imply large sizes for the active
regions, there are clear indications that, at least in some occasions, the radiating region is very compact.
Fast variability with timescale of the order of several minutes has been detected at TeV energies for several Bl Lac objects (Bl Lac, Mrk 501, Mrk 421, PKS 2155-304),
and for the FSRQ PKS 1222+216. For the BL Lac objects, the observed variability time is too short when compared to the light crossing time of the horizon of the events \citep{begelman2008}, and for PKS 1222+216, the variability
timescale is much shorter than that expected from a jet active region radiating  at $>$ 0.1 pc from the SMBH.\\
To overcome this problem,
several mechanisms are proposed to produce fast flares:
\cite{bromberg} argues that radiative cooling of the shocked outflow layer can focus and lead to reconfinement of the ejecta. \cite{marscher2013} and \cite{narayan2012} proposed that rapid flares could be produced by turbulent cells in the relativistic plasma. Alternatively, the fragmentation of the reconnection layer  in magnetic reconnection events can produce a large number of plasmoids \citep{giannios2013}. The observable features of such events are the overlapped emission of these plasmoids with timescale of the order of a day (or longer), and the occasional growth of exceptional plasmoids generating the observed fast flares.\\
 It is clear that the observations of emission states in which the spectrum extends beyond the energies expected from the absorption is particular relevant to investigate flares occurring far from the SMBH, and to study their nature. With these motivations
we have started a dedicated program, selecting flares from FSRQs with relevant emission above 10 GeV in the archival--, and in incoming--data of {\it FERMI--LAT} \citep{atwood2009}.

\section{Sample selection}
\begin{figure}
\centering
\psfig{figure=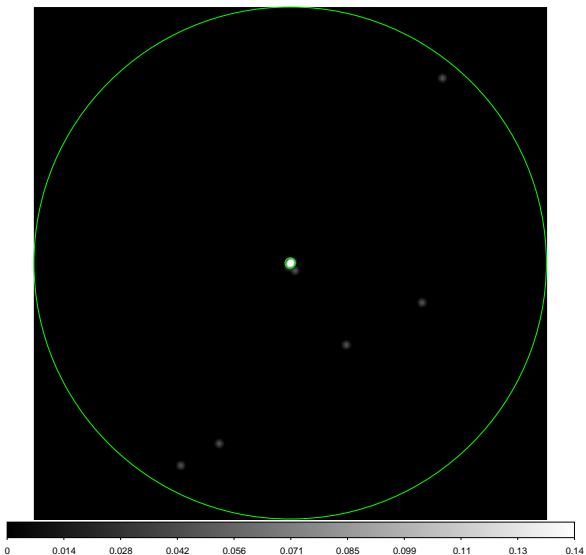,width=8cm}
\caption{Smoothed HE Counts map obtained collecting gamma-rays within a circular region of radius 20$^\circ$ around 3C 454.3 and integrating for 1.8 d during an HE flare.
The two circles have radius 0.4$^\circ$ and 20$^\circ$. In the map there are 12 HE gamma-rays, 6 of which lie within the 0.4$^\circ$ radius circular region.}
\label{fig_countshe}       
\end{figure}

\begin{table*}
 \centering
  \begin{tabular}{lllrrrrr}
  \hline
   FSRQ name    &  z   &    HE activity period       &  $\frac{\Delta t_{HE}}{1+z}$                & \# High        & chance                 & TS                     & Highest\\  
                &     &                              &      (d)          & Energy          &  prob.                 &(E$>$10 GeV)            & energy  \\ 
                &     &                              &                   & Gamma           &  (\%)$^*$              &                        & photons \\            
                &     &                              &                   & ray             &                        &                        & (GeV) \\  \hline              
PKS 0250-225    &  1.49   &  2009-02-12 11:00 -- 2009-03-07 20:00 &   9.4 &     3      &     0.36/8.4      &  26 & 18.4, 16.2 \\
PKS 0454-234    &  1.00   &  2012-11-24 09:00 -- 2012-12-13 12:00 &   9.6 &     5      &     0.055/1.8                    &  71 & 25.2, 19.3  \\
PKS 1502+106    &  1.84   &  2009-04-10 05:00 -- 2009-05-14 15:00 &  12.1 &     9      &     4.6$\times$10$^{-4}$/8.6$\times$10$^{-3}$      & 108  & 19.9, 15.7 \\
B2 1520+31      &  1.49   &  2009-04-10 14:00 -- 2009-04-27 02:00 &   6.6 &     4      &     0.18/6.4                     &  44  & 27.1, 16.0\\
4C +38.41       &  1.81   &  2011-07-02 10:00 -- 2011-07-13 09:00 &   3.9 &     3      &     0.031/1.7                                     &  34  & 13.9, 10.5\\
B2 1846+32A     &  0.80   &  2010-10-16 08:00 -- 2010-10-29 14:00 &   7.4 &     5      &     2.6$\times$10$^{-5}$/1.2$\times$10$^{-3}$      &  39  & 25.4, 23.5\\
PMN J2345-1555  &  0.62   &  2013-04-15 23:00 -- 2013-04-29 21:00 &   8.6 &     5      &     6.6$\times$10$^{-3}$/0.29                      &  68 & 96.8, 37.4 \\
CTA 102         &  1.49   &  2012-09-18 12:00 -- 2012-10-03 21:00 &   6.2 &     9      &     1.7$\times$10$^{-9}$/7.0$\times$10$^{-8}$      &  136 & 21.8, 20.1\\
PKS 0805-07     &  1.84   &  2009-05-14 13:00 -- 2009-05-23 14:00 &   3.2 &     9      &     2.3$\times$10$^{-9}$/1.6$\times$10$^{-7}$      &  112 & 23.2, 20.6\\
3C  454.3       &  0.86   &  2013-09-23 10:00 -- 2013-09-25 07:00 &   1.0 &     6      &     6.6$\times$10$^{-7}$/2.2$\times$10$^{-4}$      &  101 & 35.8, 28.4\\
\hline
\end{tabular}
\caption{List of FSRQs of our sample, together with the HE activity period, the duration of the activity period in the host galaxy frame, and the number of HE gamma rays (E$>$10 GeV) coming from a circle centered on the source position and of radius 0.4$^{\circ}$. $^*$ We give both the chance probability for a bunch of photons within the integration time, and the chance probability for the same bunch of photons within the integration time and occurring during an activity period at lower gamma-ray energy (for sake of simplicity, we assumed that all the sources were in an activity period for 1/3 of the whole {\it FERMI--LAT} operations, except for 3C 454.3 where we assumed the source in an active state for 1/2 of the whole {\it FERMI--LAT} operations).
TS significance is estimated through {\it FERMI--LAT} standard analysis recipes. Last column reports the energy of the most energetic photons during the HE activity period.}
\label{tab_he_sample}
\end{table*}
We searched for flares in the {\it FERMI}--{\it LAT} data archive
from all the FSRQs in the second {\it FERMI}--{\it LAT} catalog \citep{second_lat_cat}
with statistically relevant signal above 10 GeV.
Hereafter, we refer to these flares as High Energy (HE) flares.\\
We note that at high energy the point spread function (PSF) of the {\it FERMI--LAT} is particularly narrow
($\sim$ 0.3$^{\circ}$ at 10 GeV, 95\% containment, \citealt{lat-psf};
see also the updated PSF information reported in the {\it FERMI--LAT} web page\footnote{http://www.slac.stanford.edu/exp/glast/groups/canda/lat\_Performance.htm}),
and the background negligible.
A rough estimation for the order of magnitude of  the background rate in a circular region of radius of 0.4$^{\circ}$
is $\sim$1.1$\times$10$^{-3}$ cts/d ($\sim$2.9$\times$10$^{-10}$ph cm$^{-2}$ s$^{-1}$ E$>$ 10 GeV) at high latitude,
with the satellite scanning the whole sky.
Disregarding differences of exposures to different regions of the sky due the scanning strategy, we note that three gamma-rays detected within a few week
from a circular region of 0.4$^{\circ}$ around  the known position of a FSRQ give a TS \citep{mattox1996}  $\sim$ 25 in the range 10 GeV -- 300 GeV.
Signal significance TS is estimated through {\it FERMI--LAT} standard recipes.
A meaningful HE counts map showing the low background level, and the small PSF is shown in figure \ref{fig_countshe}.
It has been obtained collecting gamma-rays within a circular region of radius 20$^\circ$ around 3C 454.3 and integrating during an HE flare lasting $\sim$2 d.
There are 6 source counts within a circular region of radius 0.4$^\circ$, and 6 background gamma-rays within an annular region of internal radius 0.4$^\circ$ and external radius 20$^\circ$.\\
A simple and not CPU--time consuming method to search for HE flares is a two-level algorithm. The first-level algorithm
makes a rough analysis of the {\it FERMI--LAT} HE photon list,
disregarding differences of the exposures, and background.
The small PSF, and low background level allowed us to prepare the first--level of the flare search algorithm without the use of likelihood analysis, which is indeed necessary when the PSF is large
(i.e., at lower energy for the {\it FERMI--LAT}, and when each Gamma--ray could be associated with more than one source, or with diffuse--background \citep{mattox1996}.
The first--level algorithm will search for HE flares from each known source based on the differential time between consecutive HE gamma-rays detected in a circular region
of radius 0.4$^{\circ}$ around the source position, without taking care of the exposure, and of background.
We define the flaring activity period as the period of time in which a bunch of HE gamma-rays is detected by {\it FERMI--LAT} 
with differential times between consecutive gamma rays which is less than a
predefined quantity $\bar{\Delta t}$. $\bar{\Delta t}$ is defined to be $\frac{1}{3}$ of the mean differential time of HE
gamma rays from the source evaluated from the whole {\it FERMI--LAT} archive (i.e., we assert the first-level trigger for the period of time for which
the instantaneous photon rate from the source is at least three times the mean rate from the source).
We re-define the flaring activity period by adding to both the edges of it
half of the typical mean differential time between consecutive events during flare. We point out that this is only a rough pre-trigger, aimed at
accept as much HE flares as possible, with an high discrimination factor and with the minimum CPU usage.\\
The first-level trigger
generates 0.14 false first-level triggers/year for a source with a mean HE counting rate of 0.01 d$^{-1}$, and accepting a detection with 3 gamma-rays.\\
The second-level algorithm is based on the refined analysis of the {\it FERMI--LAT} data and obviously accounts for effective area, source exposure, satellite scanning strategy, and background:
it performs the standard analysis of the data within the already defined flaring activity period. The second-level runs only
at the occurrence of the first-level trigger, considerably reducing the CPU time needed to search for HE flares.
The second--level trigger is asserted once the source is detected with a TS $\ge$25 in the complementary {\it FERMI--LAT} energy band, 0.1--10 GeV.\\
This two-levels algorithm
has been applied to both archival--, and incoming--data, downloaded twice every day.
It provides a non-complete sample. In fact it selects for HE flares starting from their detectability,
which still depends on the instrument pointing strategy and Earth occultation.\\
We found HE flares from about 60 FSRQs, and for more than a dozen of sources we found archival multiwavelength observations during HE flares, or we awarded ToO observations with {\it Swift}
\citep{gehrels2004} during flares.
We investigate here HE flares of FSRQs with sufficient multiwavelength coverage and for which we were able to derive the disk luminosity of the source from BLR spectroscopy.\\
The source list, the flaring activity period, the number of HE gamma rays (E $> 10$ GeV) collected during the activity period, are reported in Table \ref{tab_he_sample},
together with the estimated chance probability for a bunch of gamma-rays to be detected within the reported integration time and simultaneous to an activity period at lower gamma-ray energies.
In table \ref{tab_he_sample} we report also the  the likelihood signal significance TS estimated through {\it FERMI--LAT} standard analysis recipes for E$>$ 10 GeV.
\\

\section{Multiwavelength observations}
In  Table \ref{tab_timeline} we report the timeline of the observatory campaigns for our sample of flares.
{\it Swift--UVOT} observations with all optical--UV filters were performed simultaneous to the {\it Swift--XRT} observations we report,
unless otherwise specified. All observations at Guillelmo Haro were performed with near IR K$_s$, H, J filters,
and all observations with SMARTS were performed with the optical--NIR filters B, V, R, J, K, unless otherwise specified.
Dates in Table \ref{tab_timeline} and  everywhere in the paper are reported in UTC. \\
We report also some peculiar cases:
for PKS 0805-07 the NIR observations where performed 10 days before the activity period in gamma-rays.
The results of a multifrequency monitoring of 4C +38.41
covering almost 4 years of activity from Radio to Gamma-rays, and including the activity period that we study in this paper, is reported in \cite{raiteri2012}.
\begin{table*}
 \centering
  \begin{tabular}{lll}
  \hline
source & X-ray observing period & optical--NIR observation/observatory \\ \hline
PKS 0250-225    & 2009-02-20 15:53 -- 2009-02-22 11:14   &  \\ 
PKS 0454-234    & 2012-12-04 12:25 -- 2012-12-04 19:30   & 2012-12-05 03:51/SMARTS  \\ 
PKS 1502+106    & 2009-04-09 12:47 -- 2009-04-09 15:02 $^{*}$   & 2009-04-09/KANATA (V, J)\\       
                &                                              & 2009-05-03 20:10/Guillermo Haro \\       
B2 1520+31      & 2009-04-24 13:09 -- 2009-04-25 02:13$^{**}$ & 2009-04-20 10:44/Catalina--survey (V) \\  
4C +38.41       & 2011-07-09 13:48 -- 2011-07-09 14:07   & 2011-07-10 16:48/Mount Maidanak (R) \\
                &                                              & \citep{atel3483} \\
B2 1846+32A     & 2010-10-27 20:48 -- 2010-10-27 22:44   & 2010-10-27 15:32/Guillermo Haro \\       
PMN J2345-1555  & 2013-04-29 14:59 -- 2013-04-29 18:26   & 2013-04-30 10:48/SMARTS  (R, J)\\
CTA 102         & 2012-09-25 10:18 -- 2012-09-25 10:38   & 2012-09-26 20:10/Guillermo Haro \\ 
PKS 0805-07     & 2009-05-18 20:53 -- 2009-05-18 21:27   & 2009-05-05 15:07/Guillermo Haro \\ 
3C  454.3       & 2013-09-23 21:12 -- 2013-09-23 21:29   & 2013-09-25 03:07/SMARTS     \\ \hline
\end{tabular}
\caption{Timeline of the multiwavelength campaigns on the sources. All X-ray observations where performed with Swift(except for ${^*}$ performed with Chandra). SMARTS observations where performed with B, V, R, J, K filters, unless otherwise specified; observations at Guillermo Haro where performed with K$_S$, H, J filters.  $^{**}$Swift observed with UV filters only.}
\label{tab_timeline}
\end{table*}

\section{data analysis}
\subsection{FERMI--LAT analysis}
We already stated that we searched for HE flares on FSRQs of the second {\it FERMI--LAT} catalog starting from the HE photon list.
The alert task has been described in the sample--selection section.
The filtered photon list used in the first--level trigger has been prepared with the latest standard
{\it Fermi} Science tools (v9r27p1 until November 2013, and v9r32p5 since this date) available at the time of trigger.
We filtered events of {\em event class} 2 
with a reconstructed energy $>$ 10 GeV, and we disregarded photons from the Earth's limb with a cut at 100$^{\circ}$ in the zenith angle.\\  
We performed the second--level trigger, and the offline analysis on flaring sources
with standard {\it Fermi--LAT}
Science tools v9r32p5, using the Pass 7 (P7REP\_SOURCE\_V15) response functions.
We disregarded photons from the Earth's limb, adopting  a cut at
100$^\circ$ in the zenith angle, and we allowed only events of {\em event class} 2.
Light curves and spectra were obtained performing the unbinned likelihood
analysis inside a region of radius 20$^\circ$ around target--sources.
Galactic diffuse and Extragalactic isotropic backgrounds were modeled using gll\_iem\_v05 and iso\_source\_v05, respectively.
Non target sources were taken from the second {\it Fermi}--LAT catalog \citep{second_lat_cat}.\\
We extracted light curves with energies $>$300 MeV
in order to process data with a smaller PSF and reduce
background Gamma-rays from neighbour sources.
This is a convenient choice for the massive analysis of temporal bins.
and it causes negligible reduction of signal significance.\\
\begin{figure*}
\centering
\begin{tabular}{cc}
\psfig{figure=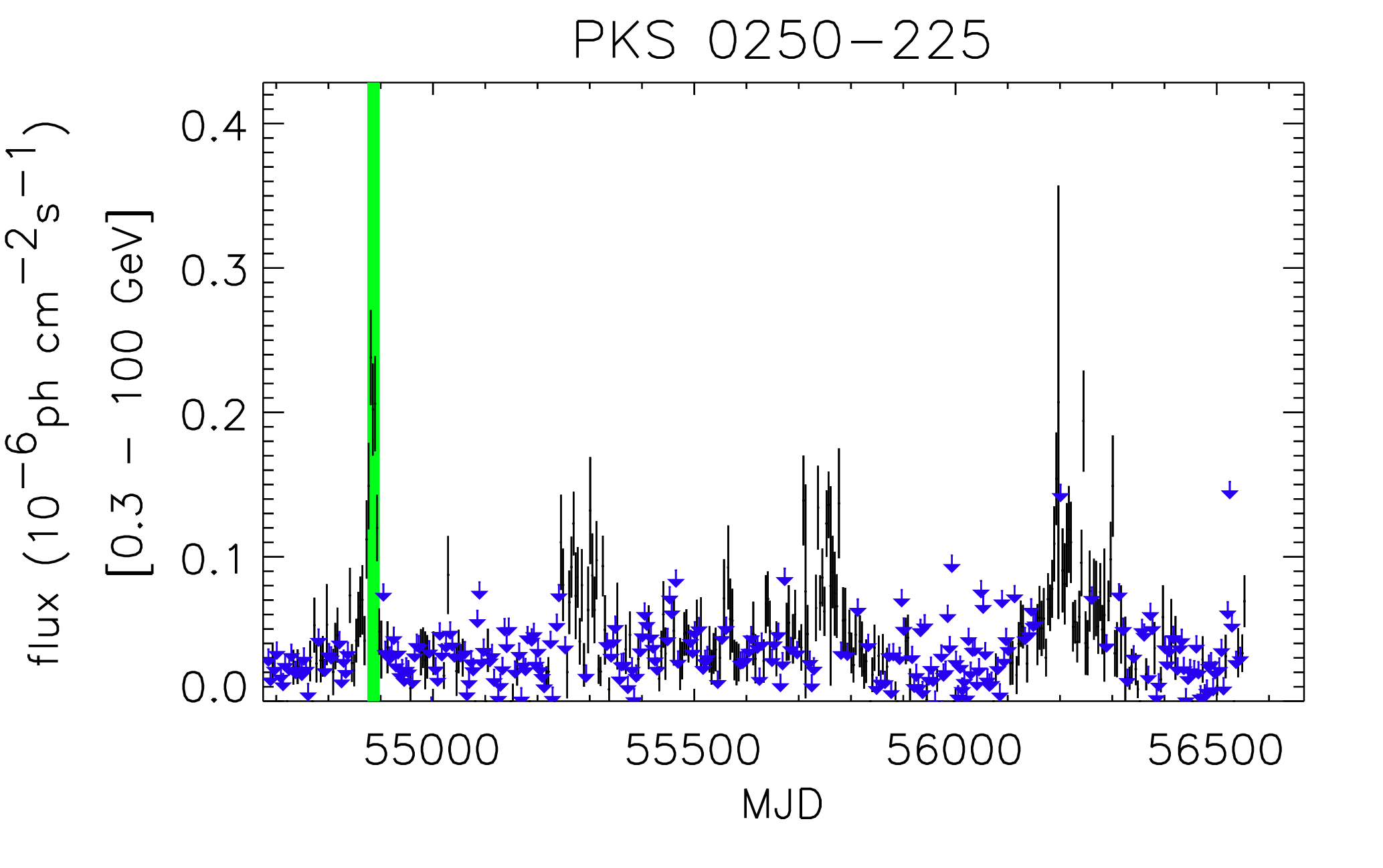,width=8cm} & \psfig{figure=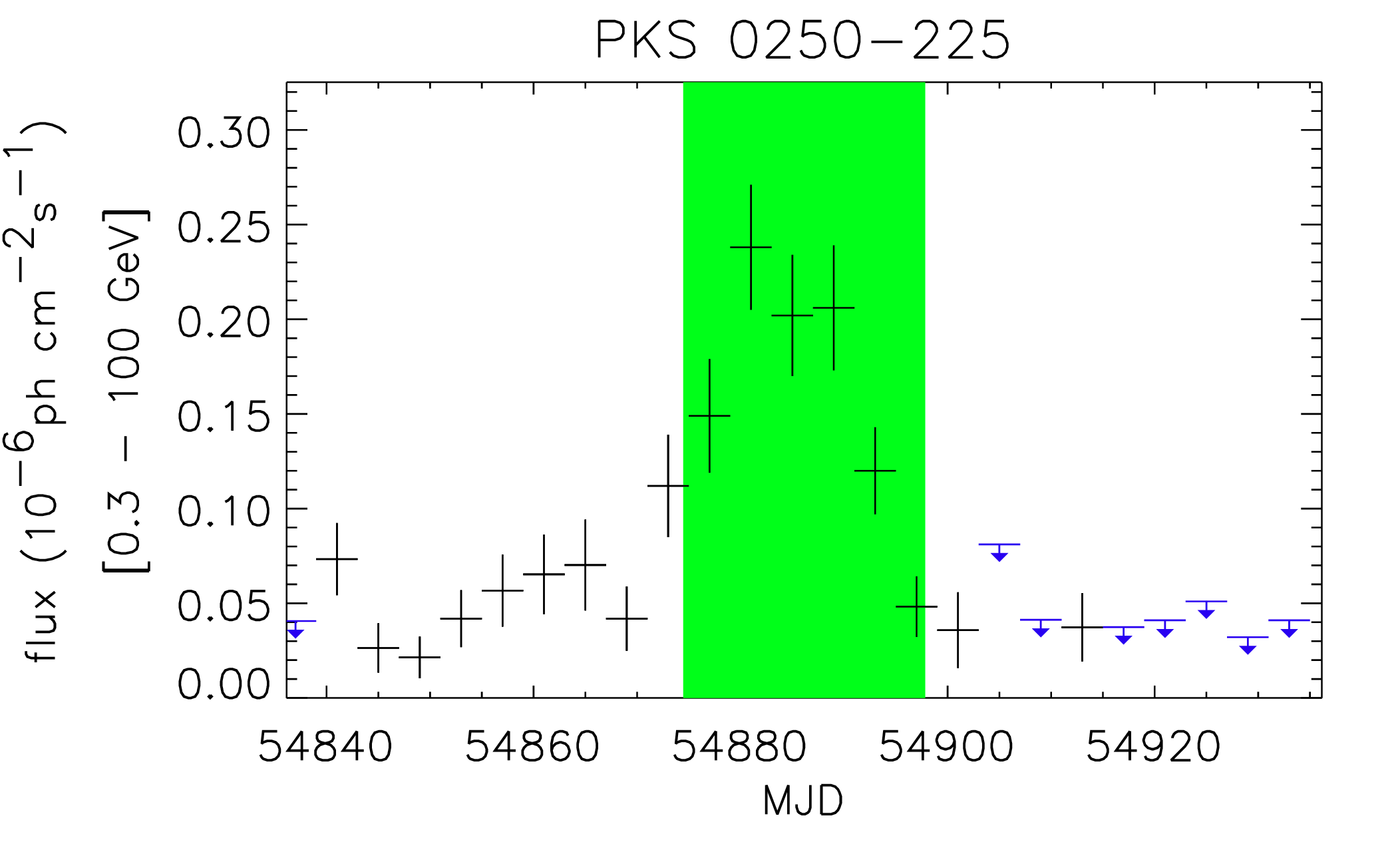,width=8cm} \\
\psfig{figure=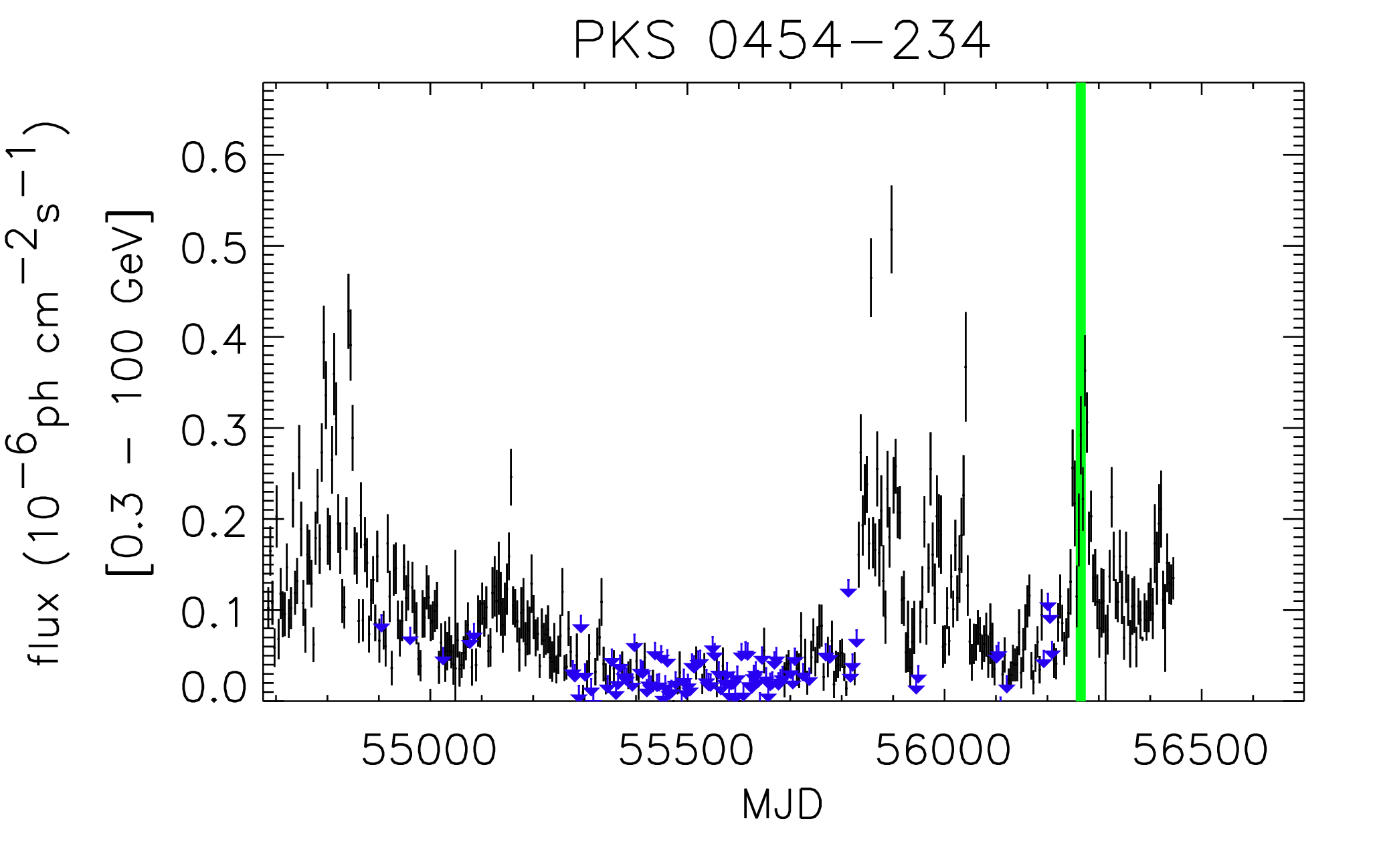,width=8cm} & \psfig{figure=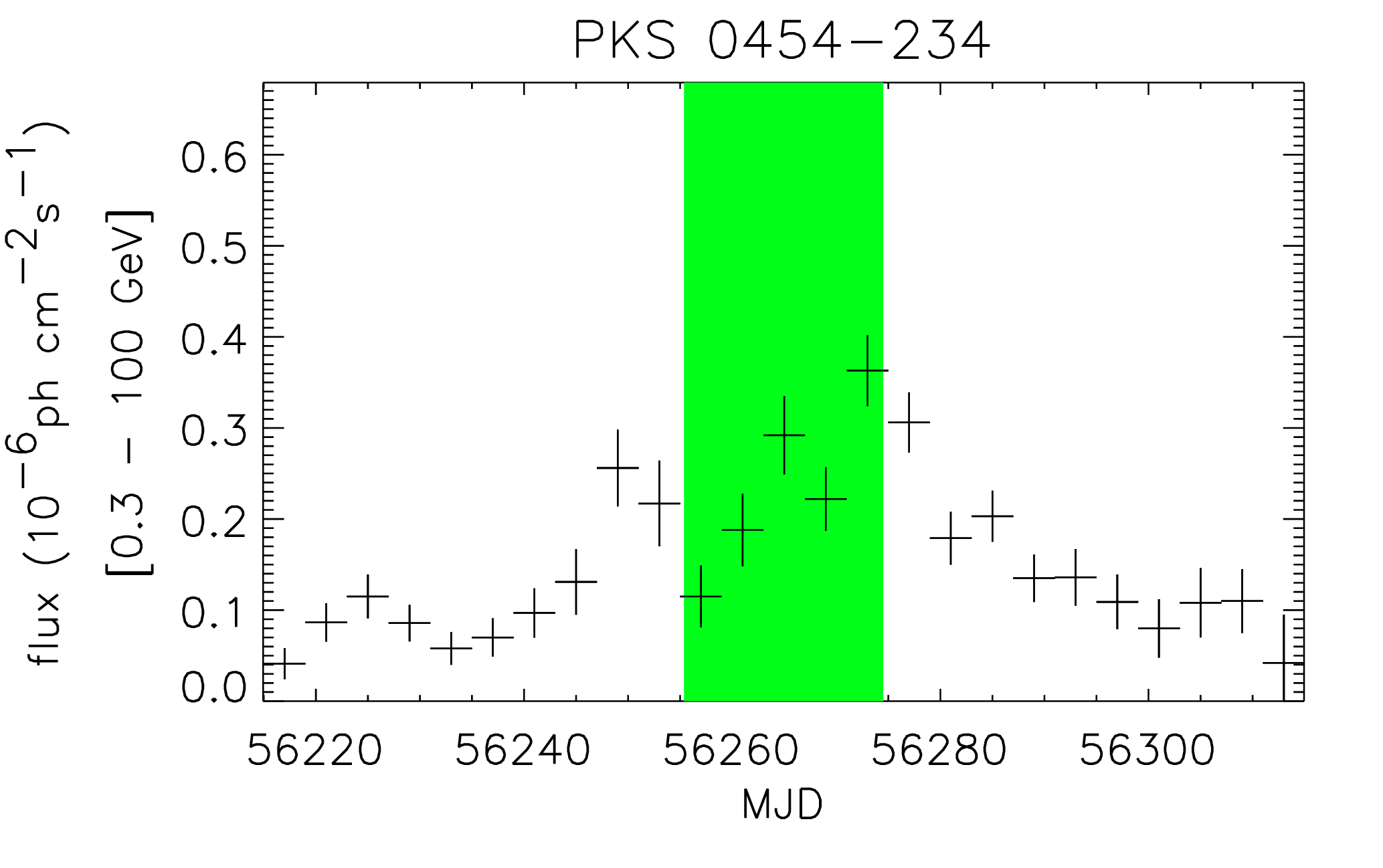,width=8cm} \\
\psfig{figure=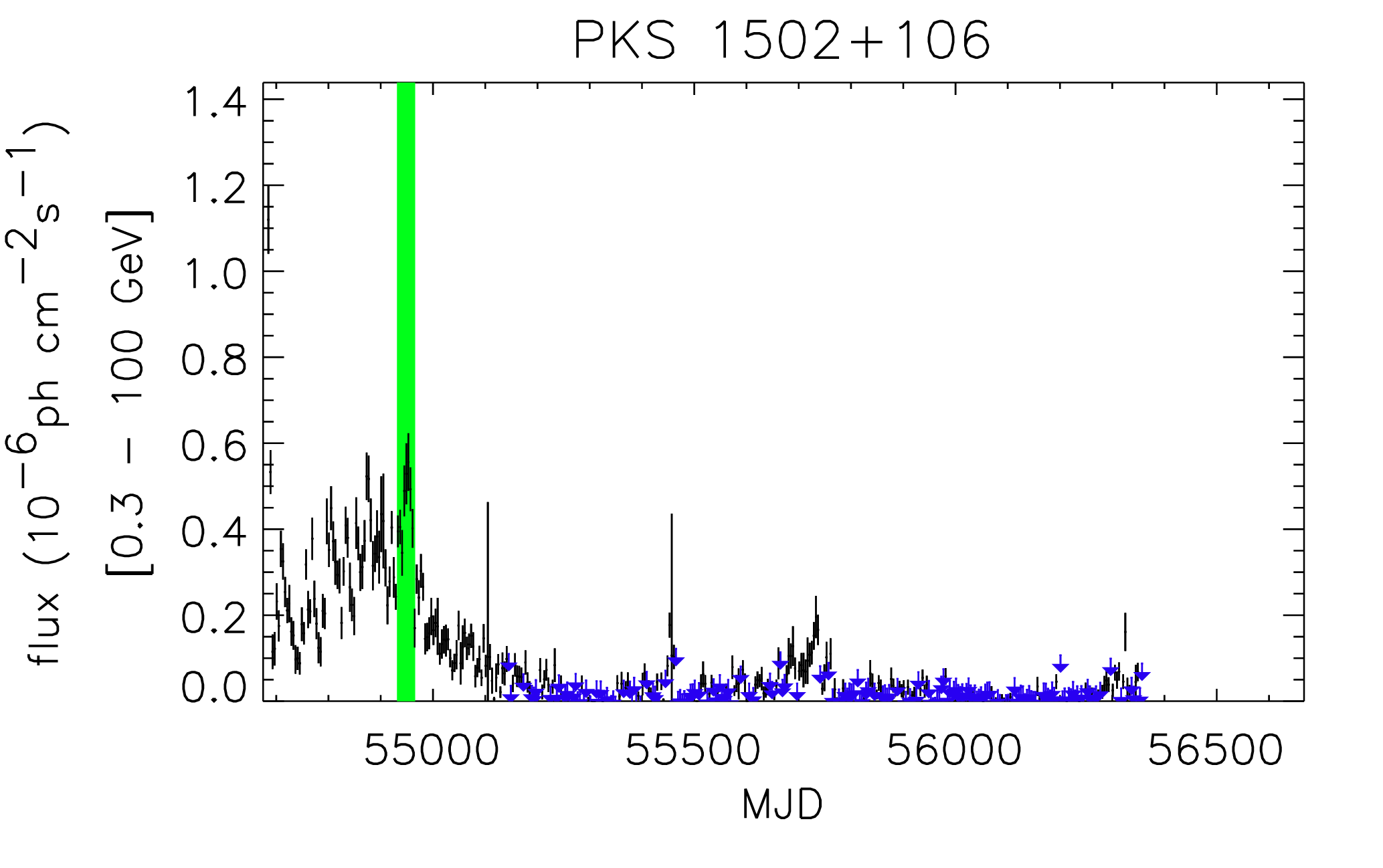,width=8cm} & \psfig{figure=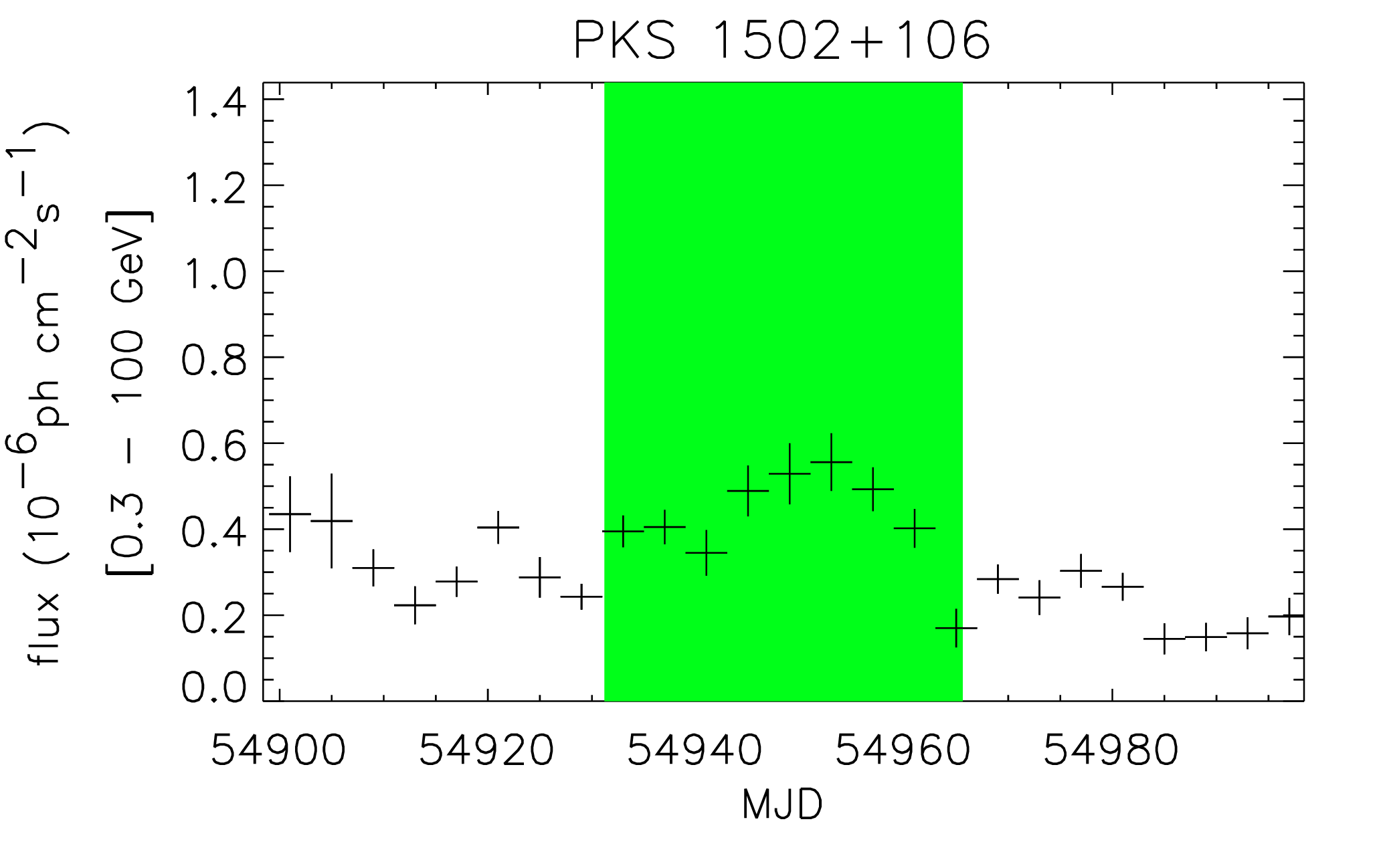,width=8cm} \\
\psfig{figure=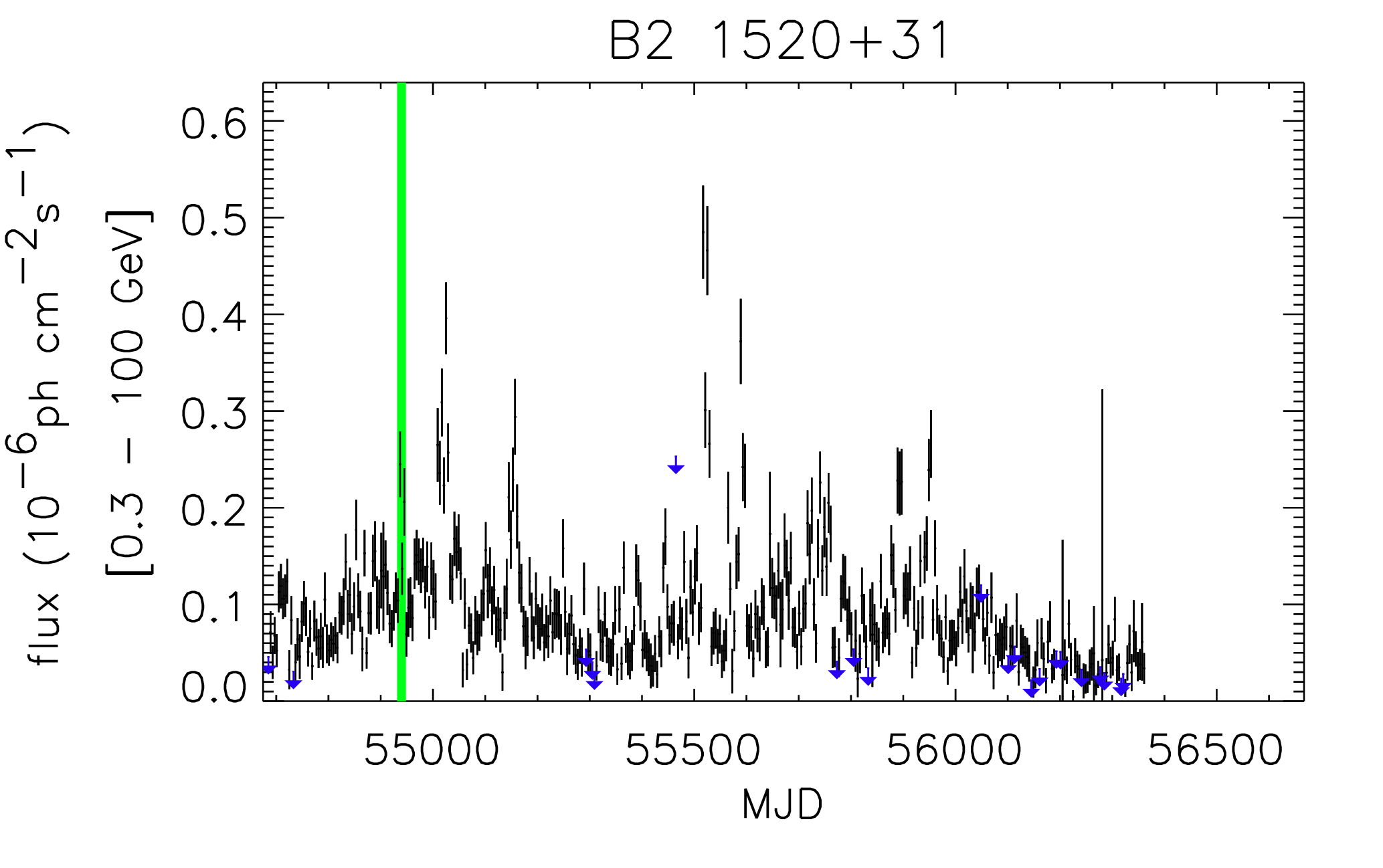,width=8cm} & \psfig{figure=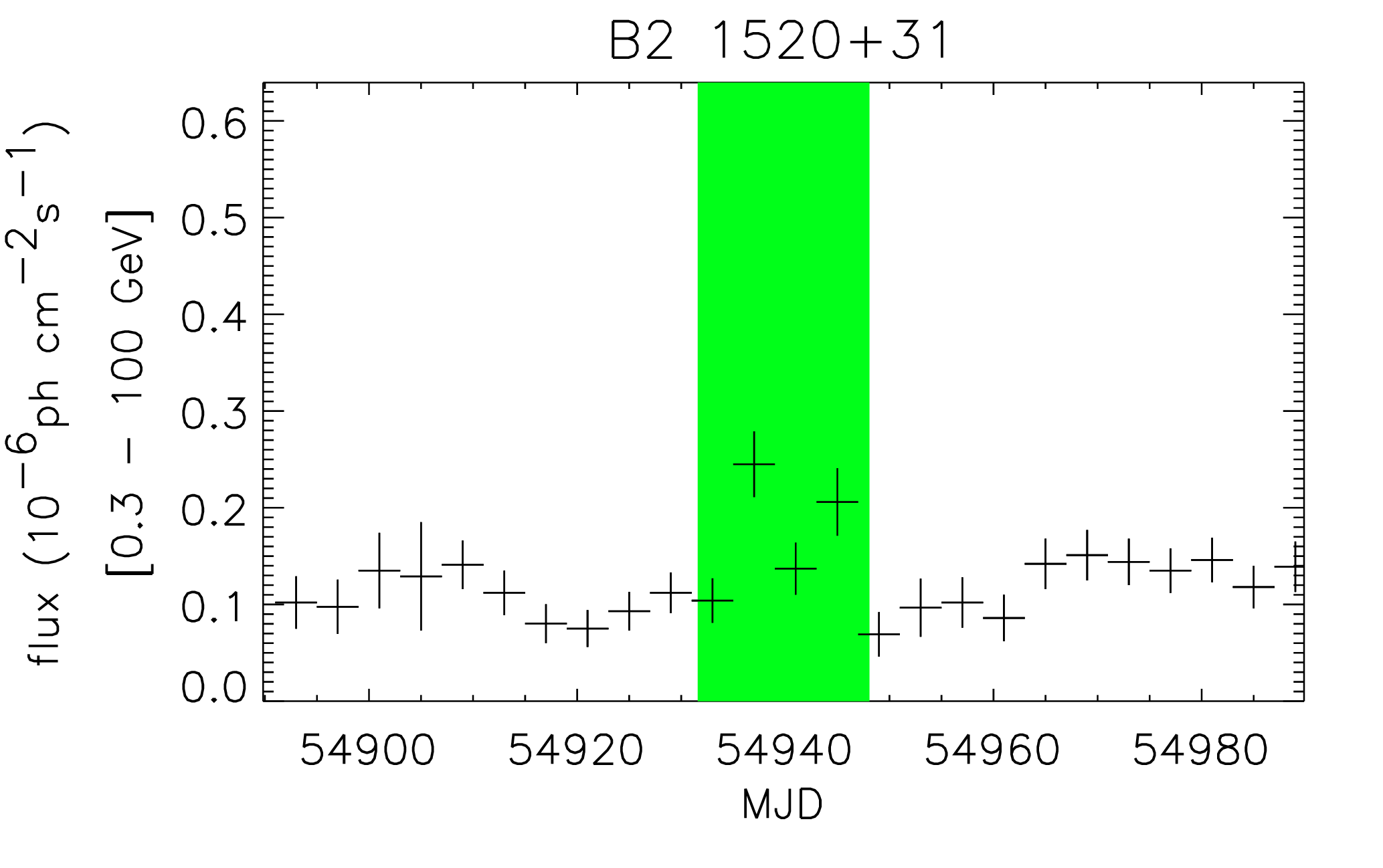,width=8cm} \\
\end{tabular}
\caption{Light curves of PKS 0250-225, PKS 0454-234, PKS 1502+106, B2 1520+31 in gamma rays. left: long integration, right: zoom around the reported flare. Binsize is 4 d.}
\label{fig_lcall1}       
\end{figure*}
\begin{figure*}
\centering
\begin{tabular}{cc}
\psfig{figure=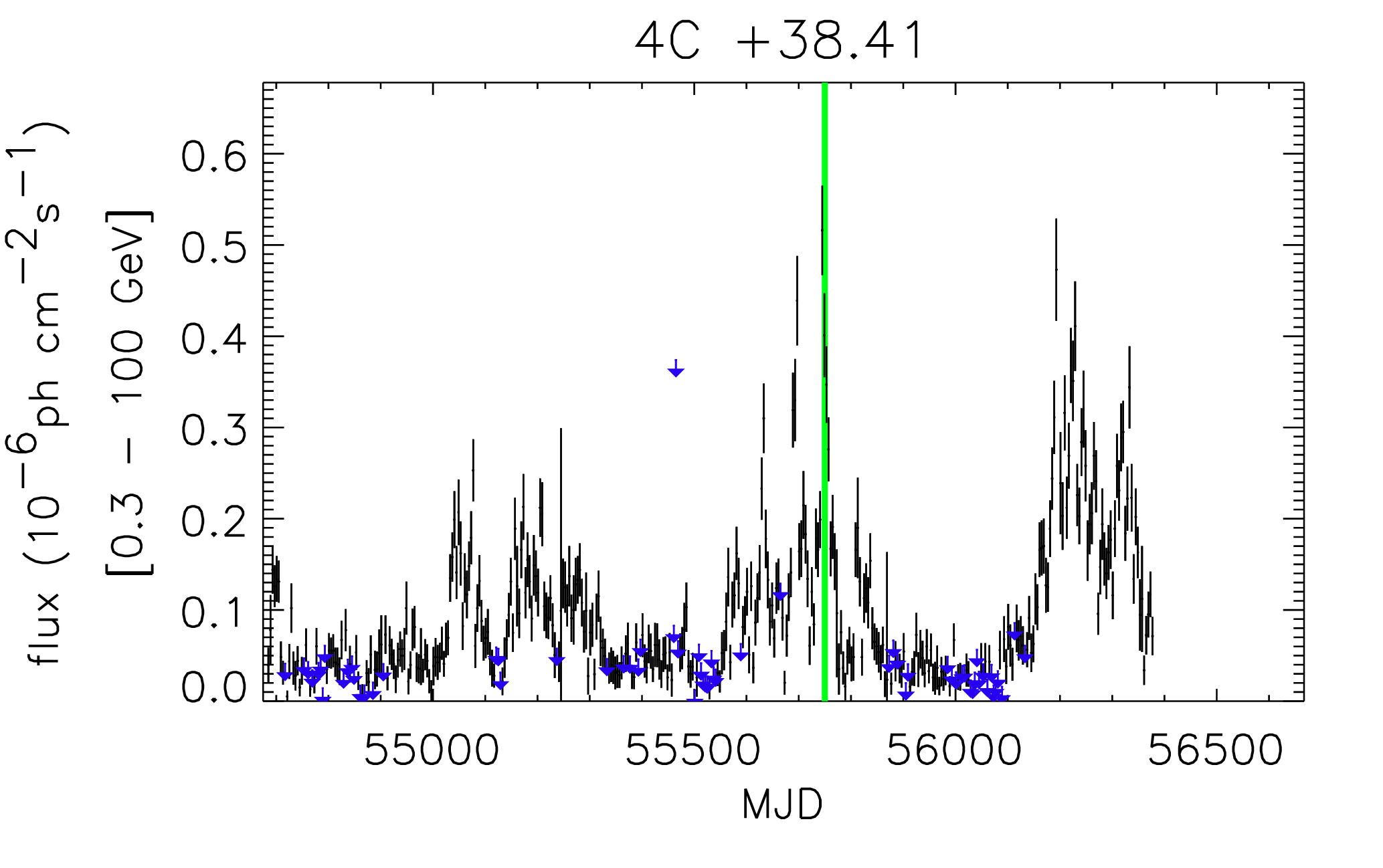,width=8cm} & \psfig{figure=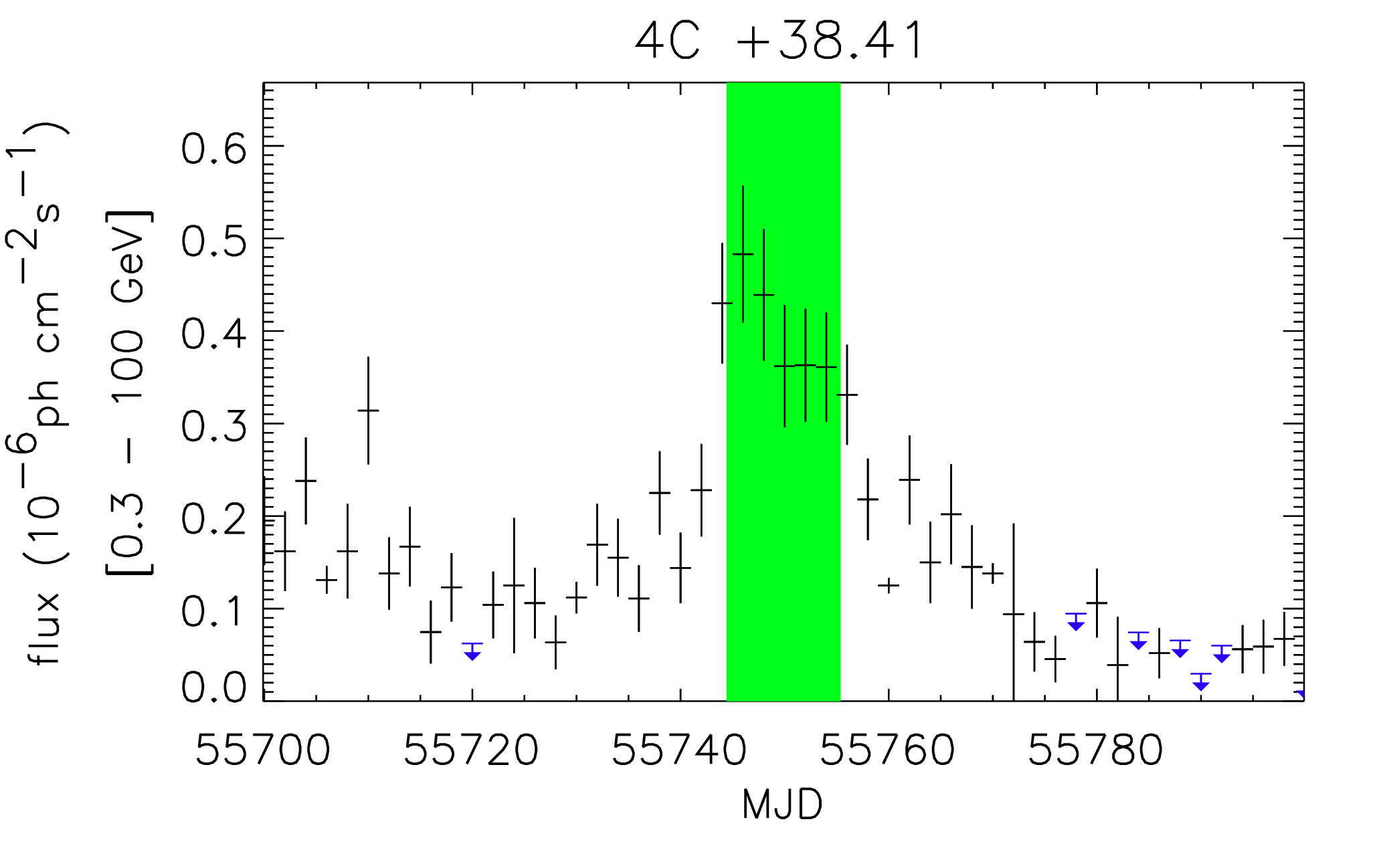,width=8cm} \\
\psfig{figure=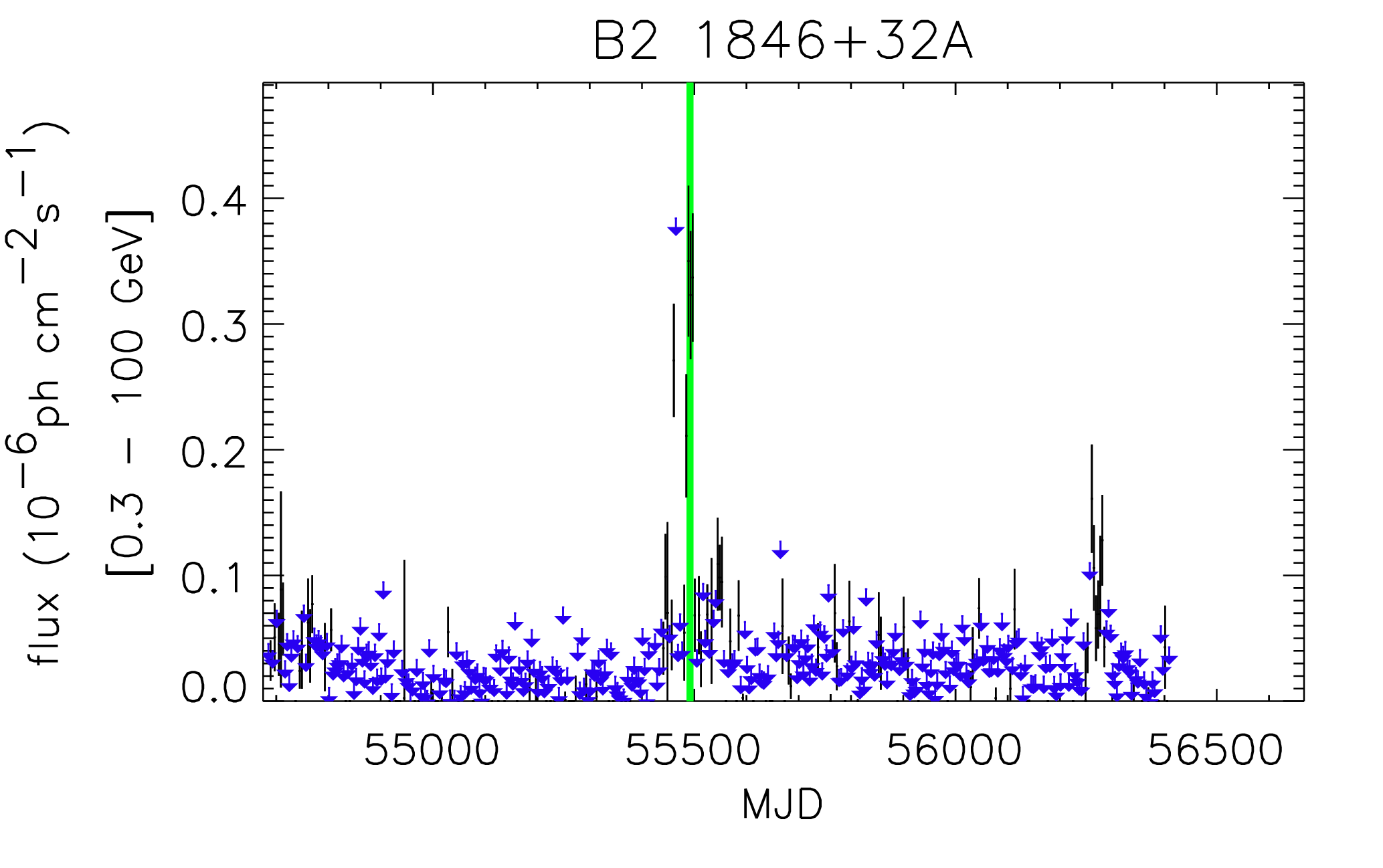,width=8cm} & \psfig{figure=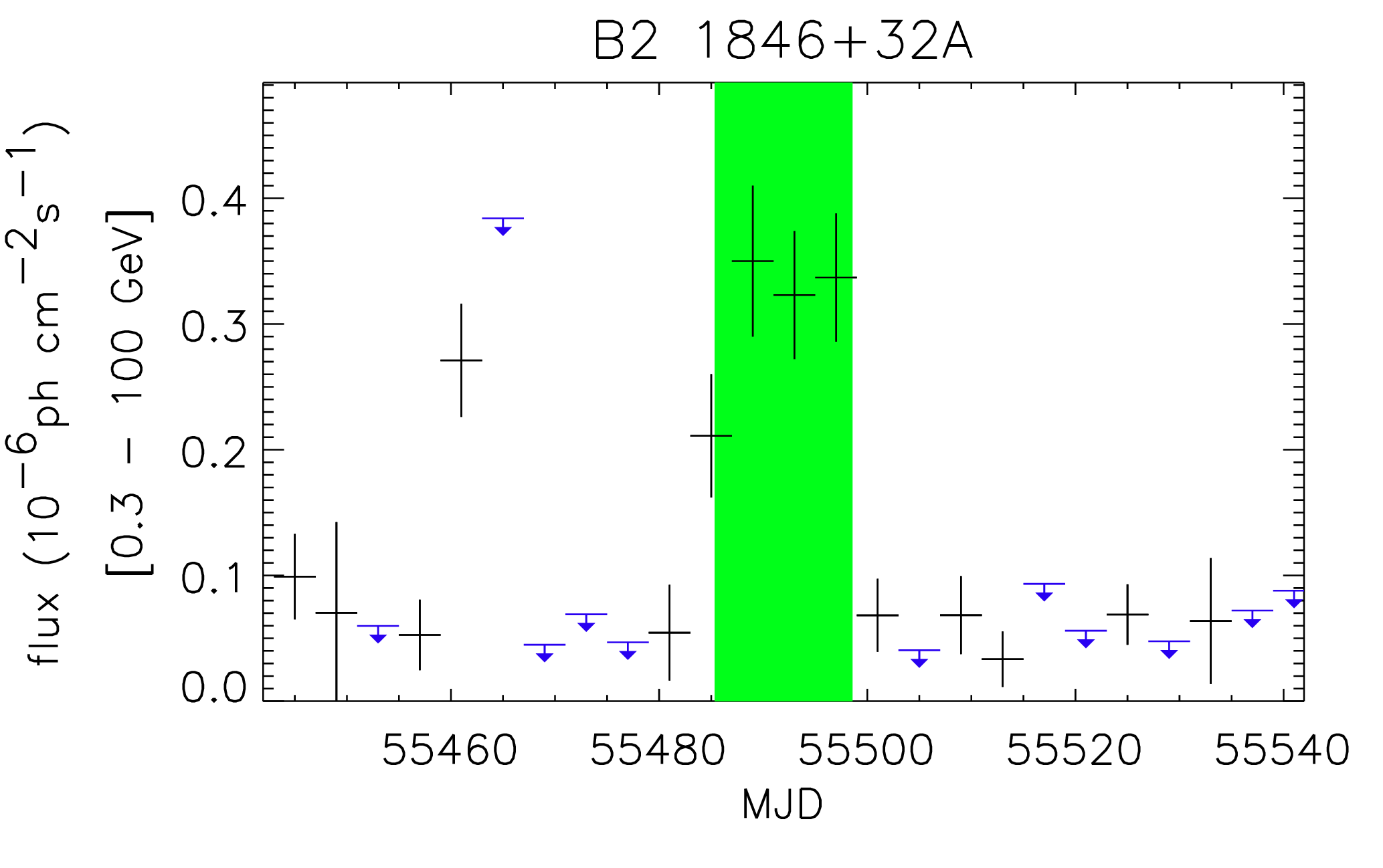,width=8cm} \\
\psfig{figure=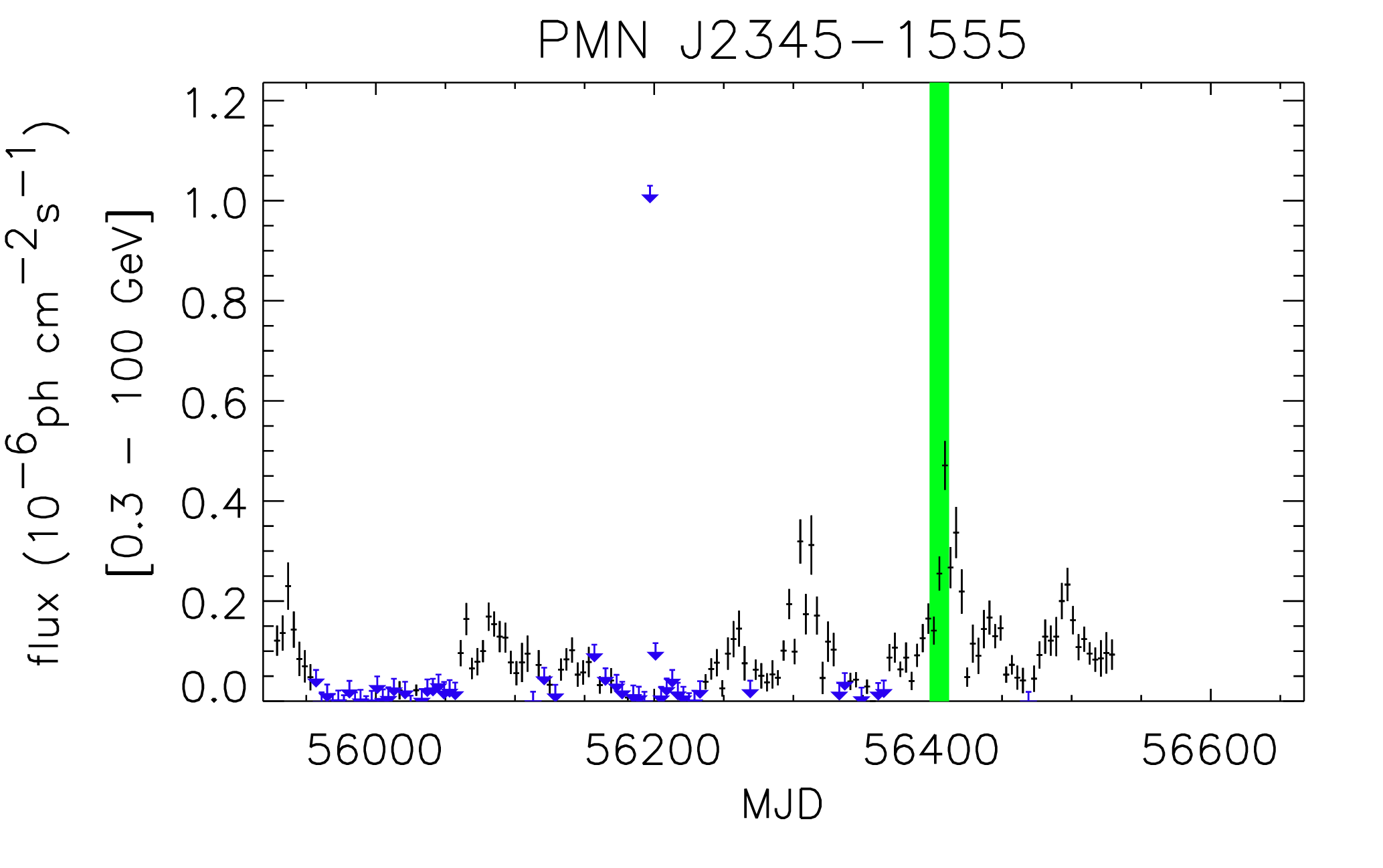,width=8cm} & \psfig{figure=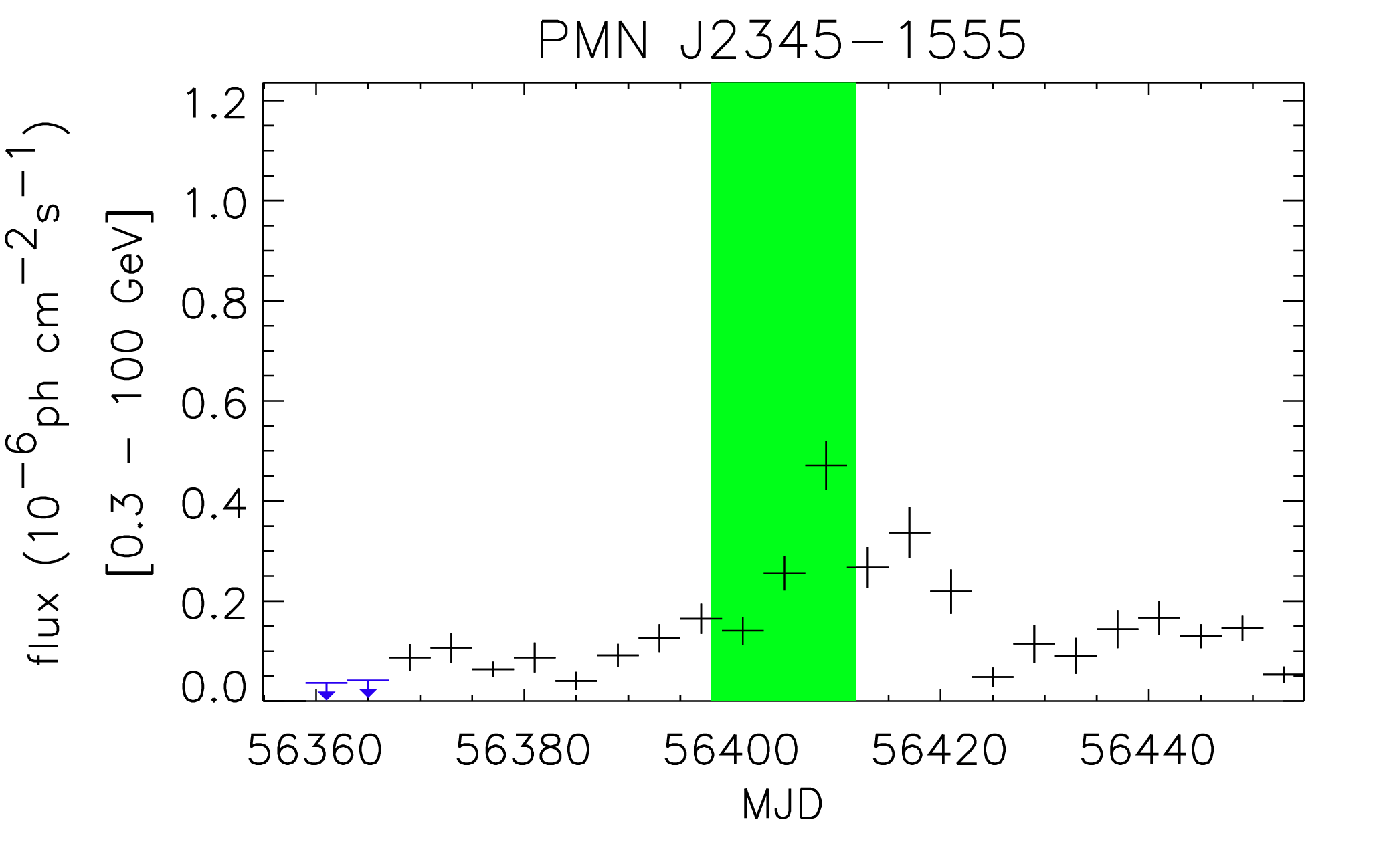,width=8cm} \\
\end{tabular}
\caption{Light curves of 4C +38.41, B2 1846+32A,
PMN J2345-1555 in gamma rays. left: long integration, right: zoom around the reported flare. Binsize is 4 d, except for the right panel for 4C +38.41 with binsize of 2 d.}

\label{fig_lcall2}       
\end{figure*}
\begin{figure*}
\centering
\begin{tabular}{cc}
\psfig{figure=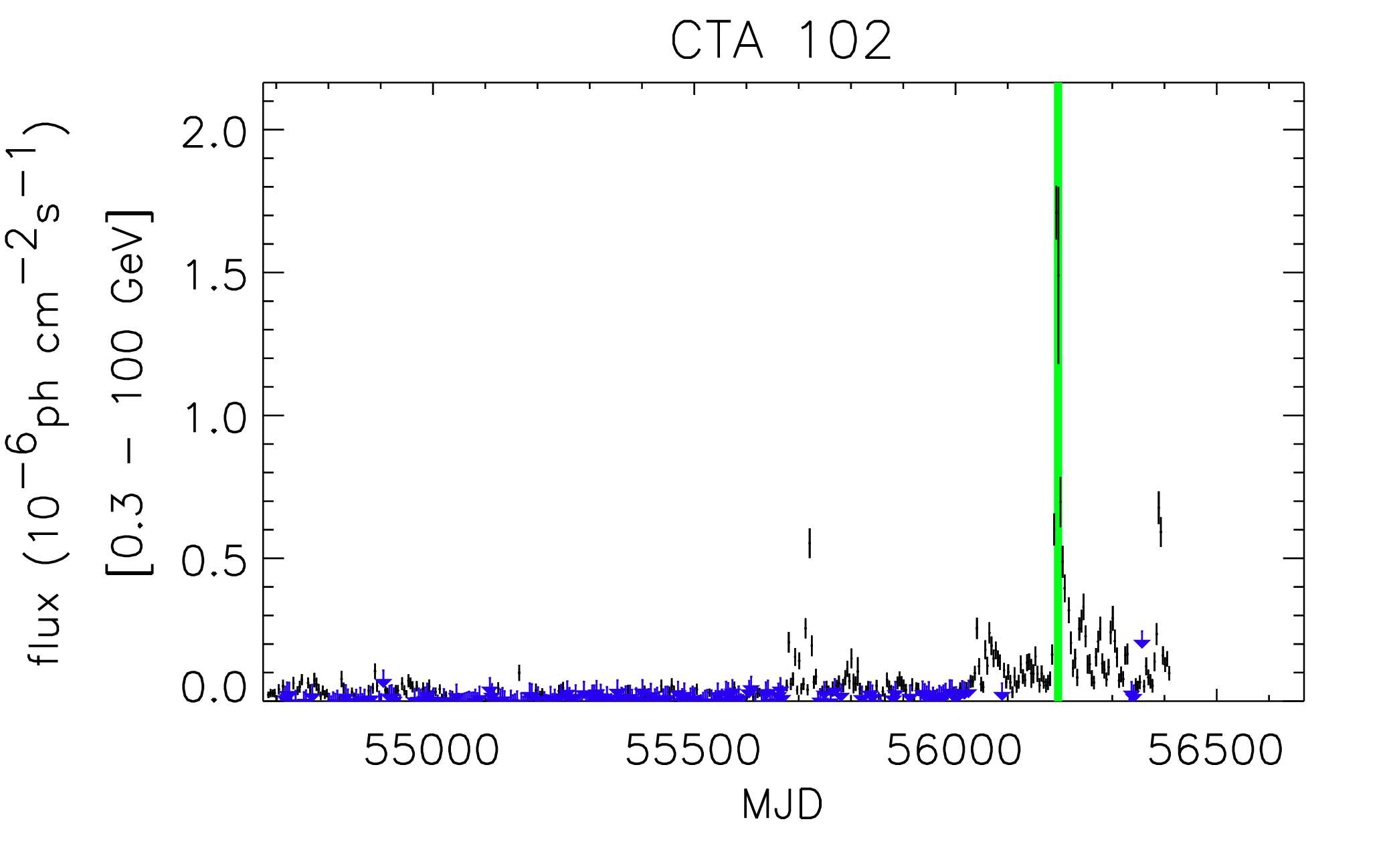,width=8cm} & \psfig{figure=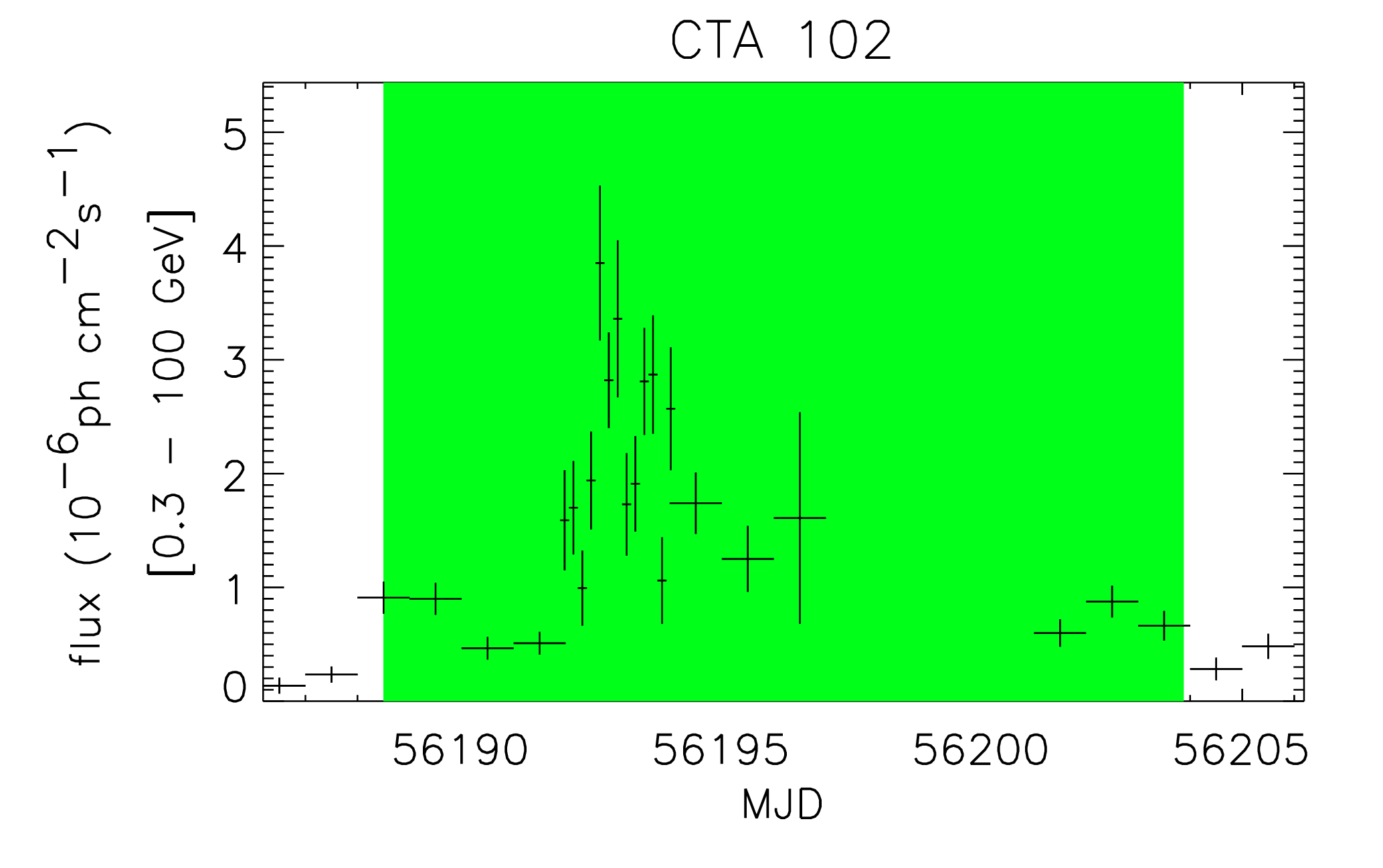,width=8cm} \\
\psfig{figure=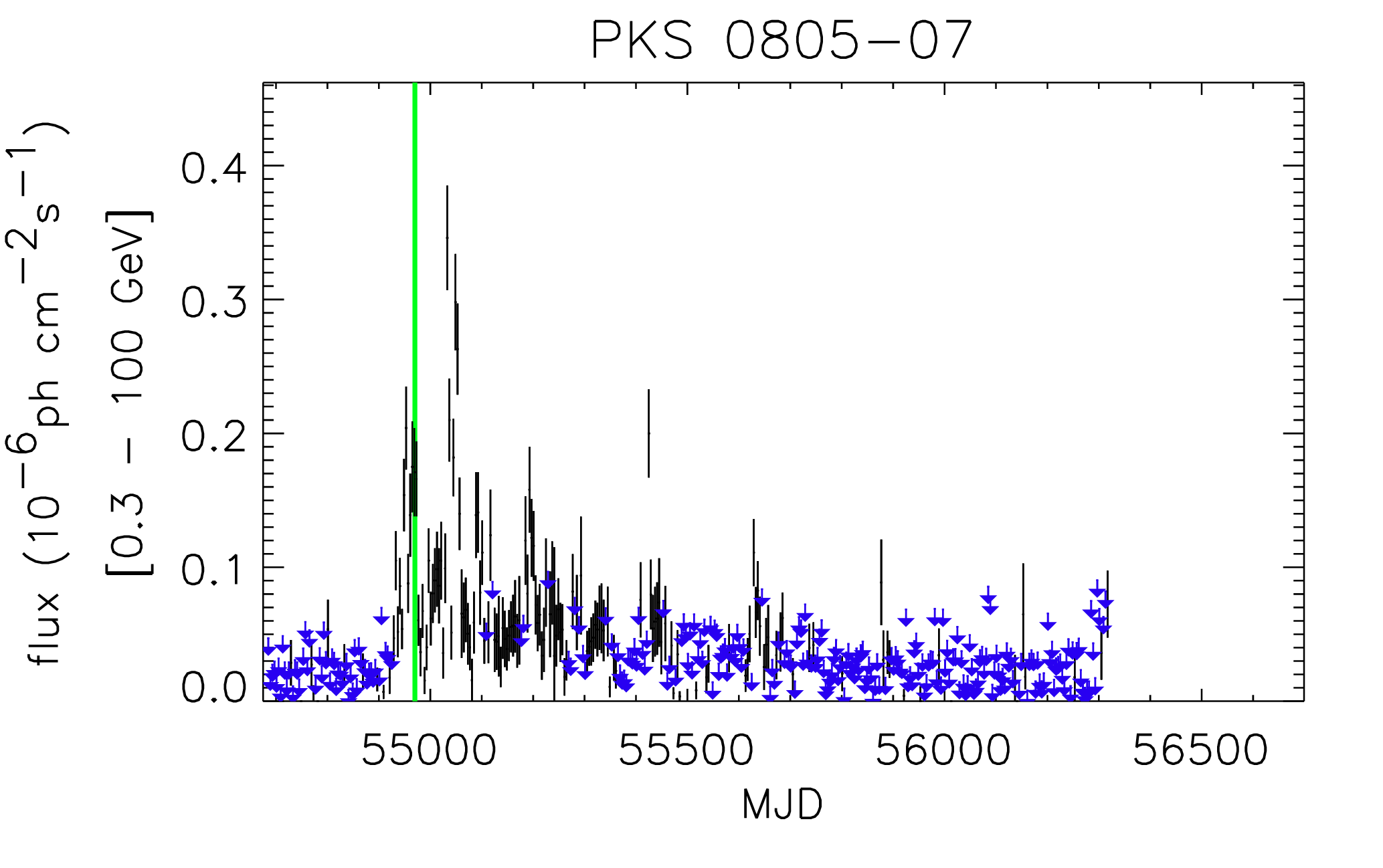,width=8cm} & \psfig{figure=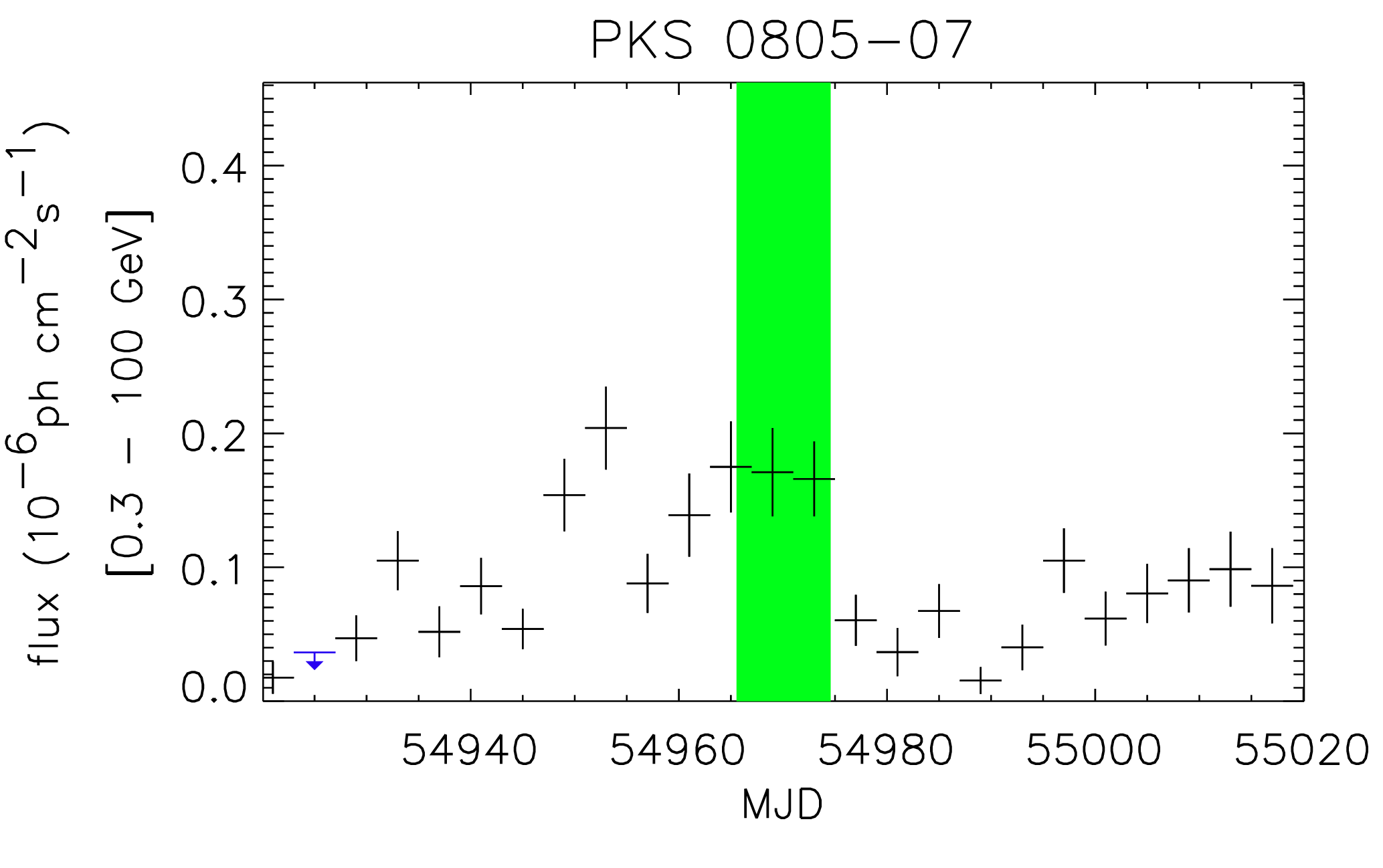,width=8cm} \\
\psfig{figure=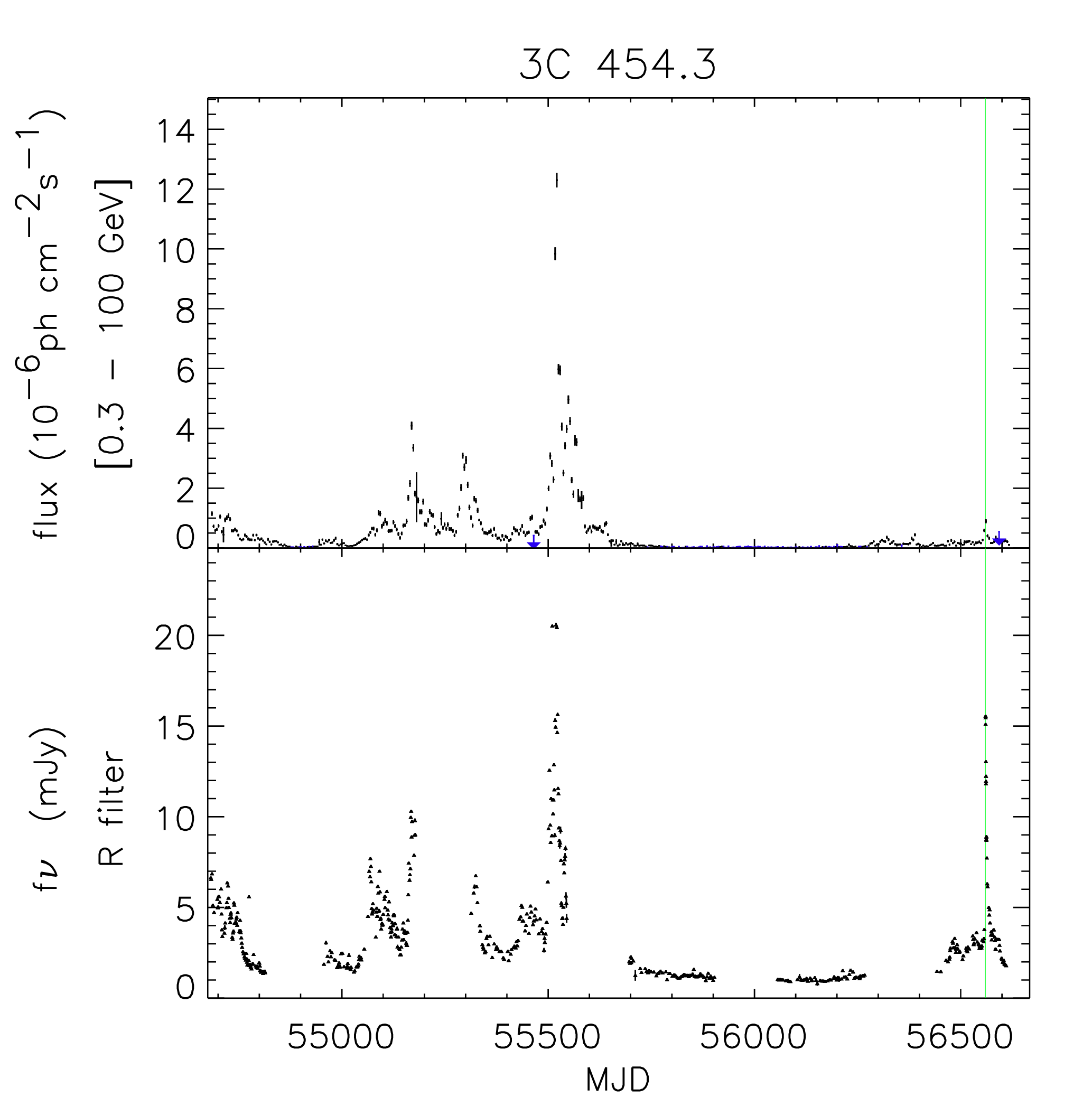,width=8cm} & \psfig{figure=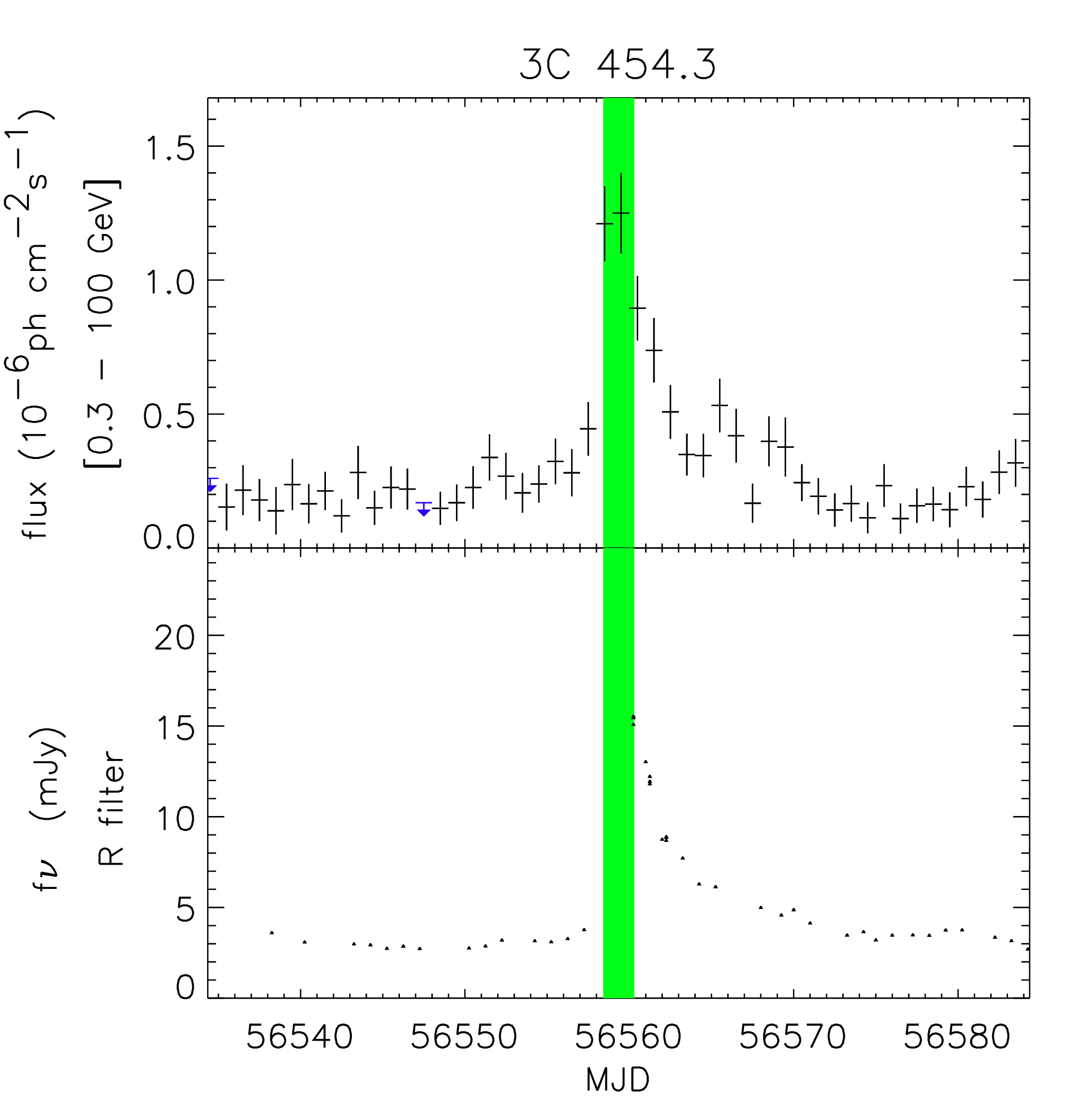,width=8cm} \\
\end{tabular}
\caption{Light curves of CTA 102, PKS 0805-07, 3C 454.3 in gamma rays. left: long integration, right: zoom around the reported flare. Binsize is 4 d, except for the right panel for 3C 454.3 with binsize of 1 d, and for the core of the emission for CTA 102, with binsize of 0.17 d.  For 3C 454.3 we show the optical light curve with R filter taken within the Yale-SMARTS monitored blazars program.}
\label{fig_lcall3}       
\end{figure*}
\subsection{X-ray data analysis}

Chandra \citep{Weisskopf2002} observed PKS 1502+106 in X-rays with the back-illuminated S3 CCD of the Advanced CCD Imaging Spectrometer (ACIS, \citealt{garmire2003}). We reduced data and performed the analysis making use of the Chandra Interactive Analysis of Observation
(CIAO, \citealt{fruscione2006}) version 4.3 software, with calibration version (CALDB) 4.1.3. We extracted the source spectrum using a circular region with a 5 arcsec radius centered on the source optical position.
The Background was taken from a nearby circular region with a 15 arcsec radius.
We produced Response Matrix (RMF) and created the Ancillary Response File (ARF).\\

We reduced {\it Swift--XRT} \citep{burrows2005} data with xrtpipeline version 0.12.3 software, and analysed it with standard tools, using the most recent calibration files available.
We selected events with grade 0--2 for window timing data, and with grade 0-12 for photon counting mode. We created the Ancillary Response files using {\it xrtmkarf}.\\

We used an absorbed power--law to fit model to the sources X-ray spectra, with absorption fixed at the galactic values \citep{kalberla2005}. 
Data for PKS 1502+106 show a statistically significant excess ($\sim$6 standard deviations)  below 0.5 keV with respect to the model.\\

In the following sections we will show the X-ray spectra of sources corrected for the galactic absorption using Wisconsin cross-section \citep{morrison1983}.

\subsection{UV--optical--NIR data analysis}
The {\it Swift--UVOT} \citep{roming2005} data analysis was performed with the FTOOLS tasks uvotimsum and uvotsource.
magnitudes were evaluated through aperture photometry within a circular region or radius 5 arcsec around the source positions.
Backgrounds were estimated from nearby source--free regions, of radius 9 arcsec.\\

We performed simultaneous $V$ and $J$-band imaging-polarimetry of PKS~1502$+$106
using the TRISPEC instrument attached to the {\it Kanata} 1.5-m telescope
\citep{trispec}. A unit of the polarimetric observing sequence consisted
of exposures at four position angles of the half-wave plate: 0, 45, 22.5,
and 67.5 deg. Typical exposure times were 200 and 20~s for each $V$ and $J$-band
image. The data were reduced according to the standard procedure of CCD images.
The differential photometry was performed with a comparison star located at
R.A.$=15:04:36.51$, Dec.$=+10:28:47.0$ (J2000.0) having magnitudes of $V=15.335$ and
$J=14.117$. The observation was a part of a large program for monitoring blazars
with Kanata, and more detailed information about the data reduction is summarized
in \cite{Ikejiri11}.\\

The SMARTS optical and near-infrared aperture photometry was performed
using the PHOT task in IRAF. Non--variable comparison stars with comparable
magnitudes to the blazars were chosen in each field.  The raw magnitudes
were calibrated using zero points obtained on photometric nights of optical
\citep{landolt1992} and near-infrared 
\citep{Persson1998}
secondary standards with the ANDICAM instrument on the 1.3m telescope. A
more comprehensive description of the photometric data reduction process
for SMARTS blazars can be found
in \cite{bonning2012} and
\cite{buxton2012}.
Optical and near-infrared light curves for the Yale-SMARTS monitored blazars
program can be found on the website.\footnotemark \footnotetext{
http://www.astro.yale.edu/smarts/glast/home.php}\\

NIR observations of  PKS 1502+106, B2 1846+32A, CTA 102, PKS 0805+07 
were carried out with  INAOE's 2.1m {\it Guillermo Haro}\footnote{http://astro.inaoep.mx/observatorios/cananea/}
telescope equipped with CANICA, a near IR camera.
Standard NIR differential photometry was obtained for 5 arcmin squared
images, centered on the objects of interest. The adopted reference local
photometric standards were those objects listed in the 2Mass point source
catalog \citep{2mass}.\\

NIR--optical--UV  photometry is de-reddened using the interstellar extinction curve proposed in \cite{fitzpatrick1999}.\\

For sources at redshift 0.9 or more, the absorption lines from neutral Hydrogen of the intergalactic medium (IGM) enters the band of the {\it Swift--UVOT} UVW2 filter
(see, e.g., \citealt{rau2012}).
We used the method proposed in \cite{prochaska2009}, to extrapolate at z$<2$ the evaluations reported in \cite{ghisellini2010},
ad we obtained that the effect is $<$ 10\% for the photometry with UVW2, UVM2 and UVW2 filters 
at redshift of 1.1, 1.4 and 1.7 respectively. These reported values are pessimistic because they are evaluated using the proper mean free path at 912 $\AA$ ($\lambda^{912}_{mfp}$)
for sources at redshift 2, instead that redshift $<$2.
Corrections are uncertain due to the poor knowledge of the neutral Hydrogen column density of the IGM.
Our sample contains sources with redshift up to 1.84. We do not try to correct optical--uv photometry for our sample. We will show that for all the sources but one
the corrections will not affect our results.

\section{Results}
\subsection{Gamma-ray light curves}
We anticipated some of the results in Table \ref{tab_he_sample}: the HE activity for our sample lasts from a couple of days to about one month.  
The gamma-ray light curves for our sources are shown in Figure \ref{fig_lcall1}, \ref{fig_lcall2}, \ref{fig_lcall3}, where we also highlight the HE activity period.
We note that for the entire sample, the HE activity period corresponds to high activity also at lower energy.
For CTA 102
the core of the activity period is reported with temporal bins of 0.17 d
in order to show the fast variability of the source during the period of interest.\\
In the last years 3C 454.3 exhibited flaring activity with peak flux exceeding 10$^{-5}$ ph cm$^{-2}$ s$^{-1}$ for E$>$ 100 MeV \citep{vercellone454,pacciani454,bonnoli454}.
The comparison of the gamma ray and optical light curves for 3C 454.3 reveals one of the peculiarity of the HE flare that we are investigating:
it exhibited a gamma ray flux of  350$\times$10$^{-8}$ ph cm$^{-2}$ s$^{-1}$ (E$>$ 100 MeV) which is an order of magnitude fainter than during the brightest
gamma-ray flares, but in optical the flux is comparable to the peak emission observed during the brightest gamma-ray flares.

\subsection{Spectral Energy Distribution modeling}
\begin{figure*}
\centering
\begin{tabular}{cc}
\psfig{figure=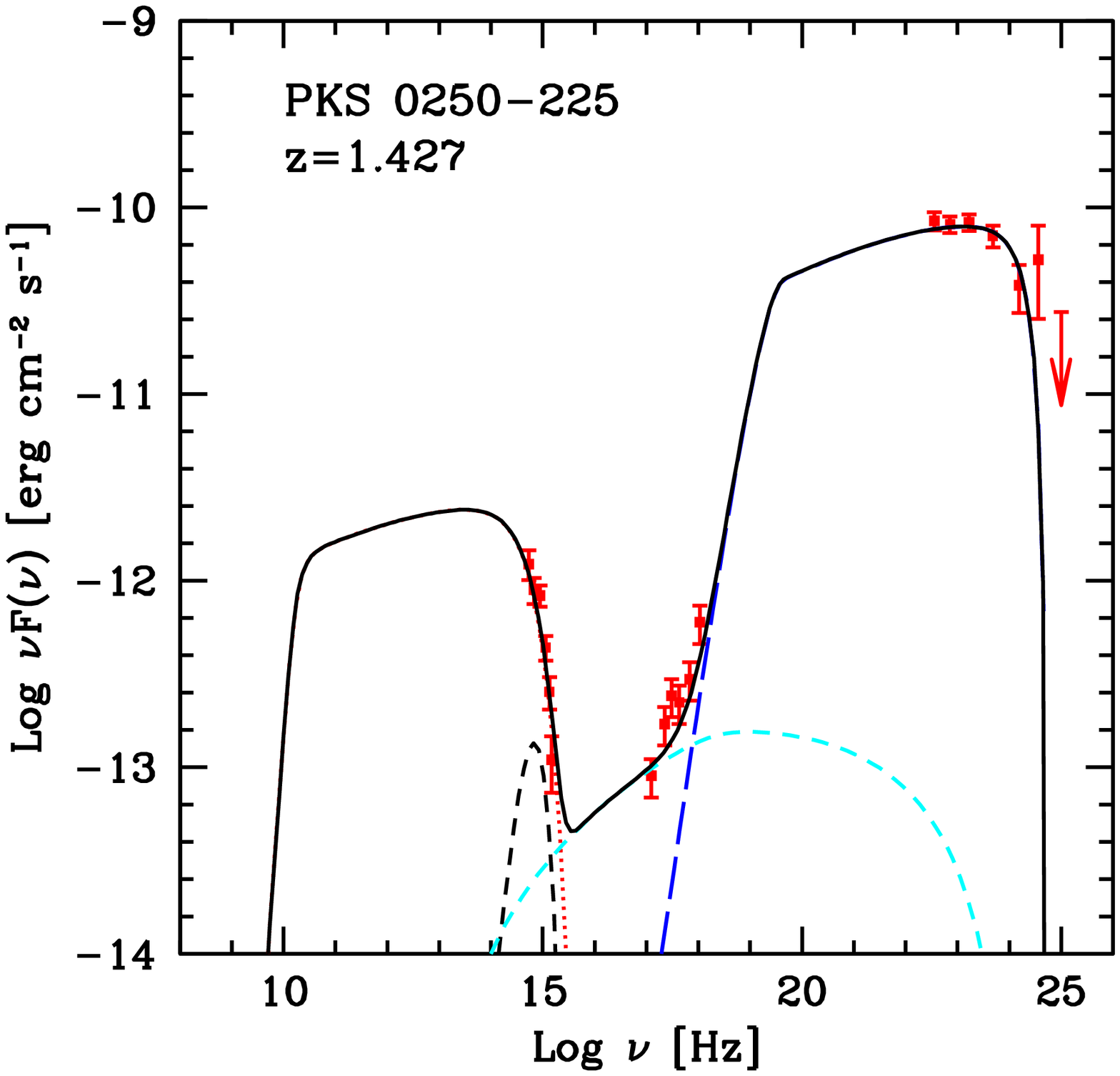,width=7.5cm}&
\psfig{figure=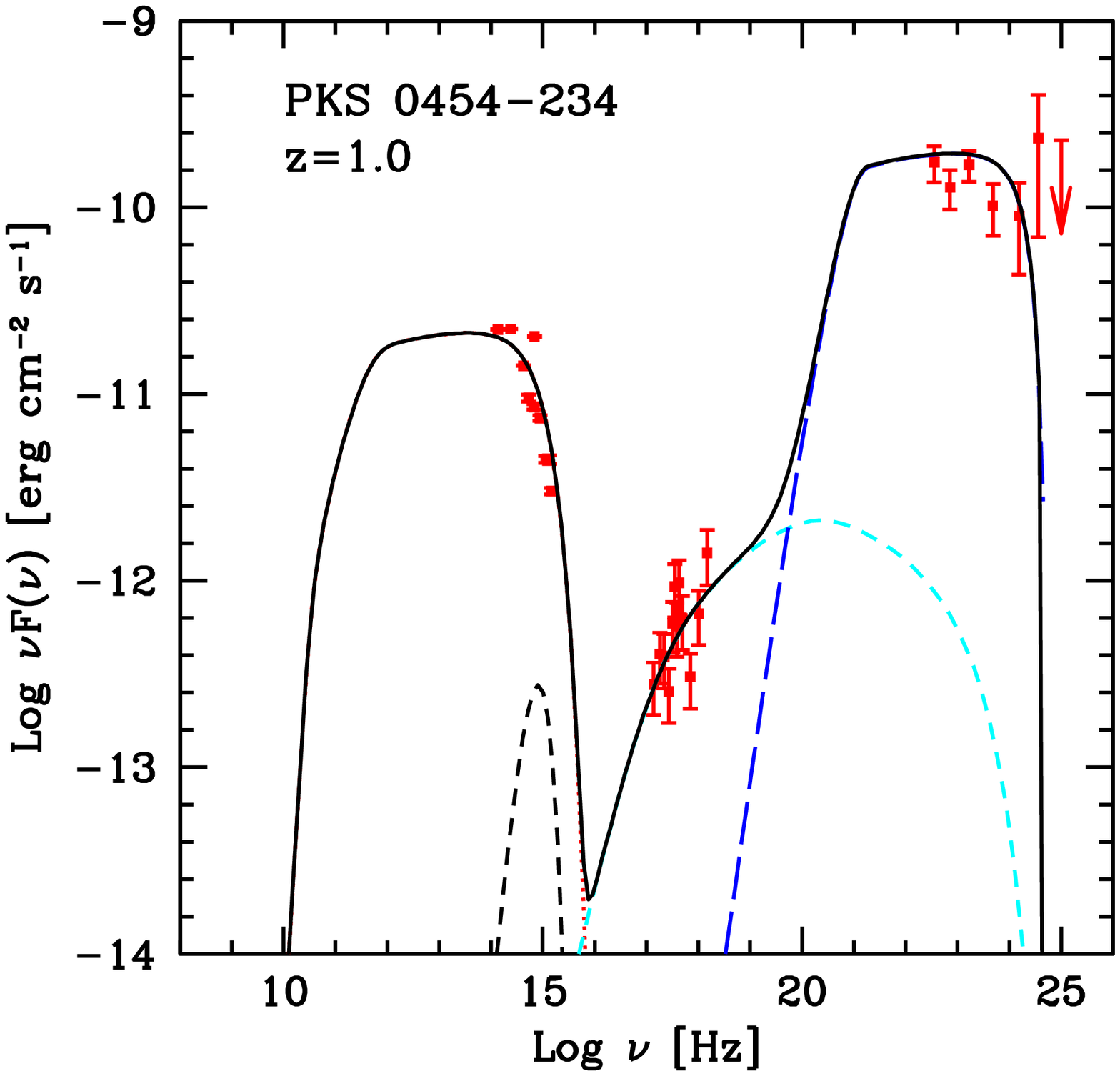,width=7.5cm}\\
\psfig{figure=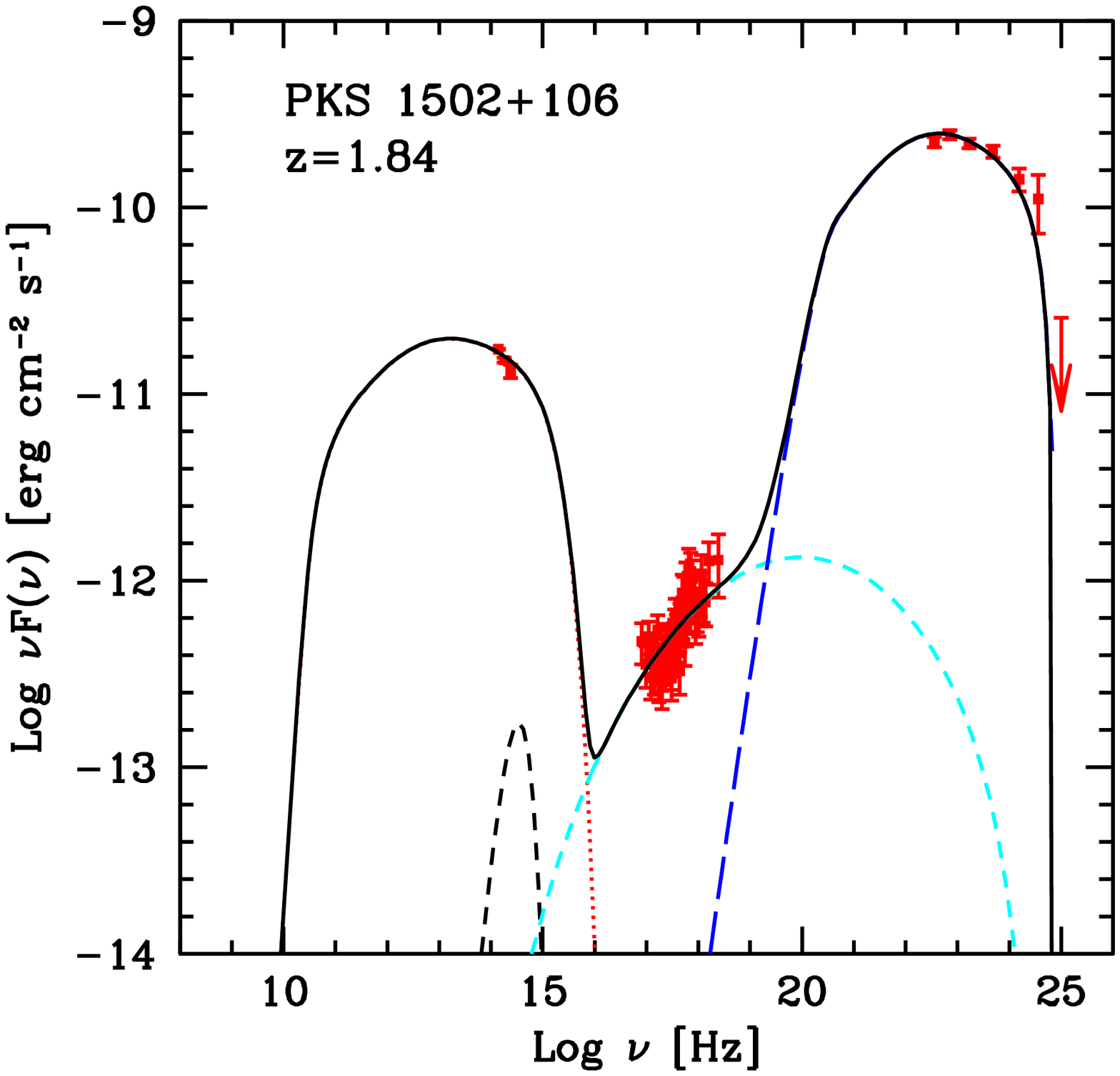,width=7.5cm}&
\psfig{figure=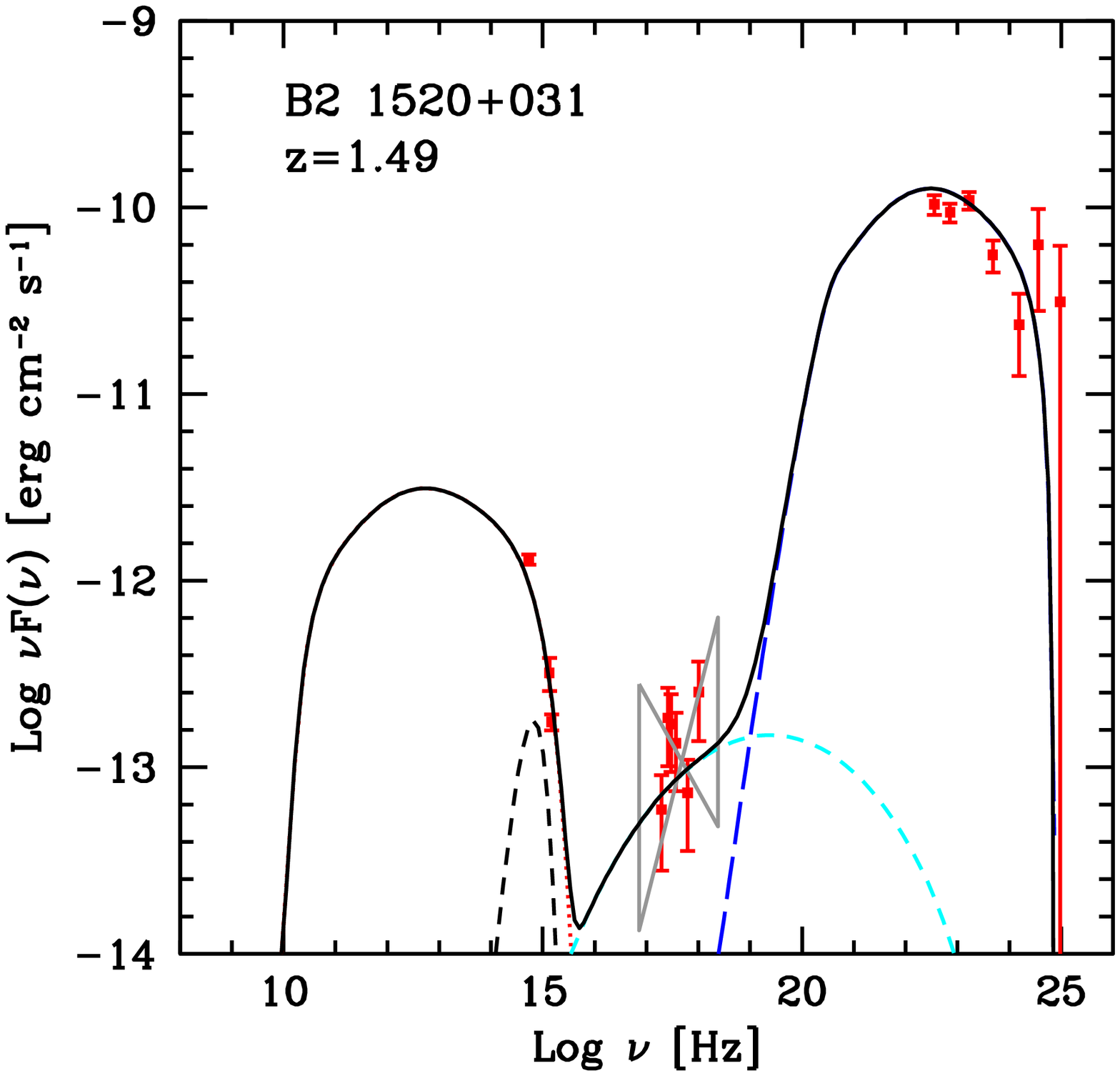,width=7.5cm}\\
\psfig{figure=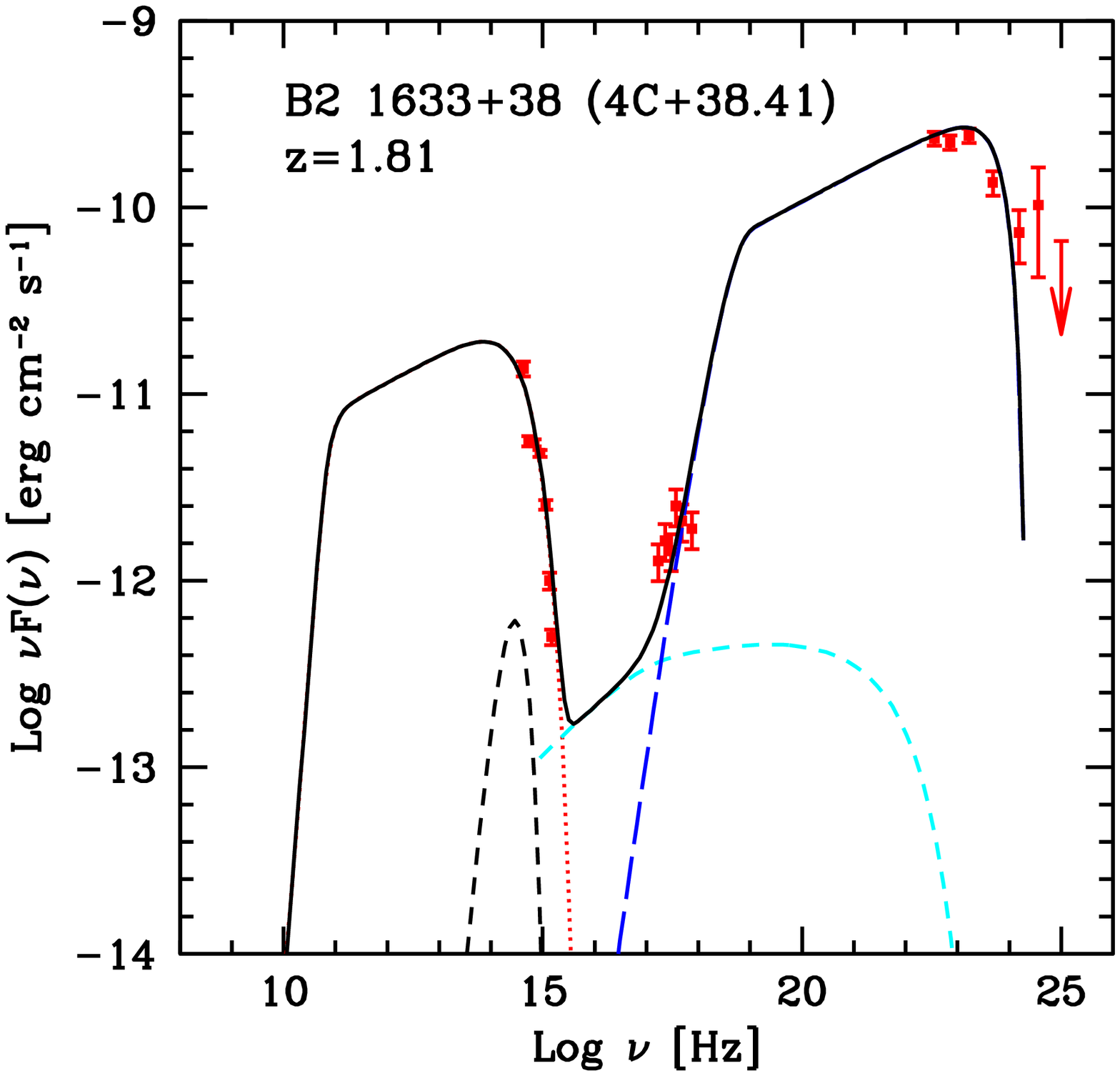,width=7.5cm}&
\psfig{figure=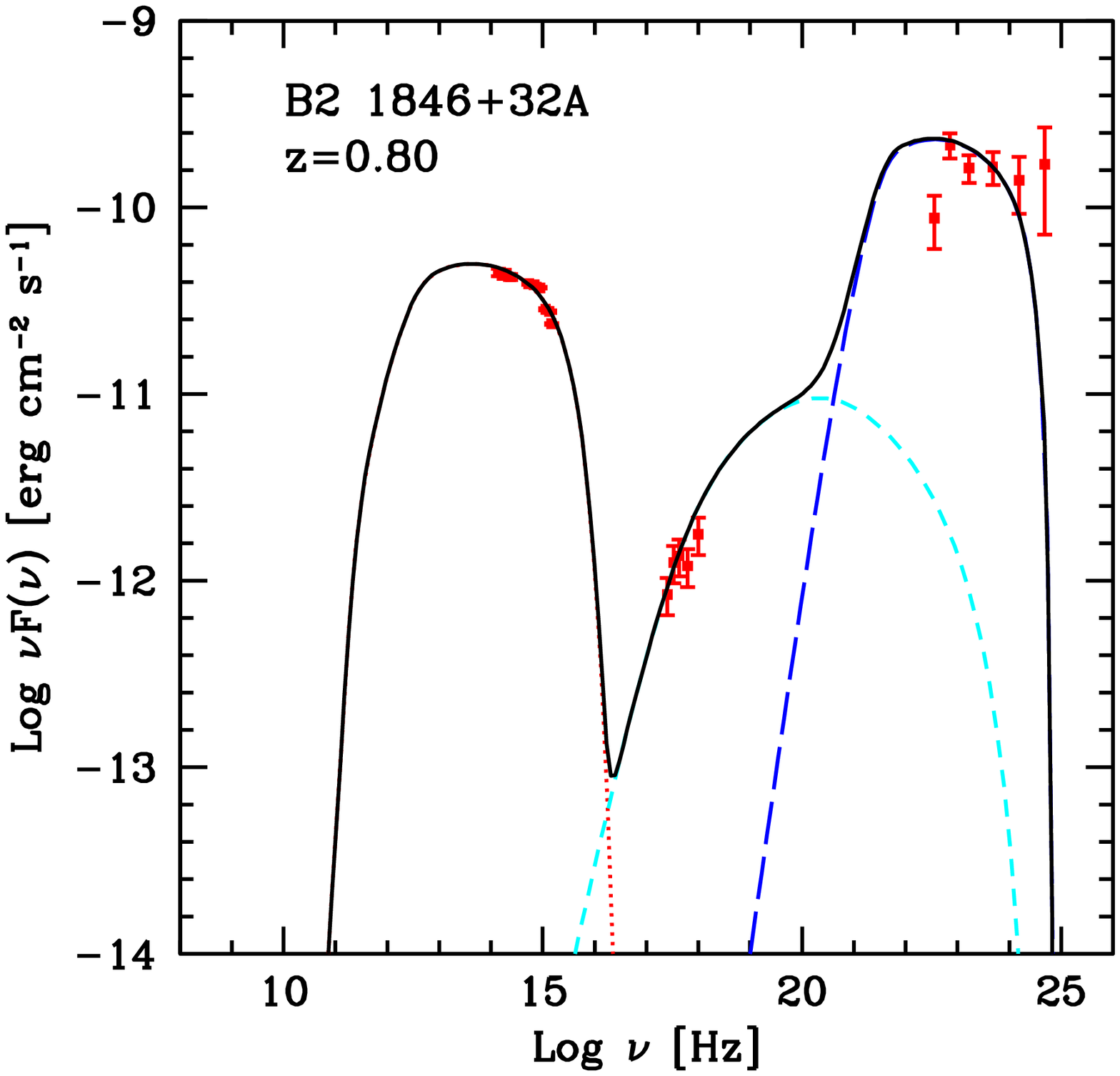,width=7.5cm}\\
\end{tabular}
\caption{SEDs for the reported flares of PKS 0250-225, PKS 0454-234, PKS 1502+106, B2 1520+31, 4C +38.41, B2 1846+32A.  Dotted lines represent Synchrotron emission, short dashed lines represent the disk emission and the SSC, long dashed lines represent the EC on torus photon field.}
\label{fig_sedall1}       
\end{figure*}

\begin{figure*}
\centering
\begin{tabular}{cc}
\psfig{figure=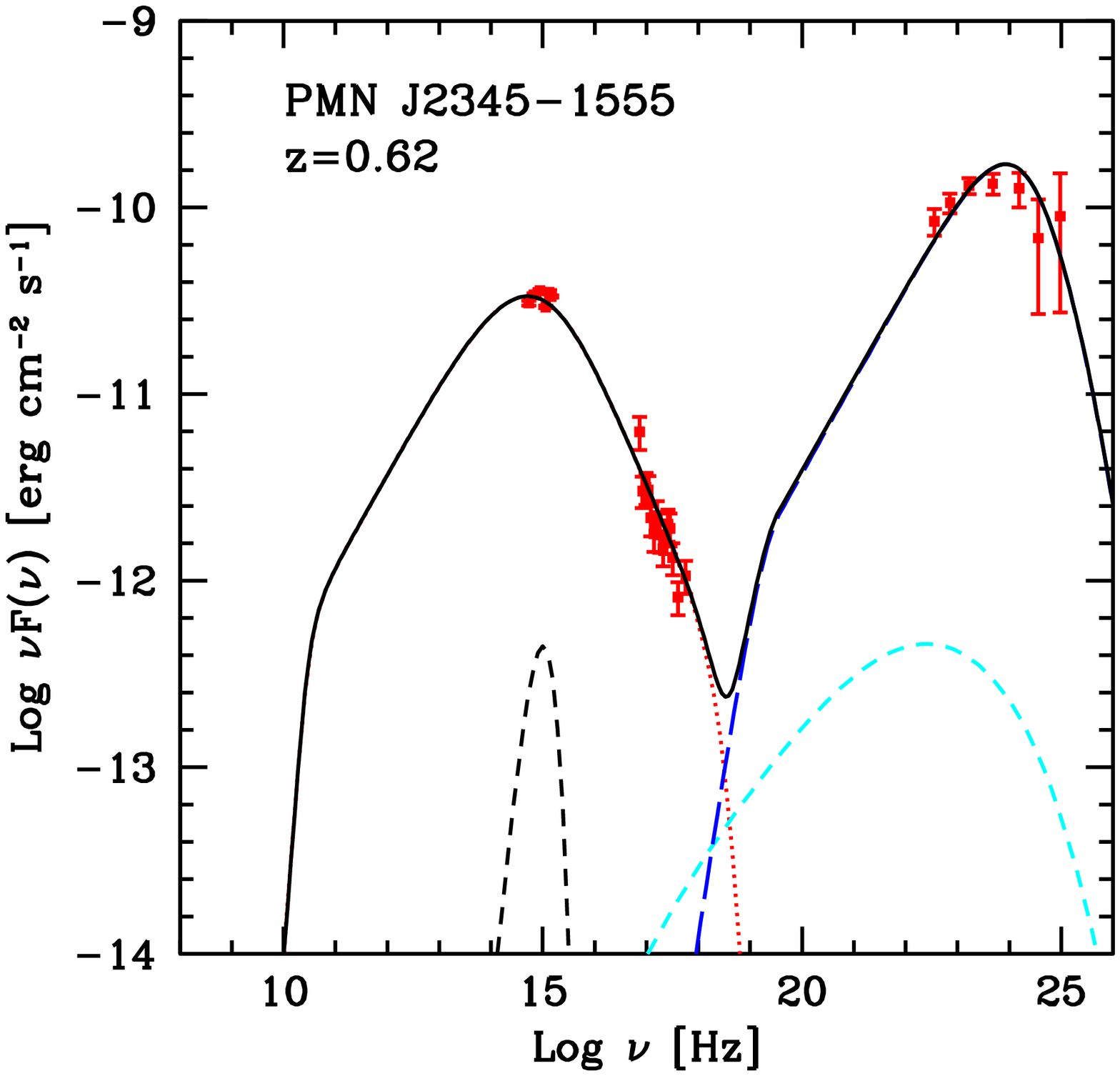,width=7.5cm} &
\psfig{figure=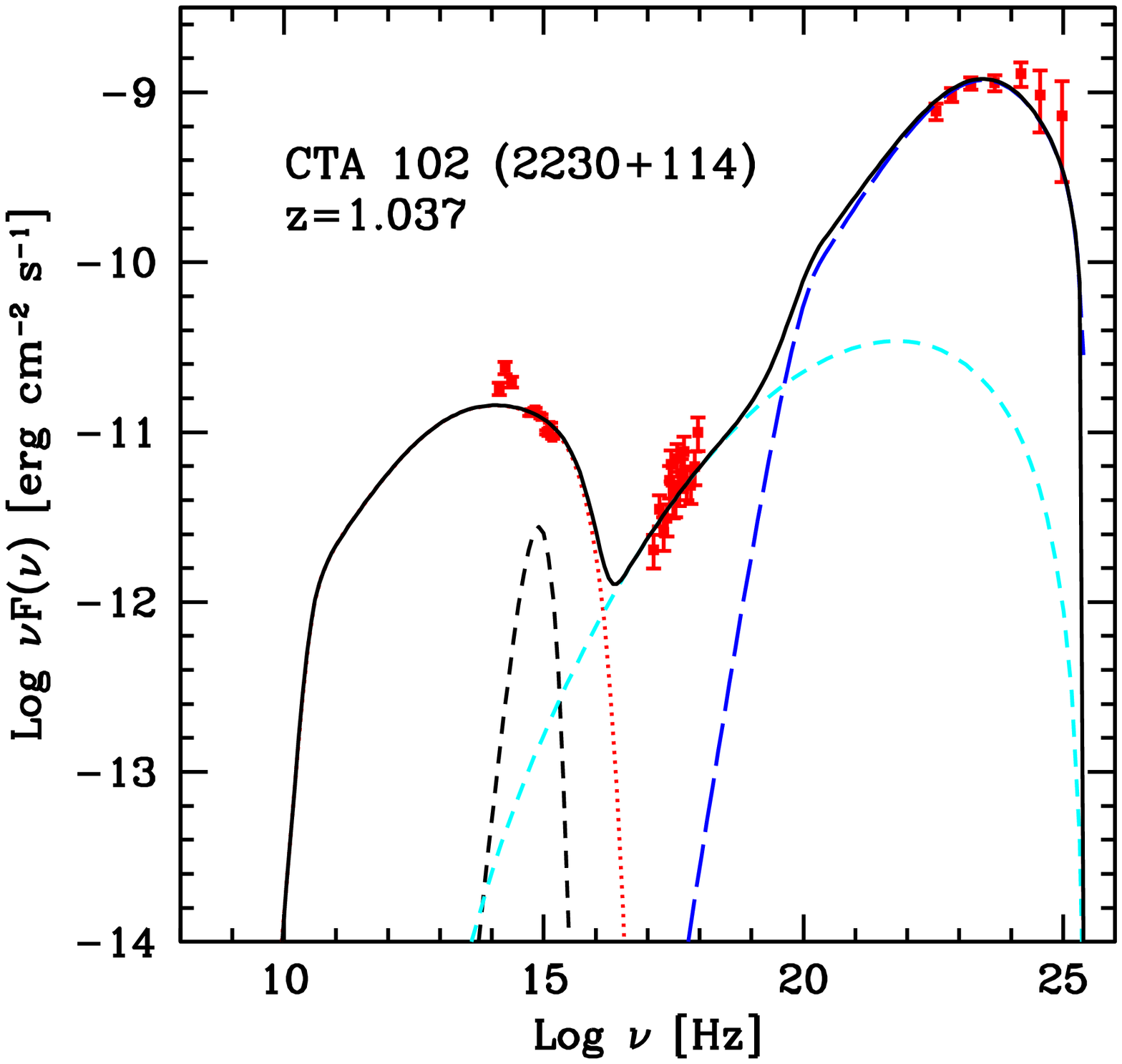,width=7.5cm}\\
\psfig{figure=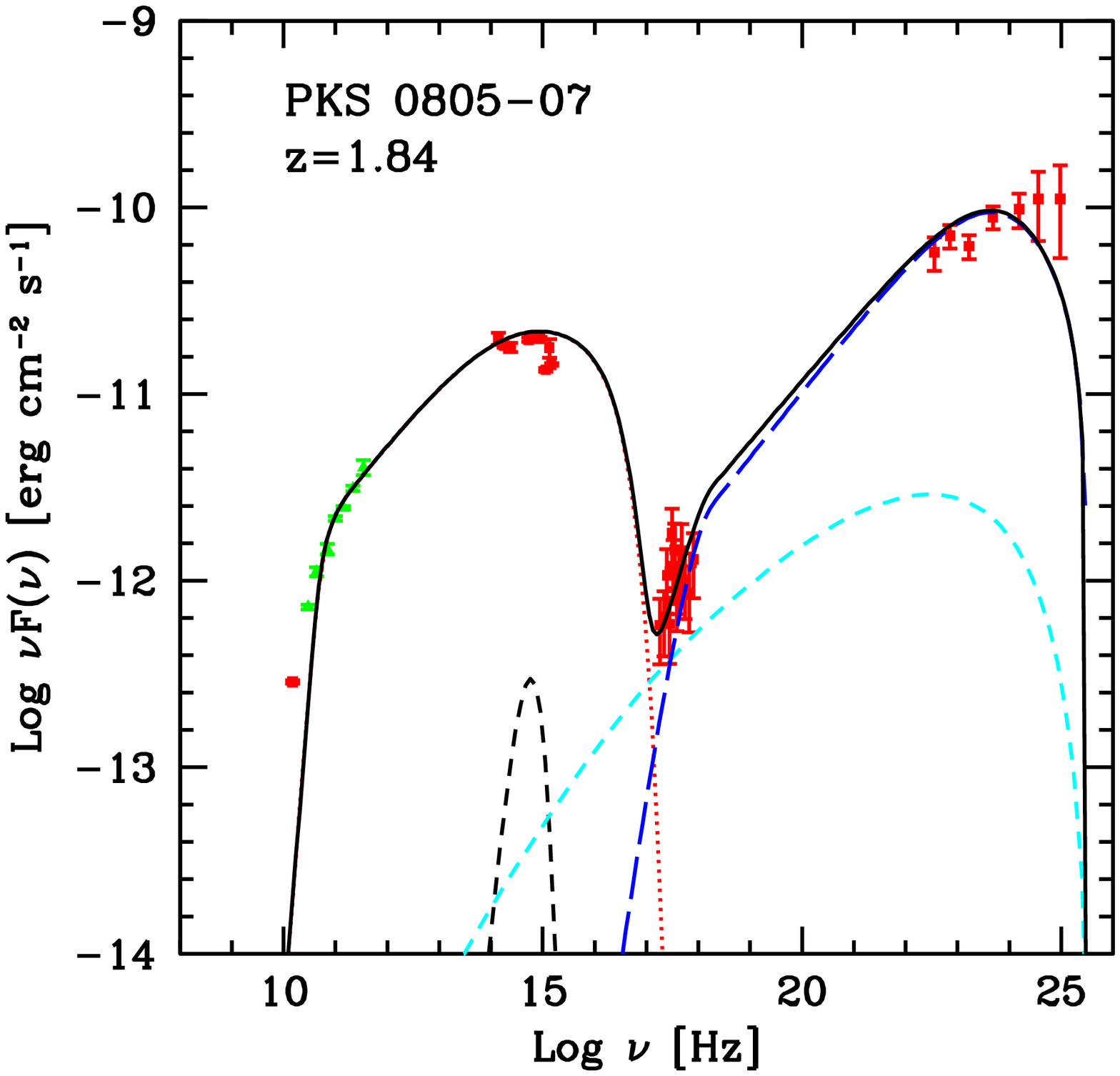,width=7.5cm} &
\psfig{figure=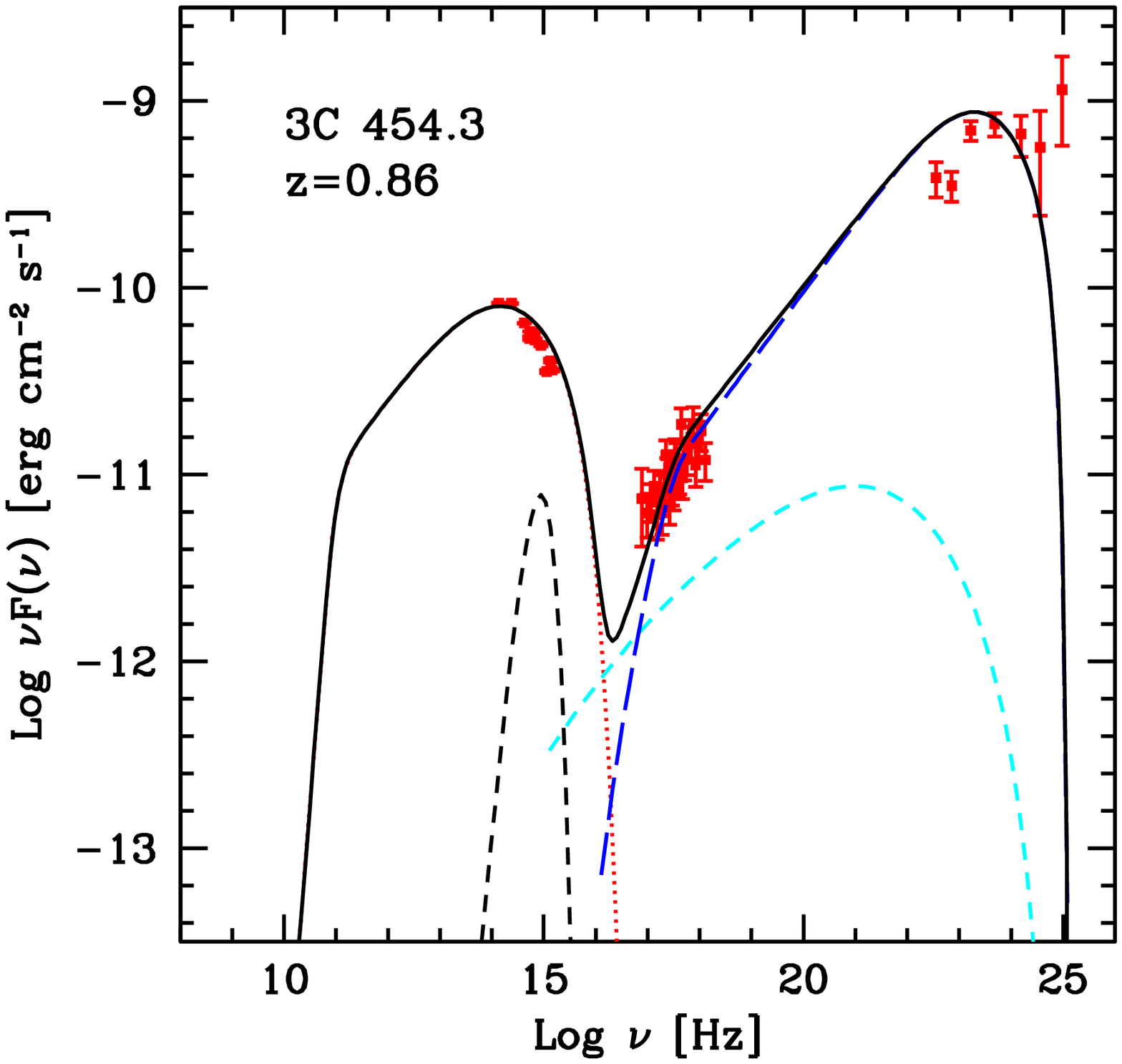,width=7.5cm}\\
\end{tabular}
\caption{SEDs for the reported flares of
PMN J2345-1555, CTA 102, PKS 0805-07, 3C 454.3. Dotted lines represent Synchrotron emission, short dashed lines represent the disk emission and the SSC, long dashed lines represent the EC on torus photon field.}
\label{fig_sedall2}       
\end{figure*}

The gamma-ray spectra during flares of the sources we investigated are reported together with the other multiwavelength simultaneous data in Figure \ref{fig_sedall1} and Figure \ref{fig_sedall2}.
For all sources we integrated the gamma-ray data for the whole period of HE emission, except for  B2 1846+32A with gamma-ray data integrated for 4 days around the NIR and optical-UV observations,
and for CTA 102 and PKS 0454-234 showing variability within the HE activity period.
We integrated gamma-ray data between 2012-09-22 02:00 and 2012-09-24 14:00 for the former, and between 2012-12-02 00:00:00 and 2012-12-06 00:00:00 for the latter.\\
For PMN J2345-1555 we observed a peculiar flare similar to that reported in \cite{ghiseblue} for the same source, with synchrotron emission extended up to X-rays.\\
We obtained a multiwavelength SED for an interesting flare of 3C 454.3, which reached a flux of 350$\times$10$^{-8}$ ph cm$^{-2}$ s$^{-1}$ (E$>$ 100 MEV),
with a flat gamma-ray spectrum up to 40 GeV and a photon index 1.82$\pm$0.06 which is rather different from the previous flares                      
reported from the {\it FERMI--LAT} data (see., e.g., \citealt{abdo454,finke454}),                                                                    
showing soft gamma-ray spectra with a break or a cutoff in the GeV range.\\
We found a similar recent flare in the archival data of 3C 454.3. Integrating {\it FERMI--LAT} data between 2013-04-05 17:00 and 2013-04-12:00,
we obtained a flux of $\sim$80$\times$10$^{-8}$ ph cm$^{-2}$ s$^{-1}$ (E $>$ 100 MeV), and a gamma-ray photon index of 1.93$\pm$0.09.
We did not found in the {\it FERMI--LAT} archive other activity periods of the source showing the same spectral characteristics.\\
By definition our sample is biased towards flares showing relevant emission above 10 GeV. We showed in Table \ref{tab_he_sample} that the collected gamma rays with energy $>$ 10 GeV can not be explained by chance coincidences,
thence the observed emission at HE is intrinsic to flaring state of the sources.
All the spectra show no, or at least negligible evidence of absorption,
with the possible exception of 4C +38.41 (which will be discussed later in this paper), and B2 1520+031. If the gamma-ray emission were produced inside the BLR cavity, we expect absorption from $\gamma \gamma$ absorption with the BLR photons:
at the threshold energy (E$_{thr}^{Ly\alpha}$=$\frac{25\ {\rm GeV}}{1+z}$) for
$\gamma \gamma$ absorption with the H Ly$\alpha$ target photons, the optical depth is of the order of $\tau\ \sim \ \frac{\sigma _{T}}{5}n_{Ly\alpha}R_{BLR}$ \citep{tavecchio2013}
where $\sigma_{T}$ is the Thomson cross section, and the density of target photons is
$n_{Ly\alpha}$=$L_{Ly\alpha}/4\pi R^2_{BLR}ch\nu_{Ly\alpha}$, and assuming the blazar dissipation zone in the center of a spherical shell shaped BLR.
The H Ly$\alpha$ luminosity can be estimated starting from the broad lines spectroscopy, using the template reported in \cite{francis1991} and the
corrections proposed by \cite{celotti1997}. 
The accretion disk luminosities (L$_{disk}$) have been evaluated assuming the BLR luminosity to be $\frac{1}{10} L_{disk}$ \citep{baldwin1978}.
The internal radius of the BLR can be inferred from the relation connecting it to the disk luminosity as indicated by reverberation mapping studies \citep{bentz2009}.
Following \cite{ghise_tav} this can be written as: $R_{BLR}=10^{17} L_{disk,45}^{0.5}$ cm, and $R_{BLR}^{out}\sim 4 \times R_{BLR}$.
\cite{liu2006} performed a refined  evaluation of the optical depth as a function of the blazar dissipation zone (i.e., outside the BLR cavity).
We evaluate the optical opacity starting from their findings and interpolating for the disk luminosity of our sample.

\begin{figure}
\centering
\psfig{figure=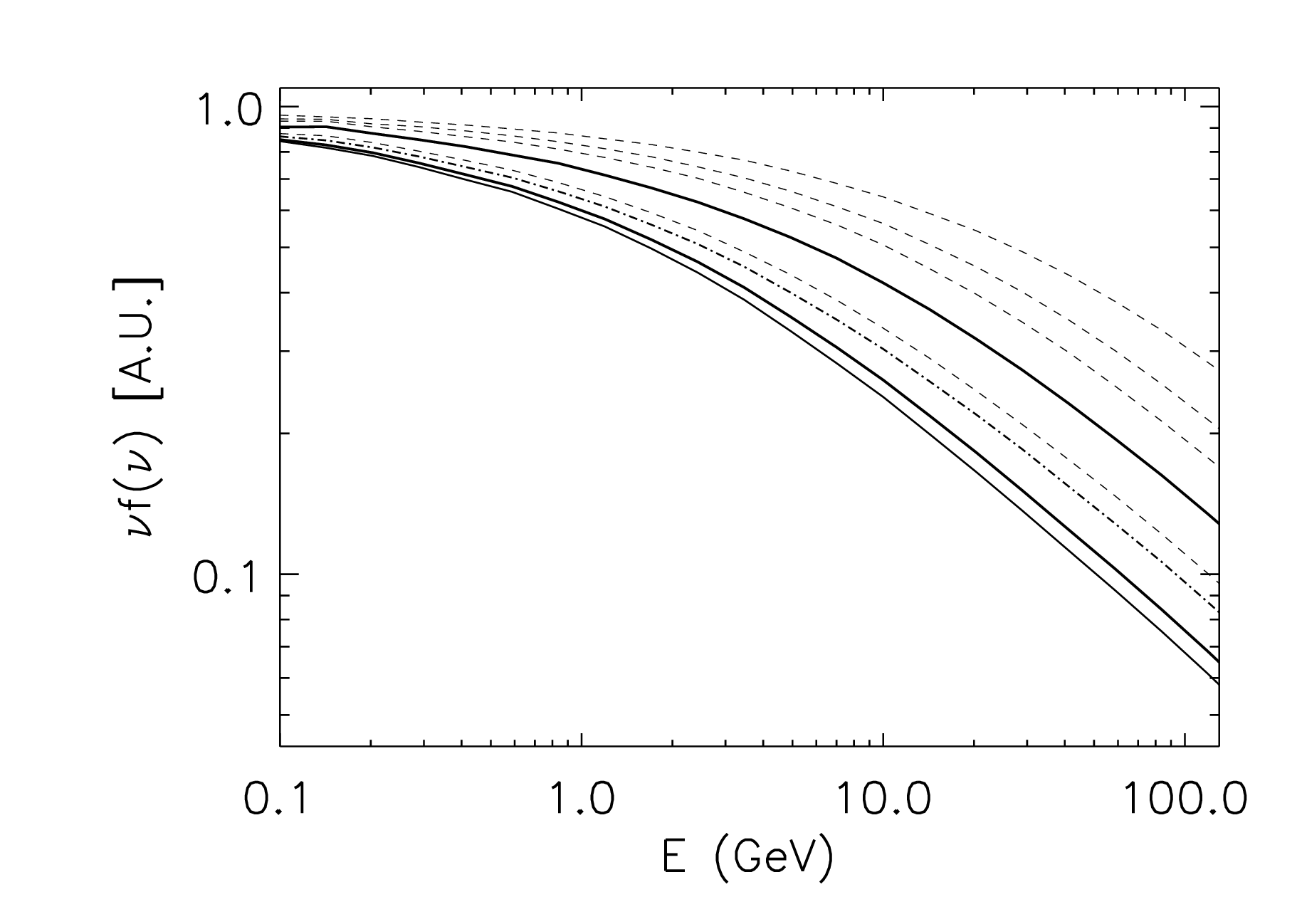,width=7.5cm,height=5.5cm}
\caption{Gamma-ray SED for the IC scattering  with seed photons from the BLR as a function of the dissipation region.
From Bottom Up, the solid curves refer to a dissipation region located at the center of the BLR cavity, at $R_{BLR}$,  $R_{BLR}^{ext}$. Dot dashed curve refers to a dissipation region at  $R_{BLR}^M$.
From Bottom Up, the four dashed curves refer to a dissipation region located at $3.8\times R_{BLR}$, $5\times R_{BLR}$, $6\times R_{BLR}$, $8\times R_{BLR}$
(corresponding to $0.95\times R_{BLR}^{ext}$, $1.25\times R_{BLR}^{ext}$, $1.5\times R_{BLR}^{ext}$, $2\times R_{BLR}^{ext}$ in our model).}
\label{fig_knsed}       
\end{figure}

\begin{table*}
 \centering
  \begin{tabular}{lccccccccll}
  \hline
source & \multicolumn{7}{c}{$\tau_{\gamma\gamma}$} & L$_{disk}$          &luminosity  \\ \cline{2-8}
name   &  \multicolumn{3}{c}{at $R_{BLR}$} & & \multicolumn{3}{c}{at $R_{BLR}^M$}  & $10^{45}$ & estimator \\ \cline{2-4} \cline{6-8}
       & at E$_{thr}^{Ly\alpha}$ & at 35 GeV & at 50 GeV & & at E$_{thr}^{Ly\alpha}$ & at 35 GeV & at 50 GeV  & erg/s &  \\ \hline
 PKS 0250-225  &  3.7 &  6.3 &  7.8 & & 0.6 &  1.1 &  1.6  &  5.3  & \ione{Mg}{II} \citep{shaw2012} \\
 PKS 0454-234  &  3.1 &  5.3 &  6.5 & & 0.5 &  0.9 &  1.3  &  3.7  &      \ione{Mg}{II} \citep{stickel1993}\\
 PKS 1502+106  &  6.2 & 10.6 & 13.1 & & 1.1 &  1.9 &  2.7  &  15.  & \ione{Mg}{II}, \ione{C}{IV} (\citealt{shaw2012}; \\
               &      &      &      & &     &      &       &       &  \citealt{sbarrato2012}) \\
 B2 1520+031   &  4.6 &  7.8 &  9.6 & & 0.8 &  1.4 &  2.0  &  8.   &      \ione{Mg}{II} \citep{shaw2012}\\
 4C +38.41     & 11.4 & 19.4 & 23.9 & & 2.0 &  3.5 &  4.9  &  50.  & \ione{Mg}{II}, \ione{C}{IV} (\citealt{stickel1993};\\
               &      &      &      & &     &      &       &       &  \citealt{sbarrato2012}) \\
 B2 1846+32A   &  3.0 &  5.1 &  6.2 & & 0.5 &  0.9 &  1.3  &  3.4  &      \ione{Mg}{II} \citep{shaw2012}\\
PMN J2345-1555 &  2.0 &  3.4 &  4.1 & & 0.3 &  0.6 &  0.9  &  1.5  & \ione{Mg}{II}, \ione{C}{IV} \citep{stickel1993}\\
       CTA 102 & 10.3 & 17.6 & 21.7 & & 1.8 &  3.1 &  4.5  &  41.  & $\times$10 L$_{BLR}$ \citep{pian2005}\\
PKS 0805-07    &  7.9 & 13.5 & 16.6 & & 1.4 &  2.4 &  3.4  &  24.  & \ione{Mg}{II}, \ione{C}{IV} \citep{white1988}\\
3C 454.3       &  9.2 & 15.8 & 19.4 & & 1.6 &  2.8 &  4.0  &  33.  & $\times$10 L$_{BLR}$ \citep{sbarrato2012}\\ 
\hline
\end{tabular}

\caption{
Disk luminosity and optical depth for $\gamma \gamma$ absorption evaluated at  E$_{thr}^{Ly\alpha}$, 35 and 50 GeV for photons emitted at the internal radius ($R_{BLR}$)  of a spherical shell of BLR, and for photons emitted in the middle between the internal and external radius of the shell  ($R_{BLR}^M=\frac{R_{BLR}+R_{BLR}^{out}}{2}$) for our sample of FSRQs. The opacity is evaluated interpolating the results of \cite{liu2006} for the disk luminosities of our sample.} The last column gives the broad lines used to evaluate the disk luminosity.
\label{tab_he_luminosity}
\end{table*}
The interpolated optical depth at E$_{thr}^{Ly\alpha}$, 35 and 50 GeV for all the sources is reported in Table \ref{tab_he_luminosity}, where we report also the broad line used to evaluate the
accretion disk luminosity, and the reference to the original data. The estimated $\gamma \gamma$ absorption for all the sources in our sample will produce relevant features in the
collected spectra if the emitting region is inside the cavity of the BLR ($<R_{BLR}$).
Moreover, for the 5 objects with the highest disk luminosities, the BLR opacity will produce relevant features for an emitting region at $R_{BLR}^M$ or beyond.
The opacity argument could not be used to exclude emitting regions located toward the outer edge of the BLR, as far as the opacity becomes negligible, and its effect on the gamma-ray spectra undetectable. \\
In the framework of leptonic models, flares dissipating inside the BLR cavity will emit gamma-ray photons subject to the Klein-Nishina suppression above $\sim$ 9 GeV (see, e.g. \citealt{tavecchio2008}).
We evaluated the Klein-Nishina suppression on the IC scattering on BLR photons as a function of the dissipation distance from the SMBH. We used the
Klein-Nishina Cross Section evaluated in \cite{aharonian1981} for a monoenergetic photon beam on isotropically distributed electrons, the parametrization proposed in \cite{liu2006} for the 
lines and continuum BLR emissivity, and the BLR spectrum computed with CLOUDY and proposed in \cite{tavecchio2008}. The jet bulk Lorentz factor $\Gamma_{bulk}$ is 15, and we assume a power-law electron distribution
with slope p=3 (corresponding to a photon spectral index $\alpha$=1 in  Thomson regime). The line of sight forms an angle $\frac{1}{\Gamma_{bulk}}$
with the jet bulk motion.
Results are reported in figure \ref{fig_knsed}, normalized to the SED in the Thomson approximation.
In the jet frame, the upstream boosted BLR energy density prevails in the SED until the jet approaches $R_{BLR}$. From about this dissipation region, the downstream
BLR energy density contribution is no more negligible. For a dissipation region at the edge or beyond the BLR, all the BLR photons come from behind.  
The net effect is that starting from $R_{BLR}$, the center of mass energy of the BLR photon - electron scattering starts to decrease for each direction of the incoming electrons, thence
progressively mitigating the Klein-Nishina suppression. For a jet dissipating at $R_{BLR}^{out}$, the  Klein-Nishina suppression is  3.7, 4.3, 5.2, 6 for a gamma-ray energy of 25, 35, 50, 80 GeV respectively.\\
Combining the two arguments (the model independent $\gamma \gamma$ absorption, and the Klein-Nishina suppression for the IC emission with seed photons from the BLR in leptonic models),
the gamma-ray spectra we showed could only be explained assuming a dissipation region at the outer edge of the BLR or beyond.
Our SED modeling, and time-resolved spectra will give further clues on the location of the gamma-ray dissipation region.\\

We modeled the SED for each epoch in the framework of leptonic
models, and with the parametrization of the photon field originating from the BLR and dusty torus provided in \cite{ghise_tav} and already used in \cite{tavecchio2013}.
The emission region (the ``blob"), assumed to be spherical with radius $R$ and moving with bulk Lorentz factor $\Gamma_{bulk}$, carries a magnetic field with intensity $B$ and a population of relativistic electrons.  The electron energy distribution is assumed to follow a smoothed broken power law function  between $\gamma_{min}$ and $\gamma_{max}$, with slopes $n_1$ and $n_2$ below and above a break at $\gamma_{break}$ and normalization $K$. The blob velocity is assumed to form an angle $\theta _{v}$ with respect to the line of sight, so that relativistic amplification effects can be fully specified by the relativistic Doppler factor $\delta$.

Given the disk luminosity of the FSRQ, the external radiation fields interacting through External Compton to produce gamma rays can be fully parametrized in terms of distance of the radiating ejecta from the SMBH.\\
As detailed in \cite{tavecchio2013}, with the suggested parametrization, and with the information of the disk luminosity, we can directly obtain an estimate of the location of the radiating ejecta. 
We briefly recall the chain of arguments followed in \cite{tavecchio2013}. 

Let us assume the torus radiation field dominates the emission (e.g., the dissipation region is at least at the outer edge of the BLR), with a black body spectrum peaked at $3\times 10^{13}$ eV.
The position of the IC peak provides the value of the Lorentz factor of the electrons emitting at the peak, $\gamma _{\rm p}$, which, in turn can be used to derive the value of the magnetic field from the synchrotron peak frequency, $\nu_{s}\simeq 3.7\times 10^6 B \gamma _{\rm p}^2  \, \delta/(1+z)$ Hz. Since, during HE outburst the IC maximum lies at energies much larger than those usually observed in FSRQ, the corresponding $\gamma _{\rm p}$ will be larger, directly implying a low magnetic field. A further step can be done exploiting the observed IC/synchrotron luminosity ratio, the so-called Compton dominance, directly proportional to the ratio of the external radiation and magnetic energy densities in the jet frame,
$L_C/L_s=U^{\prime}_{\rm ext}/U^{\prime}_B=U_{\rm ext}\Gamma_{bulk}^2/U^{\prime}_B$. Since the magnetic field is known, the Compton dominance allows us to infer the radiation field energy density and therefore, thanks to the link with the distance from the SMBH, the position of the emission region.\\
If, contrary to our previous assumption, the BLR radiation field dominates the emission over the torus IR field, then the magnetic field must be enhanced by a factor $\frac{\nu_{BLR}}{\nu_{IR}}\sim$100 (really more than this value on account of the KN suppression).
To reproduce the observed Compton-dominance, we have to rise the $U^{\prime}_{ext}$ (that now corresponds to $U^{\prime}_{BLR}$) by a factor $\frac{\nu_{BLR}}{\nu_{IR}}\times\frac{\tau_{IR}}{\tau_{BLR}}\sim$600.
This consideration makes it hard to model an SED for a dissipation region outside the BLR (in the dusty torus photon field with energy density $U_{IR}$)  making use of the seed photon field from the BLR.
In fact we have to rise  $U^{\prime}_{BLR}$ by  a factor 600 with respect to $U^{\prime}_{IR}$.
Plausible values for the maximum of the ratio $\frac{U^{\prime}_{BLR}}{U^{\prime}_{IR}}$ are $\sim$100 (\citealt{ghise_tav}, that adopted $\tau_{IR}=0.6$); and $\sim$600 (\citealt{sikora2}, that adopted $\tau_{IR}=0.1$).
Then we have also to choose the lower value for $\tau_{IR}$ (e.g., $\tau_{IR}$ $\le$ 0.1) in the modeling. With this choice
the  maximum of the ratio $\frac{U^{\prime}_{BLR}}{U^{\prime}_{IR}}$ could be obtained putting the dissipation region at $\sim R_{BLR}$ or below,
but this range of parameter $R_{diss}$ is excluded by the lack of $\gamma \gamma$ absorption and KN suppression
in the observed gamma-ray spectrum.\\

More generally speaking, in leptonic SED modeling
the characterization of the blazar zone as located outside the BLR (and irrespective of the blob radius, electron density, and $\Gamma_{bulk}$),
is a quite stable result, once the region inside the BLR cavity is excluded.
In fact, the optical-UV data show the synchrotron high-energy tail in the SEDs of all the flares, while the gamma-ray data
can be subdivided into two subsets, showing  {\it (a)} the EC High Energy tail, or {\it (b)} a flat EC spectrum with no fall at High Energy (with the exception of  PKS 0250-225, B2 1520+031, 4C +38.41
for which the observations do not detail the synchrotron emission). 
In case {\it (a)}, these data bind the B/$\Gamma_{bulk} \nu_{ext}$, and the B$^2$/$U_{ext}\Gamma_{bulk}^2$  ratios in the modeling,
thence the estimate of the $U_{ext}$ and B/$\Gamma_{bulk}$ parameter does not depend on the estimate of $\Gamma_{bulk}$ (once we have identified the prevailing seed photon field),
and does not depend on electron energy density and radius of the emitting blob. 

Once we have established that the dissipation region is at a distance $R_{diss}>R_{BLR}$ (outside the BLR cavity),
there are other two options to consider: {\it i)} the
BLR prevails on the IR radiation field, then $U'_{ext} \sim U^{\prime}_{BLR}$, which rapidly varies with $R_{diss}$ ($U^{\prime}_{BLR}$ drops of a factor $\sim 100$ at $R_{BLR}^{ext}$ with respect to $R_{BLR}$,
and a further factor 10 at $\sim 1.5\times R_{BLR}^{ext}$).
This circumstance makes the localization of $R_{diss}$ a quite stable parameter.
{\it j) IR prevails on the BLR radiation field}, then $U'_{ext} \sim U'_{IR}$, $U_{IR}$ is constant until $R_{diss} < R_{torus}$, and $\nu_{ext} = \nu_{IR}$.
Thence the estimate of the $U_{ext}$ parameter does not depend on the estimate of $\Gamma_{bulk}$. Furthermore, assuming SSC emission dominates the X-ray range, we can consider the ratio of SSC
to Synchrotron emission to further constrain the model components: the product $R_{blob}\times n_{eo}$ (see, e.g., \citealt{dermer1997} for the formula and for the definition of $n_{eo}$) must be held constant in the modeling,
due to the observational constraints. Now, using the formula for the power spectral density in \cite{dermer1997}, we obtain that the product $\Gamma_{bulk}^5 \times R_{blob}^2$ must be held constant (using the observational constraints
already obtained for $R_{blob} \times n_{eo}$, $B/\Gamma_{bulk}$, and on $U_{IR}$). Reasonable values for  $\Gamma_{bulk}$ are in the range 10-50 which are about a factor 2 around the value we chose for our modeling (see below).
This means that we could obtain reasonable models varying $R_{blob}$ in the opposite direction with respect to $\Gamma_{bulk}$
(and at most by a factor $2^{5/2}\sim$6 around the value we chose, corresponding to a change of a factor $2^{-1}$ of $\Gamma_{bulk}$), and maintaining the constraints from the measured quantities.
But the decreasing of the parameter $R_{diss}$ meets almost  a barrier  to low values (at $R_{diss} \sim R_{blr}^{ext}$), where $U_{blr}$ rapidly varies and prevails on  $U_{IR}$. Thence $R_{diss}>R_{blr}^{ext}$.
Remembering that we chose $R_{blob}\sim 0.1\times R_{diss}$, thence $R_{diss}$ could be pushed further out with respect to our modeling, at the expense of a corresponding lowering of $\Gamma_{bulk}$,
or accepting a lower value for the ratio $\frac{R_{blob}}{R_{diss}}$.
Summing up, in case {\it (j)} $R_{diss}$ is constrained at lower values ($R_{diss}>R_{blr}^{ext}$), and could be pushed further out the value we chose in our modeling by a factor 6 at most.\\
We discussed case {\it (b)} in \cite{pacciani}. In this case, we obtain from the optical-UV and gamma-ray data an upper limit for the  B/$\Gamma_{bulk} \nu_{ext}$ ratio, and a lower limit for $R_{diss}$.\\
For two of the flares with poor optical coverage of the synchrotron emission (PKS 0250-225 and 4C +38.41), the optical data give at least a lower limit for the synchrotron peak height, and we were able to derive a lower limit for
$R_{diss}$ from the SED modeling).\\
The previous considerations apply only for one-zone leptonic models. We explicitly do not tried to model the SEDs with other modeling, such as the spine-sheath or hadronic models.

\begin{table*}
 \centering

  \begin{tabular}{p{2.3cm}p{0.5cm}p{0.4cm}p{0.4cm}p{0.6cm}p{0.4cm}p{0.3cm}p{0.3cm}p{0.5cm}p{0.3cm}p{0.43cm} p{0.6cm} p{0.6cm} p{0.5cm} p{0.6cm}p{0.6cm}p{0.4cm}p{0.4cm}}
  \hline

source name & L$_{disk}$          & B        & $\gamma_{min}$ & $\gamma_{break}$ & $\gamma_{max}$ & n$_1$  & n$_2$ & K        & $\delta$ & $\Gamma_{bulk}$ & L$_{kin}^p$ & L$_{kin}^e$& $\frac{L_{jet}}{L_{disk}}$&R$_{BLR}$ & R$_{torus}$& r$_{blob}$ & d$_{blob}$ \\  
            & 10$^{45}$           &10$^{-2}$ &               &     10$^3$      &      10$^3$    &        &       &10$^3$    &           &              &  10$^{45}$   & 10$^{45}$  & &10$^{17}$ & 10$^{18}$ & 10$^{17}$  & 10$^{18}$  \\
            &  erg/s             & G        &               &                 &               &        &        &cm$^{-3}$ &           &              &   erg/s      &  erg/s    & & cm      &   cm      &  cm       &    cm    \\  
            &  [1]               & [2]      &      [3]      &      [4]        &      [5]      &   [6]  &  [7]   &    [8]   &    [9]   &     [10]      &  [11]       &  [12]      & &[13]      & [14]      &  [15]      &   [16]  \\ \hline 
  PKS 0250-225 &     5.3 &    2.5 &   55 &   8 &   20 &   2.7 &   3.2 &   10 &  26 &  25 & 330   & 29   & 60 & 2.30 &  5.76 &  8    &  9\\
  PKS 0454-234 &     3.7 &    3.5 &  350 &   4 &   20 &   2.7 &   3.2 &   20 &  26 &  15 & 110   & 9.5  & 30 & 1.92 &  4.81 &  8    &  7.6\\
  PKS 1502+106 &      15 &     11 & 100  & 0.9 &   17 &     2 &   3.35& 0.24 &  26 &  15 &  24   & 5.4  &  1 & 3.87 &  9.68 &  6    & 6\\
   B2 1520+031 &       8 &    6.3 &  100 & 0.9 &   15 &     2 &    3.5&  0.3 &  25 &  14 &  13.3 & 3.1  &  2 & 2.83 &  7.07 &  4    & 4\\   
    4C +38.41 &       50 &     26 &   13 & 0.15 &  6.3 &   2.7&   2.7 &   16 &  25 &  17 &  79   &  1.7 & 1.6& 7.07 & 17.68 &  4. &  4.\\
   B2 1846+32A &     3.4 &    59  &  235 & 0.27&   90 &   2.2 &   3.4 &  47  &  22 &  20 &   5.1 &  1.9 & 1.5& 1.84 &  4.61 &  0.5  &  0.7\\
PMN J2345-1555 &     1.5 &     21 &   18 &   7 &  300 &     2 &   4.3 & 0.03 &  26 &  21 &  4.6  & 0.42 &  3 & 1.22 &  3.06 &  2.   &  2\\
CTA 102        &      41 &   14.3 &  1.7 & 3.3 &   30 &     2 &  3.55 &  5.6 &  20 &  14 &  920  & 8.5  & 20 &  2.03&  5.09 &  1    &  1\\    
   PKS 0805-07 &      24 &   31.5 &    5 &   8 &   50 &   2.3 &   3.5 &  2.4 &  20 &  15 & 460   & 6.0  & 20 & 4.90 & 12.25 & 2.5   & 2.5\\    
       3C454.3 &      33 &     28 &  2.5 &   4 &   14 &  2.25 &   4.  &  4.3 &  24 &  15 &   294 & 4.6  & 8.8&  5.7 & 14.4  & 1.5   & 2.3\\
\hline
\end{tabular}

\caption{
Parameters of SED modeling of our sample of flares. [1] disk luminosity ($10^{45}$erg/s), [2] magnetic field (mG) , [3] minimum random Lorentz factor of electrons, [4] break random Lorentz factor of electrons (10$^3$), [5] maximum random Lorentz factor of electrons (10$^3$),
[6] low energy slope of the electron population, [7] high energy slope of the electron population, [8] electron density (10$^3$ cm$^{-3}$), [9] Doppler factor, [10] bulk Lorentz factor,
[11] kinetic power of protons (10$^{45}$erg/s, assuming one cold proton per electron), [12] kinetic power of electrons (10$^{45}$erg/s),
[13] Broad Line region Radius ($10^{17}$ cm), [14] Molecular Torus Radius ($10^{18}$ cm), [15] Blob radius ($10^{17}$ cm), [16] Blob distance from the SMBH ($10^{18}$ cm).}
\label{tab_he_sed_model}
\end{table*}

We reproduced the observed flare SED adjusting the parameters to obtain the best agreement with data.
The derived  model parameters are reported in Table \ref{tab_he_sed_model}. Note that the radius of the BLR and the torus are not free parameters but are fixed by $L_{disk}$ (which is determined by the BLR luminosity). Moreover, we reduced the number of free parameters assuming that the radius of the emitting region is roughly 1/10 of the distance from the SMBH. Source radii and Doppler factors can also be constrained by the observed duration of the gamma-ray high state, $\Delta t_{high}>t_{\rm cross}=R(1+z)/\delta c$, estimated from the light curves. Summing up, the adopted model has a total of 9 free parameters.

Our procedure naturally implies some degree of uncertainty. In this respect, the most critical point is the position of the cut-off in the gamma-ray spectrum. Higher energies implies higher electron Lorentz factors and, following the reasoning above, lower magnetic fields which, implying lower external photon energy densities, leads to infer larger distances. In all cases we try to assume the case providing the most conservative estimate of the source distance d$_{blob}<R_{torus}$, at the cost to slightly underestimate the flux at the highest energies. This is also justified in view of the fact that, although close in time, the data collected in the SED are not strictly simultaneous and (see below), the gamma-ray spectra show hints of variability at the highest energies on relatively short ($\sim$ day) timescales.
However, for two sources, PKS 0250-225 and PKS 0454-234, the shape of the SED does not allow for the solution corresponding to R$_{BLR}\ <$ d$_{blob}$ $<$ R$_{torus}$ and forces the assumption of large distances. The reason, similarly to PKS B1424-418 \citep{tavecchio2013}, is the large separation between the IC and synchrotron peak, which result in a rather low magnetic field. 
We found dissipation regions located between 0.3 and 3 pc from the SMBH, with the possible exception of B2 1846+32A. We obtained similar results for GB6 J1239+0443 \citep{pacciani} and PKS B1424-418 \citep{tavecchio2013}.\\
The IGM Lyman Complex absorption affects the {\it Swift--UVOT} photometry with UVW2, UVM2 and UVW2 filters starting at redshift 1.1, 1.4 and 1.7 respectively.
We note that all the flares we reported for sources at z $>$ 1.1, and containing UV absorbed photometry, have the synchrotron high-energy drop which is well constrained by the optical photometry alone,
except B2 1520+031, and PKS 0805-07. For the former, the correction to UVM2 photometry ($\sim17\%$) will marginally change the drop in synchrotron, and thence the modeling  parameters.
This is not the case for PKS 0805-07. For this source, we evaluated that the effective IGM optical depth is about 0.15, 0.54, and 0.84 for the UVW1, UVM2 and UVW2 filters respectively, eventually
causing a change of the peak position of the synchrotron bump from optical to UV band. In the modeling we decided not to use the UV info, and we put the synchrotron peak emission in the UV
range. The direct consequence of this choice is that $d_{blob}$ is at $\sim$1 pc. In the limit case of no relevant correction for the IGM, the synchrotron peak is in the optical band,
and $d_{blob}$ at $\sim$10 pc.\\

\subsection{Spectral Variability}
\begin{figure*}
\centering
\begin{tabular}{cc}
\psfig{figure=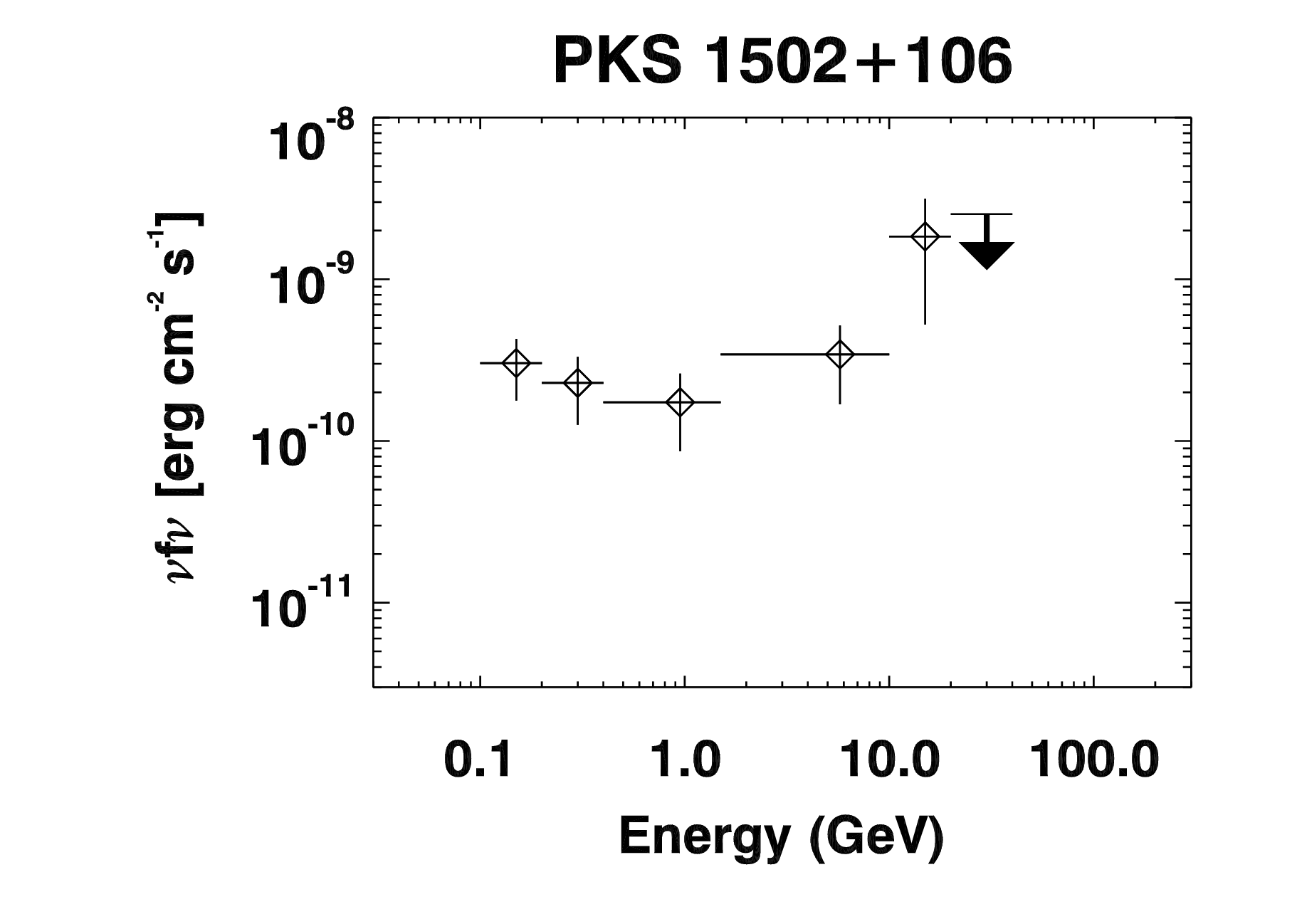,width=7.5cm,height=4.5cm}&
\psfig{figure=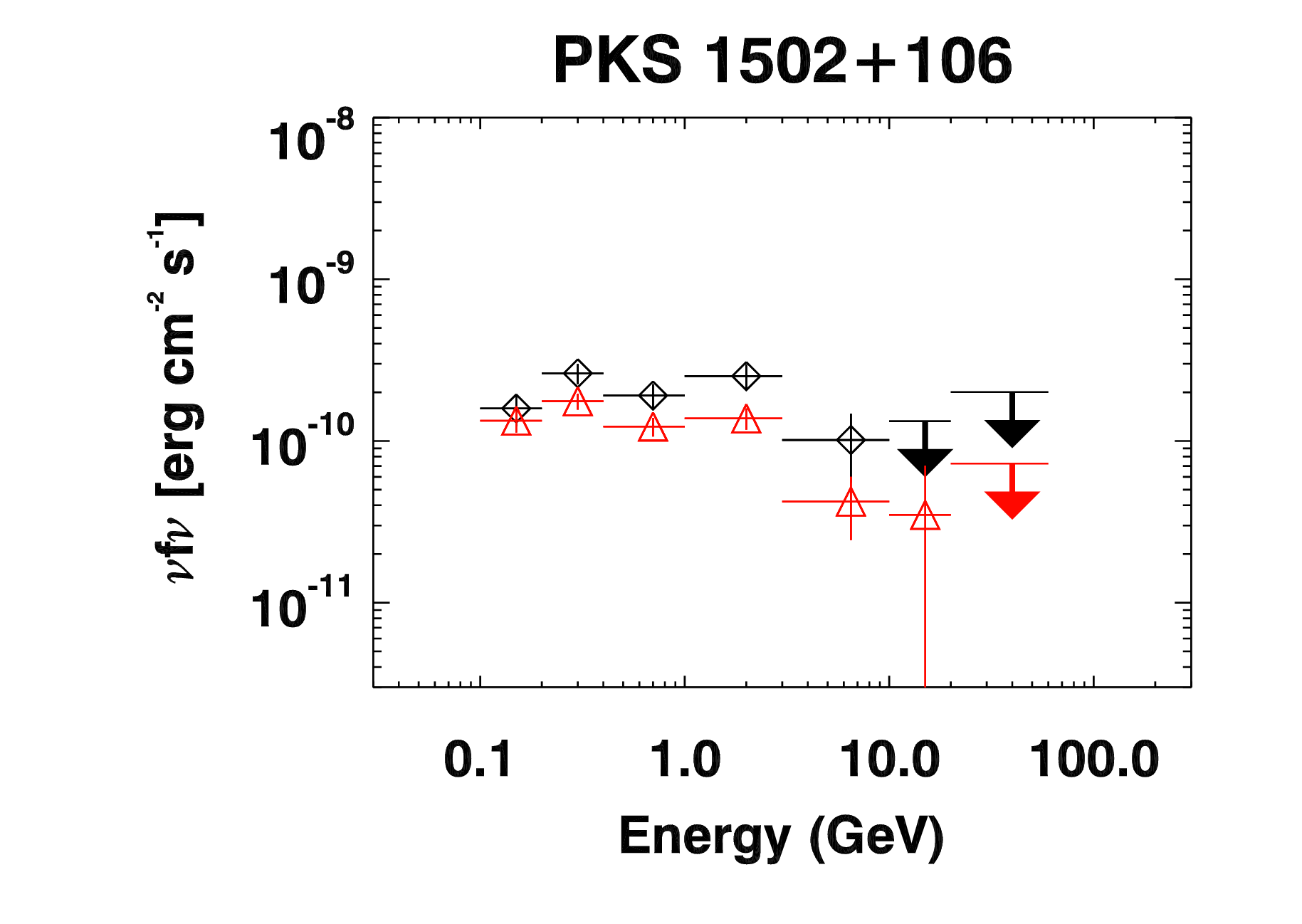,width=7.5cm,height=4.5cm}\\
\psfig{figure=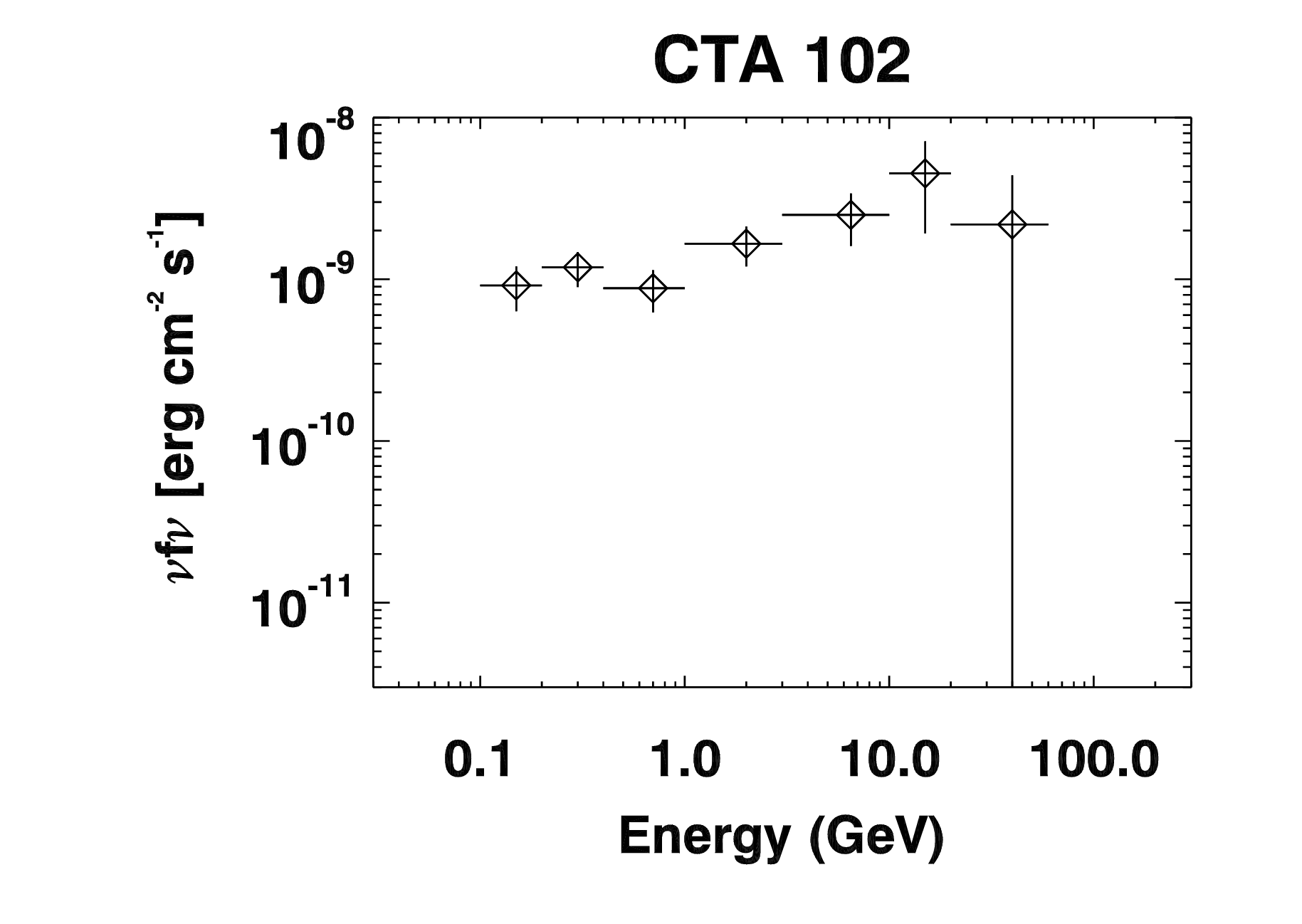,width=7.5cm,height=4.5cm}&
\psfig{figure=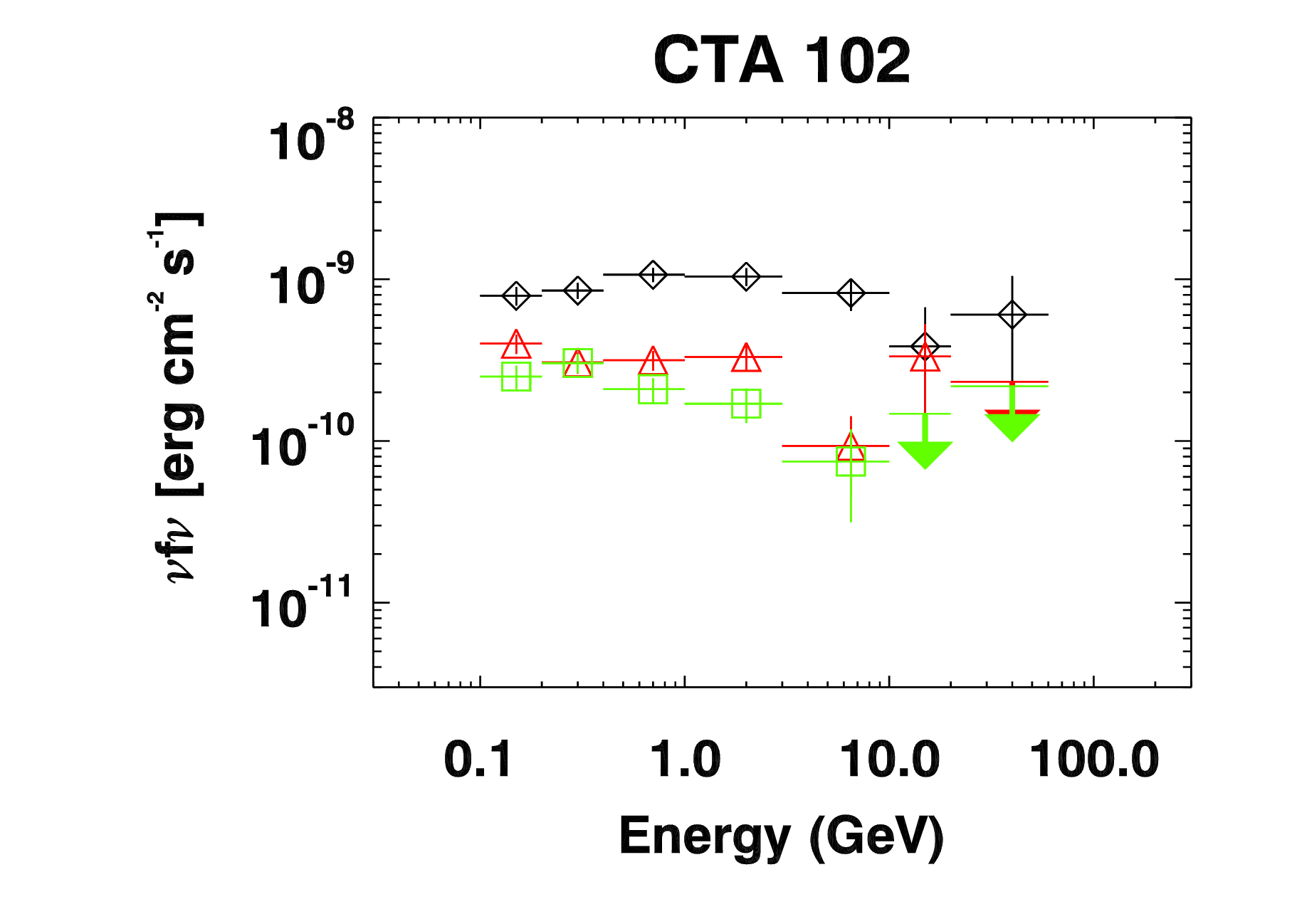,width=7.5cm,height=4.5cm}\\
\psfig{figure=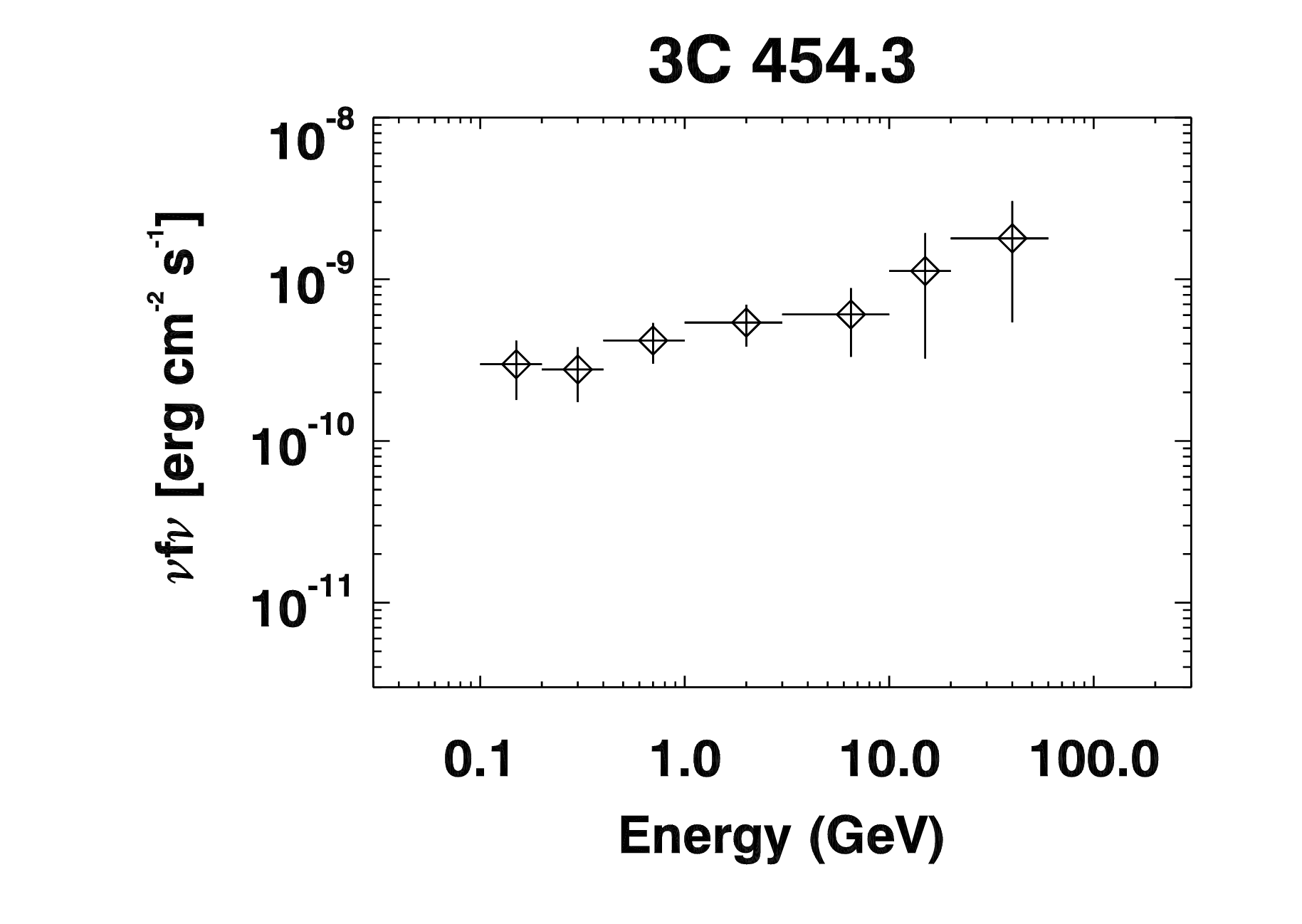,width=7.5cm,height=4.5cm}&
\psfig{figure=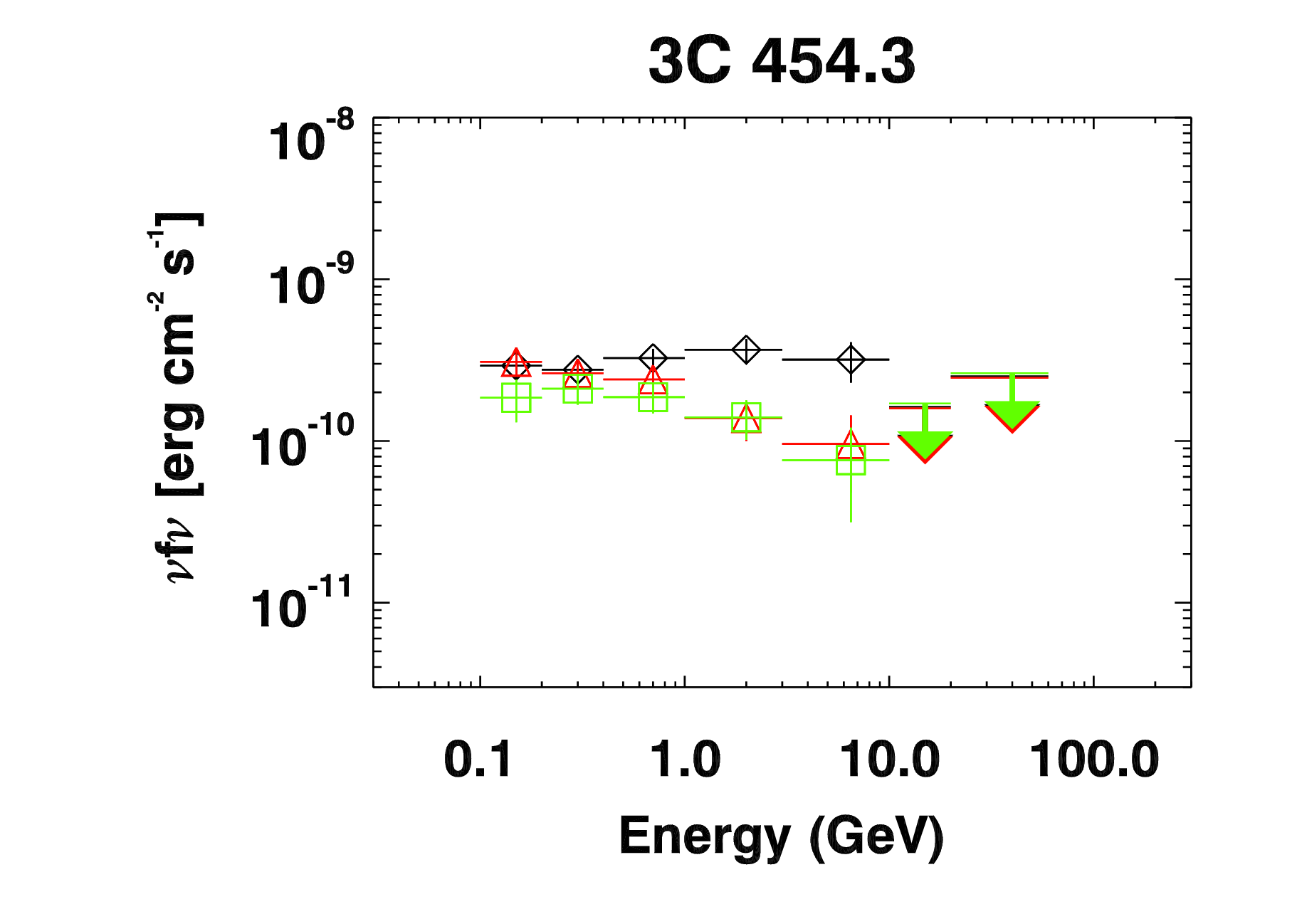,width=7.5cm,height=4.5cm}\\
\psfig{figure=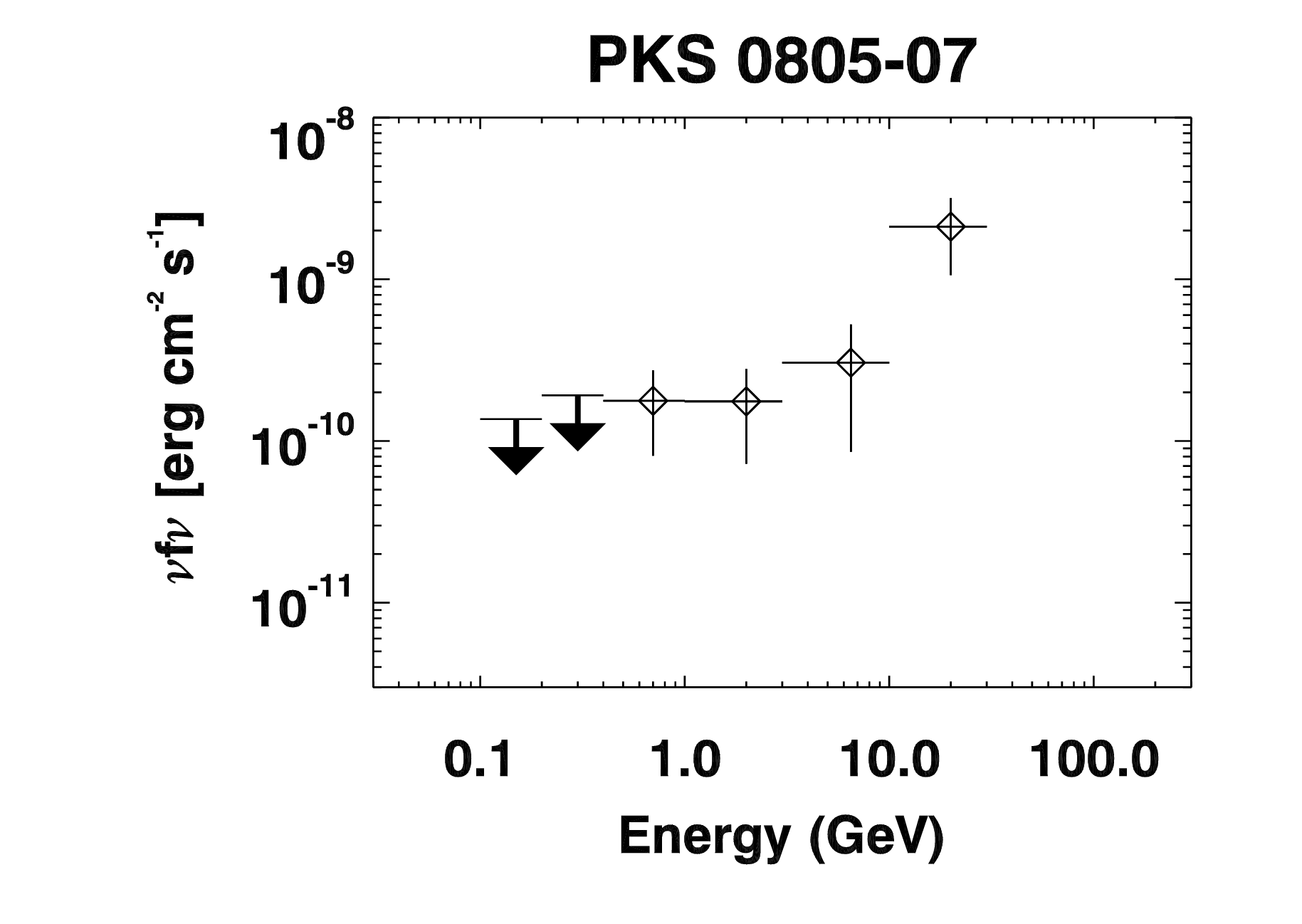,width=7.5cm,height=4.5cm}&
\psfig{figure=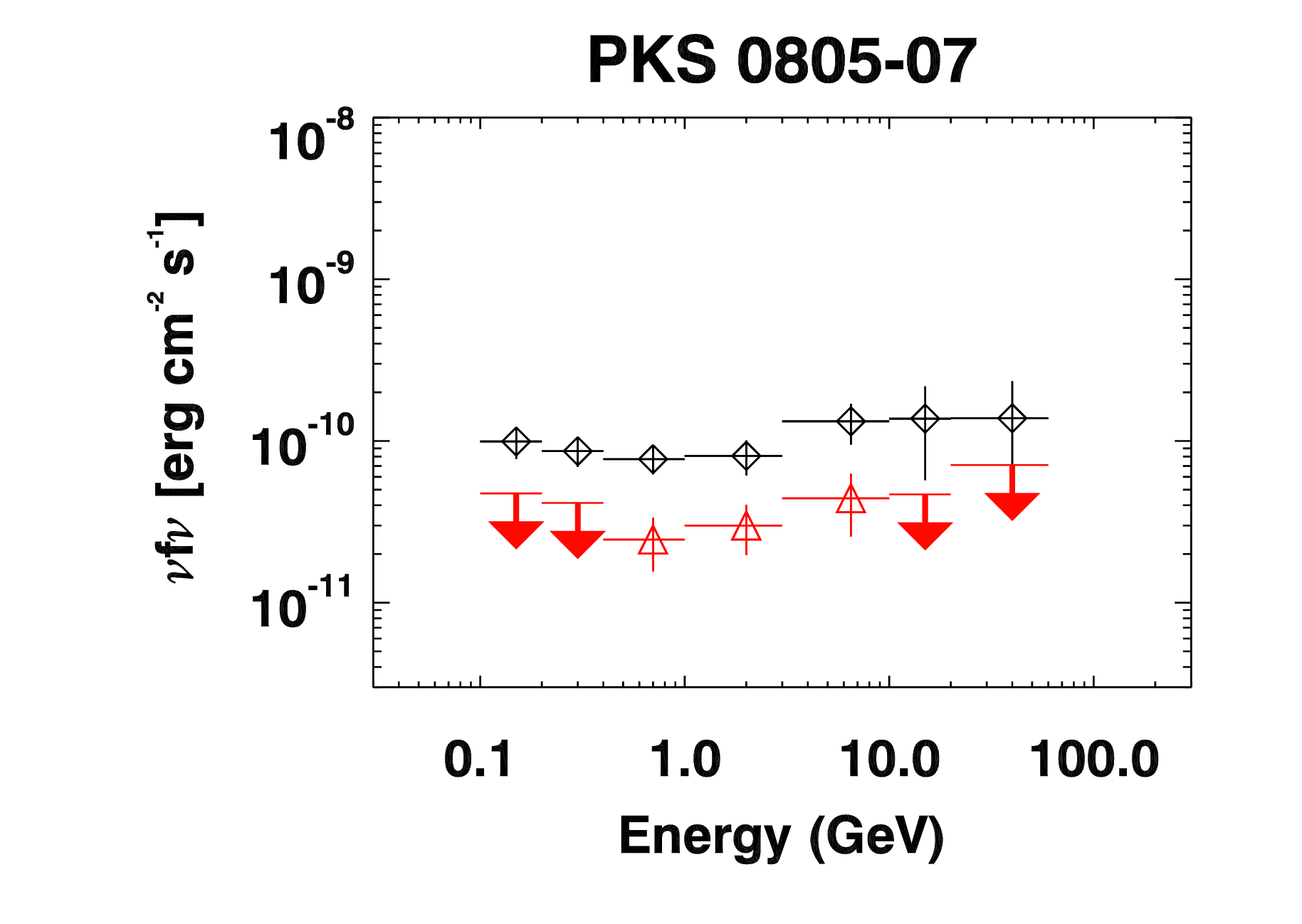,width=7.5cm,height=4.5cm}\\
\psfig{figure=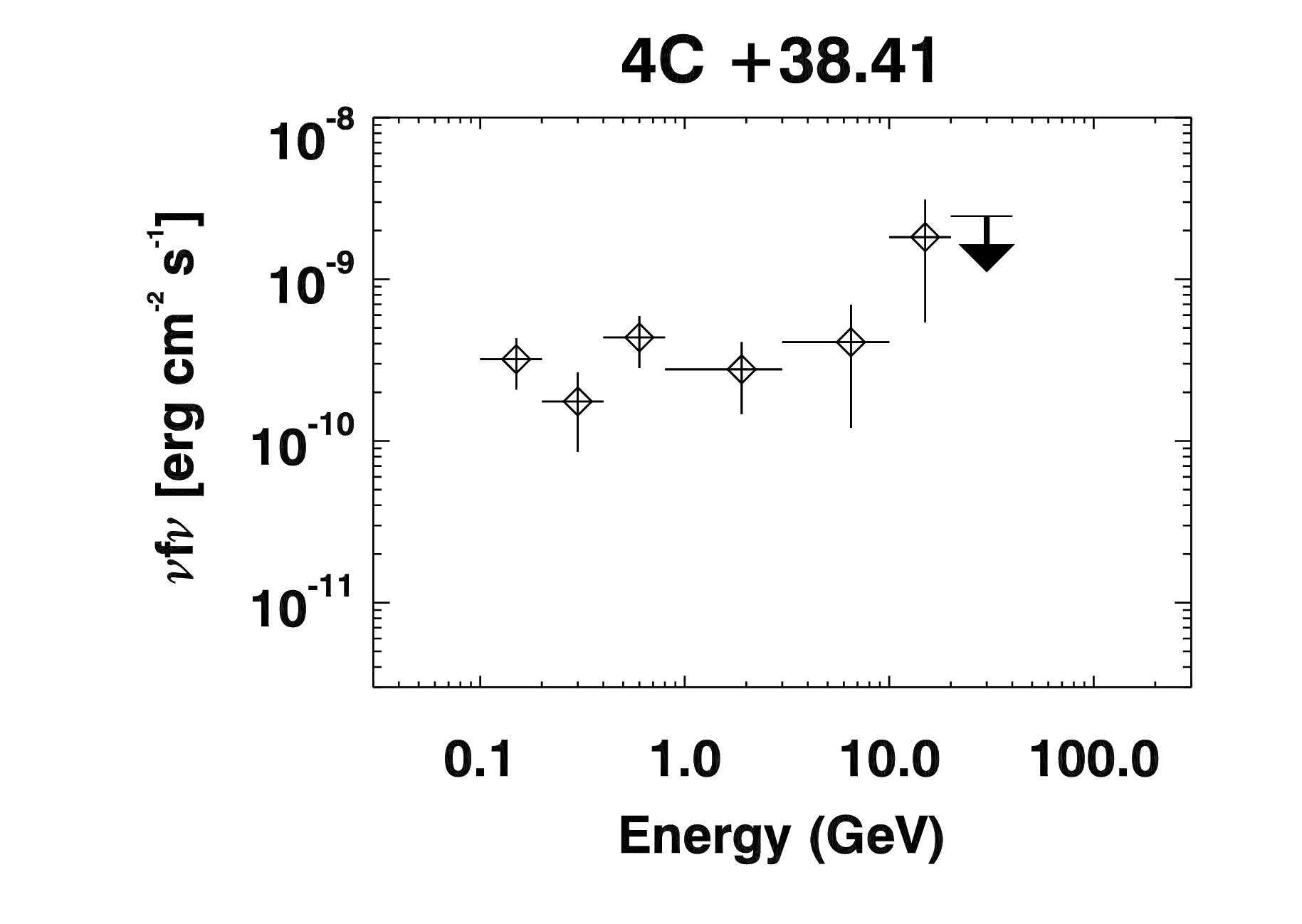,width=7.5cm,height=4.5cm}&
\psfig{figure=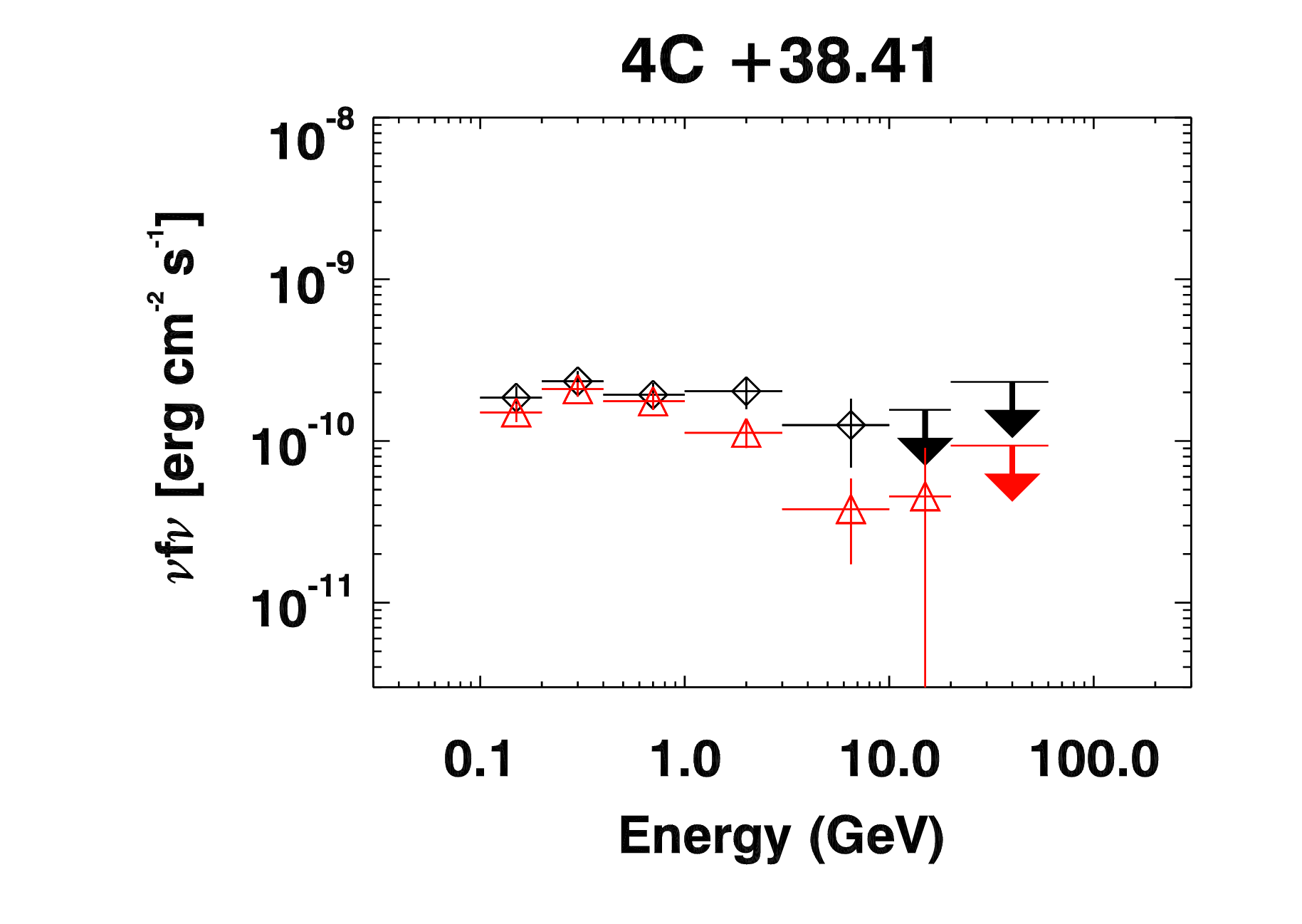,width=7.5cm,height=4.5cm}\\
\end{tabular}
\caption{Gamma-ray spectra for period A (left panels) and periods B, C, D (right panels) for the most powerful HE emitters, and for 4C +38.41.
Spectra for periods B, C, D have diamonds symbols, triangles, squares respectively (with black, red, and green colours respectively for the online version).}
\label{fig_period_a_b}       
\end{figure*}

For the most powerful HE flares (CTA 102, 3C 454.3, PKS 0805-07, PKS 1502+106)
 we tried to identify the portion of the flaring period (period A) showing the highest flux of energetic gamma-rays.\\
For the same sources we used the sub-sequent portions of the flaring period (period B, C, D) in order to investigate spectral differences, and possibly the spectral evolutions.
The integration times for period A and B of each source are reported in Table \ref{tab_period_a_b}.
For PKS 0805-07, and PKS 1502+106 we enlarged the integration time to 0.38 d in order to increase the statistics.
In the same Table we report
the number of HE gamma rays for period A,  and the chance probability for a bunch of gamma-rays to occur within the integration time
of period A, assuming that the mean gamma-ray rate is the rate of the whole activity period. We note however that we evaluated the chance probability
as a statistical fluctuation of a flat gamma-ray distribution within the whole activity period of the source. This is not the case for CTA 102. For this source, we evaluated
the mean gamma-ray rate not from the whole activity period, but from the 2.2 days in which the overall source flux was above 2$\times$10$^{-6}$ ph cm$^{-2}$ s$^{-1}$ (E $>$ 300 MeV).\\
The gamma-ray spectra for period A and B, C, D are shown in Figure \ref{fig_period_a_b}.
In the following fit procedures, we fit models to data starting from 200 MeV, in order to reduce systematics coming from the large {\it Fermi--LAT} PSF at lower energies. This choice have been applied
in \cite{hayashida}.\\
The spectral fit of period A with a powerlaw model for the sources give an hard photon index ($\Gamma_{ph}$),  reported in Table \ref{tab_period_a_b}.
We do not show fit with log-parabolic, broken--powerlaw, powerlaw with exponential-cutoff models,
because of the low statistics accumulated for period A for our sources:
there is no gain in the log-likelihood with respect to the powerlaw model, and fit do not constrain model parameters other than source flux and low--energy photon index ($\alpha$ for the log-parabola model).
For the bright flares of CTA 102 and 3C 454.3, we tried to establish if there was spectral evolution from period A to period B.
As first step, we statistically tested the hypothesis that the spectra of period A and B have the same shape.
To compare data of the two periods, we cannot obtain useful information by direct spectral comparison using the likelihood evaluated errors in each band.
If we restrict to emission below and above 10 GeV (LE and HE respectively), we can use the LE data to normalize the spectrum of
period B to the emission of period A, and compare the emission in HE band for the two periods. In HE band, we can disregard other sources, and galactic and extragalactic diffuse emission.
Thence we can use Poisson statistics to establish the probability P($shape_A=shape_B$) that a source with a given flux ratio (R$_f$=$\frac{F(E> 10\ GeV)}{F(0.1\div 10\ GeV)}$) could give rise to the observed
flux and counts during periods A and B. In this evaluation, we assume that F(0.1$\div$10 GeV) have Gaussian distribution.
We report in Table \ref{tab_period_a_b} the maximum value of P($shape_A=shape_B$) obtained varying R$_f$. 
In this evaluation, the exposure of the source in period A and B are taken into account.\\
As second step we made spectral fitting of period B with powerlaw,  broken powerlaw, log-parabola, powerlaw with exponential cutoff models.
Results are shown in Table \ref{tab_period_b_fit_complete}.
For both CTA 102 and 3C 454.3 the powerlaw model is disfavoured. The fit with all the other models succesfully constrain parameters. The fit with broken--powerlaw, powerlaw with exponential cutoff, and logparabola
models give hard low--energy photon index  ($\alpha$ for the log-parabola model) which  are consistent with the powerlaw photon index evaluated for period A and reported in Table \ref{tab_period_a_b}.\\

We could define the period A for 4C +38.41 on the basis of two HE photons only, it is reported in Table \ref{tab_period_a_b}. In this case the statistics prevent to build a statistically significant spectrum, and we enlarged the period A symmetrically on lower and higher temporal ends, to have a duration of 0.3 d. While the cumulative spectrum reported in Figure \ref{fig_sedall1} seems to show absorption
in the 1--10 GeV energy band with a drop of a factor $\sim$4 with respect to lower energy emission, the spectrum for period A does not show this feature (chance probability 1.9\%).
We remark that we made a selection above 10 GeV, thence the lack of absorption in 1--10 GeV  is not a biased result.
The drop above 1 GeV in the cumulative gamma-ray spectrum reported in Figure \ref{fig_sedall1} is not due to absorption features from the BLR, but by the integration of data on long timescales,
causing the overlap of the spectra of one or more periods like period A, and the spectra of period B.\\

The short duration of period A ($<$1d) for PKS 0805-07, CTA 102, and 3C 454.3 implies that the emission properties of the jet at the highest energies vary on timescales less than a day.

The most obvious candidate process for the spectral variability is the cooling of the high-energy electrons through the IC scattering with the IR photons.
Disregarding other emission mechanisms which are expected to give minor contribution to cooling at  parsec scale from the SMBH,
the expected variability timescale (as measured in the observer frame) due to the cooling of electrons with Lorentz factor $\gamma$, $t_{\rm obs}=t^{\prime}_{\rm cool}(1+z)/\delta$, can be written as:
\begin{equation}
t_{obs}=\frac{3m_ec\, (1+z)}{4\sigma_T U^{\prime}_{IR} \gamma \, \delta},
\end{equation}
\noindent
where the energy density of the IR torus emission is:
\begin{equation}
U^{\prime}_{IR}\simeq \frac{\tau_{IR}L_{disk}}{4 \pi R_{IR}^2c}\Gamma_{bulk}^2=2.5\times 10^{-2} \, \Gamma^2_1  \;\; {\rm erg \,\, cm^{-3}},
\end{equation}
\noindent
\noindent
in which $\tau_{IR}\sim0.6$ is the fraction of the disk luminosity reprocessed by the torus \citep{tristram2007,cleary2007,tristram2014}, and we used the scaling of Ghisellini \& Tavecchio (2009) to derive $R_{IR}$ in terms of $L_{disk}$. Inserting this into the previous equation we finally obtain:
\begin{equation}
t_{obs}=1.2\times 10^8 \frac{1+z}{\Gamma_1^2 \, \delta_1 \, \gamma} \;\; {\rm s}.
\end{equation}
\noindent
Considering an observed gamma-ray energy $E_{obs}$ and an IR photon field peaked at $\nu_{IR}$, the corresponding Lorentz factor of the emitting electrons can be derived as:
\begin{equation}
\gamma \simeq 3\times 10^{10}   \left (\ \frac{E_{obs}({\rm GeV})}{\nu_{IR}({\rm Hz})} \ \times \ \frac{1+z}{\Gamma_1 \, \delta_1} \ \right)^{1/2}.
\end{equation}
We therefore obtain for $E_{obs}$=30 GeV and $\nu_{IR}=3\times 10^{13}$~Hz:
\begin{equation}
t_{obs}= 4\times 10^3 \frac{(1+z)^{1/2}}{\Gamma_1^{3/2} \, \delta_1^{1/2}} \;\; {\rm s}.
\end{equation}
\noindent
The expected timescales, smaller than a day, are in agreement with, or even smaller than, the values reported in Table \ref{tab_period_a_b}.

\begin{table*}
 \centering
  \begin{tabular}{p{1.95cm}p{4.45cm}p{0.4cm}cp{1.3cm}p{1.7cm}p{0.25 cm}p{0.3cm}p{0.25cm}p{1.7cm}}
  \hline

   source   &\multicolumn{1}{c}{period A}     & $\Delta_t$      &\# HE           &  Chance        &    $\Gamma_{ph}$             &  \multicolumn{3}{c}{$\Delta_t$ (d)}   & Prob  \\ 
             &                                &  (d)            & photons        &  Prob.         &                         &  \multicolumn{3}{c}{for period}       & shape$_A$=shape$_B$  \\           
             &                                &                 &                &  (\%)$^{****}$      &    [0.2--10 GeV]        &    B  &  C  &   D                     &  (\%)\\ \hline    
PKS 1502+106    &     2009-05-06 05:20--2009-05-06 13:11 & 0.326  (0.38)$^{*}$ & 2 &  0.27/32.3      &  1.99$\pm$0.31   &    4         &  8         &           &     $<$3.4      \\
CTA 102         &     2012-09-22 18:12--2012-09-22 21:55 & 0.155  & 4 &  0.16/2.3$^{****}$   &  1.73$\pm$0.14   &    3         &  4$^{***}$         &  4 & 0.36 \\
3C  454.3       &     2013-09-24 15:00--2013-09-25 04:12 & 0.55   & 5 &   4.1/15.3      &  1.77$\pm$0.17 (1.84$\pm$0.08)$^{**}$ & 3 & 3    &  3     & $<$0.053\\ 
PKS 0805-07     &     2009-05-15 00:21--2009-05-15 08:26 & 0.337  (0.38)$^{*}$ & 4 &  0.028/0.82     &  1.51$\pm$0.34 (1.77$\pm$0.10)$^{**}$ &  8 & 8    &       & 0.97  \\ \hline
4C +38.41       &     2011-07-03 15:39--2011-07-03 18:56 & 0.136 (0.30)$^{*}$  & 2 &  0.038/4.1      &  1.85$\pm$0.23   &    4        &  8         &            & $<$1.4 \\
\hline

\end{tabular}
\caption{Integrations for period A and B as defined in the test for the most-powerful sources in the sample, and for 4C +38.41. We report also other integration periods (C,D) consecutive each other.
$\Gamma_{ph}$ is the photon index estimated fitting data of period A with a powerlaw model from 0.2 to 10 GeV. 
 The last column is the estimate of probability that period A and B  have the same spectral shape.
$^*$Chance probability is evaluated both for a bunch of Gamma-rays within the integration time of period A, assuming as mean rate the rate of the whole activity period, and for the same bunch of photons within the integration time and occurring during the whole activity period at HE. $^{*}$ for PKS 1502+106, PKS 0805-07 and 4C +38.41 we enlarged the period A symmetrically on lower and higher temporal ends, to have a duration of 0.38, 0.38 and 0.30 d, respectively.  Photon index is reported in the energy range 0.2 -- 10 GeV. $^{**}$photon index in the same energy range for the whole HE activity period is also reported between brackets for 3C 454.3 and PKS 0805-07. $^{***}$For CTA 102 there is a gap in the data, thence period C starts 5.08 d after the end of period B. $^{****}$To evaluate chance probability for CTA 102 we do not consider the whole activity period, but the period from 2012-09-21 21:36 to 2012-09-24 02:38, when the source was at the highest flux in Gamma-ray.}
\label{tab_period_a_b}
\end{table*}

\begin{table*}
 \centering

 \begin{tabular}{cccccrr}
  \hline

         &   \multicolumn{5}{c}{Powerlaw} \\ \hline \hline
Source   & F ($>$100 MeV)              &  $\Gamma_{ph}$     &                 &                  &  TS   \\
         & ($10^{-8}$ ph cm$^{2}$ s$^{-1}$)  &           &                 &                 &     \\ \hline
3C 454.3 & 222$\pm$ 24                 & 2.07$\pm$0.08 &                 &                 & 720.0  &  \\
CTA 102  & 565$\pm$ 37                 & 2.05$\pm$0.06 &                 &                 & 1760.2 &   \\ 

\\
\\
         &   \multicolumn{5}{c}{Broken powerlaw} \\ \hline \hline
Source   & F ($>$100 MeV)              &  $\Gamma^{LE}_{ph}$   &  $\Gamma^{HE}_{ph}$      & E$_{break}$      &  TS  & -2$\Delta L$\\
         & ($10^{-8}$ ph cm$^{2}$ s$^{-1}$)  &               &                 & (GeV)           &     \\ \hline
3C 454.3 & 205$\pm$ 24                 & 1.80$\pm$0.16 &  2.55$\pm$0.31  &  1.65$\pm$0.08   & 726.1   &  5.6 \\
CTA 102  & 536$\pm$ 37                 & 1.88$\pm$0.09 &  2.60$\pm$0.32  &  2.6$\pm$1.2     & 1768.9  &  7.8 \\ 

\\
\\

         &   \multicolumn{5}{c}{logparabola} \\ \hline \hline
Source   & F ($>$100 MeV)              &  $\alpha$   &  $\beta$      & E$_{break}$      &  TS & -2$\Delta L$\\
         & ($10^{-8}$ ph cm$^{2}$ s$^{-1}$)  &               &                 & (GeV)           &     \\ \hline

3C 454.3 & 203$\pm$ 24                 & 1.55$\pm$0.30 &  0.18$\pm$0.08  &  0.30$\pm$0.17     & 727.5  & 6.8\\
CTA 102  & 520$\pm$ 37                 & 1.66$\pm$0.18 &  0.13$\pm$0.05  &  0.30$\pm$0.12     & 1770.2 & 0. \\

\\
\\

         &   \multicolumn{5}{c}{powerlaw with exponential-cutoff} \\ \hline \hline
Source   & F ($>$100 MeV)              &  $\Gamma_{ph}^{LE}$   &  $\Gamma_{ph}^{HE}$      & cutoff         &  TS & -2$\Delta L$\\
         & ($10^{-8}$ ph cm$^{2}$ s$^{-1}$)  &               &                 & (GeV)           &     \\ \hline

3C 454.3 & 203$\pm$ 24                 & 1.66$\pm$0.18 &  1              &  4.6$\pm$2.2     & 729.6  & 9.0\\
CTA 102  & 540$\pm$ 37                 & 1.88$\pm$0.10 &  1              &  14$\pm$7        & 1767.1 & 1.6   \\

%

\end{tabular}
\caption{
Fitting parameters for Gamma-ray spectra of CTA 102 and 3C 454.3 for period B. Fitting are performed with powerlaw,  broken powerlaw, logparabola, powerlaw with exponential cutoff for energies above 200 MeV.
  $\Delta L$ is the difference of the log likelihood of the fit with respect to a single powerlaw fit.}
\label{tab_period_b_fit_complete}
\end{table*}

\subsection{Caveat}
An obvious problem of the cooling scenario (and, more generally, of the observed variability, see below) is that, assuming a single, homogeneous 
emission region, the observed variability timescales would be dictated by the much longer light-cross time $t_{cross}=R(1+z)/c\delta$, which in the most extreme cases is of the order of 1--30 days.
While (by construction) this timescale is in agreement with the  duration of the observed active phases (see Figure \ref{fig_lcall1}, \ref{fig_lcall2}, \ref{fig_lcall3}),
it is too long to explain the faster spikes in the gamma-ray light curve and the observed spectral variability.
A possibility to solve this problem is to abandon the one-zone framework and to assume, besides the large region, accounting for the long duration flare, other,
much smaller sub-regions responsible for the fast variability.  
Different realizations of this idea are implemented in the reconnection model \citep{giannios2013} and in the turbulence scenario \citep{marscher2013,narayan2012}.\\

In the magnetic reconnection framework, as presented in \cite{giannios2013}
for the case of a blob size larger than the typical reconnection site,
one can envisage the following scenario. Reconnection sites occurring in the
whole large emitting volume provide an almost stationary source of
high-energy electrons (randomly oriented envelope-plasmoids), whose integrated emission accounts for the
long-lasting high-states (this would correspond to the {\it slow-envelope}
emission in \citealt{giannios2013}).
The randomly oriented production of envelope-plasmoids troughtout the whole blob mimics the randomly oriented electron velocity distribution from the whole blob
in leptonic models. 
Occasionally, much more powerful
events produce small (compared to the whole blob) plasmoids whose emitted luminosity is comparable
or even larger than the envelope one, and oriented toward the observer ({\it monster-plasmoids}).
These events could explain the fast spikes in the light curve, and fast gamma-ray spectral variability that we showed for 4 sources.
Electron energy distribution is expected to be similar for envelope- and monster-plasmoids.\\

We just depicted a multi-zone scenario which, instead,  we modeled in the framework of single-zone homogeneous models, assuming large blob size ($\sim 1/10 R_{diss}$).
We will show that the model we provided accounts for the  envelope-emission (assuming electrons accelerated throughout the whole blob), even if we showed fast-spikes during HE activity periods:
{\it i)} the envelope- and monster-plasmoid will dissipate at the same $R_{diss}$, thence the  $\gamma\gamma$ opacity and the Klein-Nishina arguments apply simultaneously for the two-components of the model,
and the dissipation zone must be searched for at the outer edge of the BLR or beyond;
{\it ii)} in our SEDs we integrated gamma-ray data for long lasting periods including the envelope emission, and the fast spikes which are expected to have a similar intensity to the envelope emission,
and similar electron energy distribution;
{\it iii)} the X-ray, NIR, and optical-UV data where never collected during HE fast spikes. So, at most, we overestimated the gamma-ray spectra by a factor 2 for the envelope emission
(because envelope- and monster-plasmoid emission are expected to give similar intensities);
{\it iv)} a correction in SED modeling to account for the fast-spikes will eventually cause the Compton dominance to be lowered
for the envelope-emission,  and thence the $B/\Gamma_{bulk}$ to be raised of the same amount, and finally the product $\Gamma_{bulk}^5 \times R_{blob}^2$ to be lowered by a factor $\sim$2 in order to maintain the synchrotron and
SSC emission at the observed value. The last requirement could be accomplished with modest changes of $R_{blob}$ and/or $\Gamma_{bulk}$ (and thence of $R_{diss}$).\\

 A similar idea could be sketched for the turbulence scenario \citep{marscher2013,narayan2012}. Of
course this scheme needs to be quantified and specified, a task clearly beyond the aims of the present study.

\section{Discussion and Conclusion}
Our systematic study enlarges the sample of FSRQs showing evidence of  gamma-ray flares occurring outside the BLR.
The lack of the expected absorption signatures due to the interaction with the BLR radiation field constraints the dissipation region outside the cavity of the BLR for all the sources, and outside the core of the BLR
spherical shell for the 5 most powerful Broad Line emitters in our sample. This consideration does not depends on the emitting mechanism of FSRQs.
It is based on existing measurement of the Broad line luminosities, and slightly depends on the BLR geometry and emission model \citep{liu2006}.\\
In the framework of leptonic models for blazar emission, we showed that IC on BLR photon field is subject to Klein-Nishina suppression. The lack of this suppression, and of the $\gamma \gamma$ absorption in our spectra could only be explained assuming
a dissipation region at the outer edges of the BLR or further out.\\

Furthermore, the leptonic SED modeling locates, in some cases, the radiating zone at parsec scale, assuming the existence of the molecular torus for all the sources \citep{tristram2007,cleary2007,tristram2014}.
If this is not the case, the solution must be put just outside the broad line regions, because, in this hypothesis, the comoving intensity of the external radiation field drops rapidly with the distance of the moving source from the SMBH.\\
Once the region within the BLR cavity is excluded,
the location of the blazar zone is a stable result of leptonic SED modeling for solutions found within the core of the BLR (between inner and outer shell radii of the BLR) because of the rapidly varying
external radiation energy density $U_{blr}$ from the inner to the outer radius of the BLR.
The characterization of the blazar-zone as located outside the BLR is also a stable result of leptonic modeling, once the solution is found outside the external radius of the BLR.
But in this case, the refined localization depends on the chosen details in the modeling ($R_{blob}/R_{diss}$ ratio, the electron density, and  $\Gamma_{bulk}$), because the external radiation field $U_{IR}$  is constant
up to $\sim R_{IR}$.\\
Here we assumed the emitting (envelope) region to encompass the entire jet cross section, and a jet aperture angle $\theta_{\rm j}=r_{blob}/d_{blob}\simeq 0.1\approx 1/\Gamma_{\rm bulk}$. Recent VLBI observations (Clausen-Brown et al. 2013, Jorstad et al. 2005, Pushkarev et al. 2009) indicate slightly lower values, $\theta_{\rm j}\approx 0.2/\Gamma_{\rm bulk}$. However, even these lower aperture angles would imply in several cases jet cross section radius larger than $\sim 10^{17}$ cm (see Table \ref{tab_he_sed_model}), i.e. minimum variability timescales of several days.

As already noted, for several sources such long minimum variability timescales are inconsistent with the much shorter variability timescale inferred from the gamma-ray light-curves. This problem cannot be easily solved assuming that the region is located at smaller distance from the SMBH, since the variability timescale of $<1$ day can be reproduced only with radii of the order of $2.5\times10^{16}/(1+z)$ cm, which implies distances in several cases smaller than the BLR radius.
The same problem, i.e. the inconsistency between the observed variability timescales and those expected from the estimate of the source dimension,  has been encountered in several other blazars, both BL Lacs and FSRQs
(e.g. the BL Lac object Mrk 421, \citealt{gaidos1996}, or PKS 2155-304, \citealt{aharonian2007}).
For FSRQ the most extreme case is that of PKS 1222+216, which showed a doubling of the TeV flux in about 10 minutes \citep{aleksic}.\\

We point out that the location of the blazar zone in this work has been achieved within the framework of leptonic models, and we did not try other modeling such as the spine-sheath or hadronic ones.\\

We obtained periods of HE activity lasting from 1 d to about a month. The HE activity periods coincide with activity at lower energy. Within this period of activity we
have identified shorter periods with timescales of the order of 0.2--0.6 d characterized by brighter emission of HE gamma-rays.
Focusing on these short periods, a rather obvious consequence of our searching criterion is the relevant flux we derive above 10 GeV.
We indeed evaluated that chance probability of our searching method is very low for at least 2 flares 
(CTA 102 and PKS 0805-07) over 4 sources for which we can try this study. For the other two flares that we examined in details
(3C 454.3, and PKS 1502+106) the gamma-ray spectrum from 200 MeV to 10 GeV (i.e., below the energy threshold of our searching method) is rather hard.
Thence the hard gamma-ray spectra obtained below 10 GeV are not biased results.
They imply an energy spectral index $<$ 1, suggesting that the emission derives from a fresh (i.e. not cooled) population of electrons injected/accelerated in the source.

The spectral fit to gamma-ray spectra of periods B for the flares of CTA 102 and 3C 454.3 (Table \ref{tab_period_b_fit_complete}), when compared to periods A,
reveals a break, or a curvature, or a cutoff. The limited statistics does not allow us to discriminate among the different models. We note that the fit with a broken powerlaw  gives $\Delta \Gamma_{ph}$ $\sim$ 0.75$\pm$0.32 for 3C 454.3,
and 0.72$\pm$0.35 for CTA 102. One valuable option is that we are observing the progressive cooling of the fresh high-energy electrons, leading to a break with $\Delta\Gamma_{ph}=0.5$. The relatively long cooling time implied by the data supports the view that the emission occurs through the EC scattering in an environment with a low energy density of the target photons, such as that of the dusty torus.
The spectra for period A of the other sources show similar trend (with the exception of PKS 0805-07).
The gamma-ray spectra of periods C, D give a further hint of the ongoing cooling.
We note, however, that period A for CTA 102 corresponds to a fast flare in the whole {\it FERMI--LAT} energy band (not only at HE), as shown in figure \ref{fig_lcall3}.
This consideration makes it hard to correlate periods A and B for this source in terms of slow--cooling, with period A encompassing the whole development of the gamma-ray flare.\\
It's worth mentioning the gamma-ray spectrum reported in \cite{tanaka1222} for PKS 1222+216, obtained integrating FERMI--LAT gamma-ray data for 8 days around the fast TeV flare studied in \cite{aleksic}.
With this integration time, in the slow cooling dominated scenario, the cooled electron population dominates the gamma-ray spectrum. \cite{tanaka1222} reports $\Delta \Gamma_{ph} = 0.44 \pm 0.11$ and they invoke
the slow--cooling scenario for the TeV flare.\\  
Some of the gamma-ray  spectra integrated on long periods for 4C +38.41 and B2 1520+031 show peculiar structures, reminiscent of absorption features.
We performed the time-resolved study of the gamma-ray spectra of 4C +38.41 which showed the brighter HE flare.
The analysis revealed, instead, that there was at least one period in which the spectrum is flat (period A), followed by periods in which the spectrum is soft (period B). As above this behaviour could be interpreted as due to the injection and subsequent cooling of high-energy electrons in the emitting region.
The integrated spectra show both mechanisms in action,
resulting in absorption--like features. It is not always possible to time-resolve acceleration-- from cooling--dominated periods. In fact, we reported in Table \ref{tab_he_sample} that the HE gamma-rays
are emitted on long timescales, possibly causing the superpositions of accelerations and cooling phases.\\ 

We did not attempt with this study to establish if HE is a rare emission phase or not for FSRQs. From the sub-sample of 10 FSRQs reported here,
we triggered $\sim$30 HE flares in total, ranging from 1 to 8 per source within 5.3 years
of {\it FERMI--LAT} operations. We will address this study on the whole {\it FERMI--LAT} sample in a forthcoming paper.\\

Summing up, our findings picture a scenario in which, during HE flares, the emission from FSRQ occurs in regions distant from the central BH. Further, we find evidence for the likely presence of substructures, responsible for the observed relatively rapid variability (flux and spectrum). This framework is similar to that inferred from the observations of other FSRQ, especially during the emission of TeV photons. Among the possible theoretical scenarios advanced to explain such a phenomenology, we briefly discuss the reconnection model of \cite{giannios2013} and that invoking turbulence in the flow \citep{marscher2013, narayan2012}.
In the relativistic magnetic reconnection scenario envisaged in Giannios (2013), long term high states are thought as the ``envelope" of events of dissipation of magnetic field in the jet through reconnection. During these events there is the possibility that ``monster plasmoids" form, which accounts for the fast flares. The discussion in \cite{giannios2013} was tailored to the case of the TeV flare of PKS 1222+220, but the scenario should be applicable to the cases presented in our paper with small changes.

Turbulence in the flow has been invoked as a possible mechanism able to produce rapid flickering of the light-curve both for FSRQ \citep{marscher2013} and BL Lac objects \citep{narayan2012}. In this scheme, small cells of fluids characterized by fast turbulent speed can produce the rapid flares, while the long-term emission is produced by the larger active region encompassing the whole jet. In the \cite{marscher2014} model, the activity is triggered by the passage of the flow in a re-collimation shock, thought to form at parsec scale when the jet internal pressure drops below that of a confining medium. In this idea, the modulation of the injected power into the jet by the central engine accounts for the long-term evolution of the emission, while the emission from single turbulent cells produce the rapid flares. Simulation along the lines of those reported in \cite{marscher2014} could test if the scenario can reproduce the phenomenology shown by LAT.  \\

The evidence of an emitting region outside the BLR led to the obvious conclusion of an interaction of the emitted gamma-ray with the external radiation field of the IR torus. 
The consequent absorption would lead to a cut-off in the gamma-ray spectrum in the VHE (Very High Energy,  $E>100\,$GeV) range \citep{ghise_tav}, where the Imaging Cherenkov Telescopes (IACTs) systems operate.
The detection by IACTs of the FSRQs of our sample would allow to verify this hypothesis.    
The objects of this study have high redshift and the absorption by the optical and IR Extragalactic Background Light (EBL) has a non-negligible effect that must be taken into account.  
The unabsorbed fraction at 100 GeV spans from a factor 0.75 for PKS 2345-1555 at z=0.6 up to 0.1 at z=2. EBL absorption is more severe at 300 GeV, where most of the present generation of Cherenkov telescopes (MAGIC, HESS and VERITAS) have their highest sensitivity. At 300 GeV the unabsorbed fraction ranges from a factor 0.1 at z=0.6 up to $10^{-5}$ at z=2.
Therefore the possibility of detection in the VHE range relies on one side on instruments with the lowest energy threshold such as MAGIC \citep{Aleksic2012}, 
and on the other side on the high sensitivity of the Cherenkov telescopes of future generation, such as the Cherenkov Telescope Array (CTA, \citealt{actis2011}).

Assuming a sensitivity of $\sim  10^{-11}$\, erg/cm$^2$/s at 100 GeV for few hours of observations by present IACTs \citep{Aleksic2012} and an unabsorbed fraction of $\sim 0.1 - 0.01$, 
the extrapolation of the spectra shown in figure \ref{fig_period_a_b} shows that some of the sources with the hardest and highest flux, e.g. CTA~102, 3C~454.3, PMN~J2355-1555,
could be detected during a flaring episode, or could yield a sound upper limit constraining the cut-off. 
On the other hand the fast drop of the flux at higher energies would reduce the possibility to measure precisely the spectrum at few hundreds of GeV, which is the necessary condition to
disentangle the absorption of EBL from the internal absorption due to the IR radiation field.
To perform this measurement the existing IACTs should restrict to the nearest sources, while most distant ones will be the targets of the forthcoming CTA,
which promises a sensitivity up to an order of magnitude better than the present generation of Cherenkov telescopes.

\section*{acknowledgements}
L.P. and F.T. acknowledge partial financial contribution from grant PRIN-INAF-2011.
SMARTS observations of LAT monitored blazars are supported
by Yale University and Fermi GI grant NNX 12AP15G, and
the SMARTS blazar monitoring program is carried out by C.M.U.,
E.M., Michelle Buxton, Imran Hasan, J. I., Charles Bailyn and
Paolo Coppi. C.B., M.B. and the SMARTS 1.3m observing queue
receive support from NSF grant AST-0707627. We thank the Swift
team for the ToO observations. We acknowledge all agencies and
Institutes supporting the Fermi-LAT operations and the Scientific analysis
tools.
This research has made use of data obtained from the Chandra Data Archive and the Chandra Source Catalog,
and software provided by the Chandra X-ray Center (CXC) in the application packages CIAO, ChIPS, and Sherpa.
This publication makes use of data products from the Two Micron All Sky Survey, which is a joint project of the University of Massachusetts
and the Infrared Processing and Analysis Center/California Institute of Technology, funded by the National Aeronautics and Space Administration
and the National Science Foundation.
%


\label{lastpage}

\end{document}